\documentclass[longauth]{aa}
 
\usepackage[varg]{txfonts}
\usepackage{adjustbox,lipsum}

\usepackage{graphicx}   
\usepackage{amsmath} 
\usepackage{multirow}  

\begin{document}

\title{A few StePS forward in unveiling the complexity of galaxy evolution: light-weighted stellar ages of intermediate redshift galaxies with WEAVE}
\titlerunning{A few StePS forward in unveiling the complexity of galaxy evolution}

\author{L.~Costantin\inst{1,2}\thanks{luca.costantin@inaf.it}
\and A.~Iovino\inst{1}
\and S.~Zibetti\inst{3}
\and M.~Longhetti\inst{1}
\and A.~Gallazzi\inst{3}
\and A.~Mercurio\inst{4}
\and I.~Lonoce\inst{5}
\and M.~Balcells\inst{6,7,8}
\and M.~Bolzonella\inst{9}
\and G.~Busarello\inst{4}
\and G.~Dalton\inst{10,11}
\and A.~Ferr\'e-Mateu\inst{12,13}
\and R.~Garc\'ia-Benito\inst{14}
\and A.~Gargiulo\inst{15}
\and C.~Haines\inst{16}
\and S.~Jin\inst{10,11,17}
\and F.~La~Barbera\inst{4}
\and S.~McGee\inst{18}
\and P.~Merluzzi\inst{4}
\and L.~Morelli\inst{16}
\and D.~N.~A.~Murphy\inst{19}
\and L.~Peralta de Arriba\inst{19}
\and A.~Pizzella\inst{20,21}
\and B.~M.~Poggianti\inst{21}
\and L.~Pozzetti\inst{9}
\and P.~S\'anchez-Bl\'azquez\inst{22}
\and M.~Talia\inst{9,23}
\and C.~Tortora\inst{3}
\and S.~C.~Trager\inst{17}
\and A.~Vazdekis\inst{7,8}
\and D.~Vergani\inst{9}
\and B.~Vulcani\inst{21}
}

\institute {INAF - Osservatorio Astronomico di Brera, via Brera 28, I-20121 Milano, Italy
\and Centro de Astrobiolog\'ia (CSIC-INTA), Ctra de Ajalvir km 4, Torrej\'on de Ardoz, E-28850 Madrid, Spain
\and INAF - Osservatorio Astrofisico di Arcetri, Largo Enrico Fermi 5, I-50125, Firenze, Italy
\and INAF - Osservatorio Astronomico di Capodimonte, Salita Moiariello 16, I-80131 Napoli, Italy
\and Department of Astronomy and Astrophysics, University of Chicago, Chicago, IL 60637, USA
\and Isaac Newton Group of Telescopes, Apartado 321, 38700 Santa Cruz de La Palma, Canary Islands, Spain
\and Instituto de Astrof\'isica de Canarias, Calle V\'ia L\'actea s/n, E-38200 La Laguna, Tenerife, Spain 
\and Departamento de Astrof\'isica, Universidad de La Laguna, Calle Astrof\'isico Francisco S\'anchez s/n, E-38205 La Laguna, Tenerife, Spain
\and INAF - Osservatorio di Astrofisica e Scienza dello Spazio di Bologna, via Gobetti 93/3, 40129 Bologna, Italy
\and Department of Physics, University of Oxford, Keble Road, OX1 3RH, UK
\and STFC-RALSpace, Rutherford Appleton Laboratory, Didcot, OX11 0QX, UK
\and Institut de Ciencies del Cosmos (ICCUB), Universitat de Barcelona (IEEC-UB), E-02028 Barcelona, Spain
\and Centre for Astrophysics \& Supercomputing, Swinburne University of Technology, Hawthorn, VIC 3122, Australia
\and Instituto de Astrof\'isica de Andaluc\'ia (CSIC), P.O. Box 3004, 18080 Granada, Spain
\and INAF - IASF Milano, via Bassini 15, I-20133 Milano, Italy
\and Instituto de Astronom\'ia y Ciencias Planetarias de Atacama, Universidad de Atacama, Copayapu 485, Copiap\'o, Chile
\and Kapteyn Astronomical Institute, University of Groningen, Postbus 800, NL-9700 AV Groningen, The Netherlands
\and School of Physics and Astronomy, University of Birmingham, Edgbaston, Birmingham, B15 2TT, UK
\and Institute of Astronomy, University of Cambridge, Madingley Road, Cambridge CB3 0HA, UK
\and Dipartimento di Fisica e Astronomia ``G. Galilei'', Universit\`a di Padova, vicolo dell'Osservatorio 3, 35122, Padova, Italy 
\and INAF - Osservatorio Astronomico di Padova, vicolo dell'Osservatorio 5, I-35122 Padova, Italy
\and Departamento de F\'isica Te\'orica, Universidad Aut\'onoma de Madrid, E-28049 Cantoblanco, Spain
\and University of Bologna - Department of Physics and Astronomy, via Gobetti 93/2, 40129, Bologna, Italy
}

\vspace{4cm}

\abstract 
{The upcoming new generation of optical spectrographs on four-meter-class
telescopes, with their huge multiplexing capabilities, excellent spectral
resolution, and unprecedented wavelength coverage, will provide invaluable 
information for reconstructing the history of star
formation in individual galaxies up to redshifts of about 0.7.}
{We aim at defining simple but robust and meaningful physical parameters that can
be used to trace the coexistence of widely diverse stellar components:
younger stellar populations superimposed on the bulk of older ones.}
{We produce spectra of galaxies closely mimicking data from the
forthcoming Stellar Populations at intermediate redshifts Survey (StePS), 
a survey that uses the WEAVE spectrograph on the William Herschel Telescope. 
First, we assess our ability to reliably measure
both ultraviolet and optical spectral indices in galaxies of different spectral
types for typically expected signal-to-noise levels. Then, we analyze such mock spectra
with a Bayesian approach, deriving the probability density function of
$r$- and $u$-band light-weighted ages as well as of their difference.}
{We find that the ultraviolet indices significantly narrow
the uncertainties in estimating the $r$- and $u$-band light-weighted ages
and their difference in individual galaxies. These diagnostics, 
robustly retrievable for large galaxy samples even when observed at moderate 
signal-to-noise ratios, allow us to identify secondary episodes
of star formation up to an age of $\sim 0.1$ Gyr for 
stellar populations older than $\sim 1.5$ Gyr, pushing up to an age of $\sim 1$ Gyr
for stellar populations older than $\sim 5$ Gyr.}
{The difference between $r$-band and $u$-band light-weighted ages is shown to be a powerful diagnostic
to characterize and constrain extended star-formation histories and the presence 
of young stellar populations on top of older ones.
This parameter can be used to explore the interplay 
between different galaxy star-formation histories and
physical parameters such as galaxy mass, size, morphology, and
environment.}


\maketitle



\section{Introduction \label{sec:introduction}}

In the $\Lambda$ cold dark matter paradigm, galaxies present different physical properties as
the direct consequence of the multiplicity of pathways for their formation and evolution.
Photometric and spectroscopic information is widely used
to trace the stellar mass content and its assembling mechanisms 
\citep{PerezGonzalez2008, Thomas2010, Davidzon2017} and to characterize substructures in nearby galaxies 
\citep[i.e., bulges, disks, bars;][]{MendezAbreu2012, deLorenzoCaceres2013, 
Morelli2015, Costantin2017, Costantin2018, MendezAbreu2018},
allowing for insightful comparisons with predictions from numerical simulations
in a cosmological context \citep{Nelson2015, Schaye2015}.
The evolution of observed galaxy properties 
as a function of time contains clues on 
how different channels of evolution 
have affected the hierarchical growth of their stellar mass
and which environmental effects have shaped their star-formation history 
\citep[SFH;][]{Poggianti2009, FerreMateu2014, LaBarbera2014, Guglielmo2019}.
Thus, studying the evolution of observed galaxy properties allows us to explore 
the processes driving the assembly history of galaxies
and testing the predictions 
of theoretical models and numerical simulations.

In the last decades, the study of galaxies in the local Universe 
has greatly enriched our knowledge and understanding.  
Theoretical and empirical approaches are now anchored by the large, uniform, 
and complete spectroscopic measurements of the local Universe from the 
Sloan Digital Sky Survey \citep[SDSS;][]{York2010}.
Observations of colors, morphology, spectral type, 
and star formation of galaxies show a clear bimodal distribution, 
where blue star-forming late-type galaxies are separated
from red quiescent early-type galaxies \citep{Kauffmann2003, Blanton2003, Baldry2004}. 
Many observations have shown that galaxy stellar mass is one of the 
most fundamental quantities that enters to predict these different galaxy 
properties, with a relatively minor role being played by environment 
\citep{Kauffmann2004, Baldry2006, BlantonMoustakas2009, Bamford2009}.
The bimodality is also visible in spatially resolved observations 
of nearby galaxies \citep{Zibetti2017, LopezFernandez2018}, 
suggesting its local and structural origin within galaxies. 
This marked separation of two galaxy populations persists at higher redshifts, 
where the relative importance 
of the two peaks of the bimodality changes, 
as a consequence of the continuing decline
of star formation and the ensuing galaxy migration from the blue, 
star-forming to the red, passive galaxies sequence 
\citep{Haines2017}. Galaxy stellar mass is still the dominant driving factor and 
environment plays a secondary role \citep{Iovino2010, Peng2010, Hahn2015}.

In this context, a key question remains whether galaxies that quenched at early epochs 
remained passive since then or whether they experienced further star formation episodes
(the so-called rejuvenation of galaxies).  
Indeed, a totally passive evolution is not entirely consistent with currently available observations, 
especially for massive galaxies. 

A puzzling observational result keeps emerging from the different data, 
suggesting that even the most massive and apparently passive galaxies do not lead an undisturbed 
evolution after their star formation has stopped, but may have experienced new star formation episodes. 
At high redshifts ($z\sim3$) and towards the more recent past, semi-analytical models as well as observations 
from CANDELS and GAMA tell us that 31\% of quiescent galaxies have experienced at 
least one rejuvenation event \citep{Pandya2017}.
Stacked spectra of red galaxies at $z \sim 0.9$ display ages too old to be connected 
by simply passive evolution to local SDSS galaxies \citep{Schiavon2006}. 
Also, the purely passive evolution of individual massive passive galaxies at $z\sim 0.7$ 
would result in a tiny spread in the present-day age distribution, which is inconsistent with 
the observed age distribution of local massive passive galaxies \citep{Gallazzi2014}. 
In the redshift interval $0.6<z<1.0$, the survey LEGA-C \citep{vanderWel2016} is, as of today, 
the best suited to trace SFH in individual galaxies, 
targeting  $\sim 3000$ spectra of signal-to-noise ratio $SNR \gtrsim 10$ $\AA^{-1}$. 
LEGA-C observations confirm that ages of massive galaxies ($M > 10^{11}$ M$_{\odot}$) at $z \sim 0.8$  
are inconsistent with those of their local counterparts \citep{Wu2018, Spilker2018}. 
In the LEGA-C sample of quiescent $z \sim 0.8$ galaxies 
a fraction of $\sim 16\%$ have returned to the star-forming 
sequence in the interval $z \sim 0.7 - 1.5$, after having reached 
quiescence at some earlier time \citep{Chauke2019}.
	
As the evidence for episodes of rejuvenation within the passive galaxy sample grows, 
suggesting that the path to quiescence is not such an undisturbed one, 
the possible mechanisms at work and their origin (internal, external, a mixture of both) 
remain a matter of discussion. 
A possible solution invokes continuous residual star formation in individual quiescent galaxies, 
where a minority of young stars add over a base of an old stellar population \citep{Trager2000}, 
but external mechanisms (either HI gas accretion or mergers which bring in new gas) 
have also been proposed \citep{Kaviraj2009}.
In this context, the intermediate redshift range ($0.3 \lesssim z \lesssim 0.7$) 
offers an interesting niche of investigation: 
the span in cosmic time covered up to $z\sim0.7$ is nearly half the age of the Universe, 
enabling the direct observation of galaxies over a significant and continuous 
interval of their evolutionary life. 
In this redshift range, a coarse estimate of the morphological type is possible 
with data from ground-based telescopes \citep{Krywult2017};
the fraction of galaxies located in structures like groups progressively 
rises as expected in a hierarchical structure formation scenario \citep{Knobel2009}, and 
the increase of number of red massive passive galaxies 
is significant down to $z\sim 0.5$ \citep{Gargiulo2017, Haines2017}. 
However, the intermediate redshift regime is still largely unexplored by surveys 
of sufficient spectral quality, 
lying between the redshift ranges covered by SDSS and LEGA-C data. 

The upcoming new generation of spectrographs at four-meter-class telescopes 
with their huge multiplexing, extraordinary collecting capabilities,
and unprecedented wavelength coverage, offer a good opportunity to fill this observational 
gap and to provide spectral data comparable in quality to those obtained for the nearby Universe. 
A non-negligible advantage offered by the wide wavelength coverage 
(typically $3600 \lesssim \lambda \lesssim 9500$ \AA) is that at $z > 0.3$ 
the near-ultraviolet region of galaxy spectra enters the observed window. 
This is a region where low-level ongoing star formation, 
that leaves only weak imprints in the optical galaxy spectra, may be traced unambiguously \citep{Vazdekis2016}. 

In this paper, we focus on the new wide-field spectroscopic facility for the 4.2m 
William Herschel Telescope (WHT) in the Canary Islands, 
WEAVE \citep[][Jin et al. \emph{in prep.}]{Dalton2012}, and on the project StePS 
(Stellar Population at intermediate redshift Survey; Iovino et al., \emph{in prep.}), 
one of the eight surveys that will be carried out 
during the first five years of WEAVE operations (starting in 2020). 
StePS aims to obtain high resolution ($R\sim$5\,000) moderate quality 
($SNR$ $\gtrsim$ 10 \AA$^{-1}$) spectra of $\sim$25\,000 galaxies in the redshift range $0.3-0.7$, 
thus providing reliable measurements of the 
absorption features in the stellar continuum 
for a statistically large sample of galaxies (roughly ten times larger than LEGA-C).  
The targets are selected simply by magnitude (I$_{AB} < 20.5$ mag) and photometric 
(spectroscopic when available) redshift at $z > 0.3$, thus filling 
the crucial range between SDSS and LEGA-C data sets. 
The galaxy mass targeted by StePS (computed assuming the \citealt{Chabrier2003} 
initial mass function) ranges from $M \sim10^{10.2}$ M$_{\odot}$ 
at $z=0.3$ to $\sim$10$^{11}$ M$_{\odot}$ at $z=0.55$  
and $\sim$10$^{11.3}$ M$_{\odot}$ at $z=0.7$, the massive tail of galaxy mass distribution.

The main goal of this paper is to show how the values of light-weighted 
stellar ages in the two photometric SDSS $u$ and $r$ bands, and their difference, are simple but 
efficient tools for unveiling the presence of a younger stellar population coexisting with the bulk of an older one. 
To achieve this purpose, StePS expected performances are tested using realistic simulations at different redshifts and $SNR$. 
The main novelty is the use of ultraviolet indices together with more classic optical ones in the 
context of a full Bayesian analysis to infer simple but meaningful physical 
properties of past SFH of galaxies, the so-called archaeological approach.  
In our simulations, we use spectral data only 
and focus on the use of spectral indices as opposed to the full spectral fitting type of analysis. 
Spectral indices offer the advantage that one can select highly informative features to constrain 
stellar populations parameters, attempting to break the 
degeneracies among them using information
defined in small portions of the galaxy spectrum.
Our templates include only galaxy absorption features, assuming that the emission-line contribution 
has been effectively removed from our spectra.
This is a sensible choice given our main science goal, 
that is the detection of past star-formation episodes 
using spectral indices information for those galaxies where the emission-line contribution has already faded.

The paper is organized as follows. In Sect.~\ref{sec:models} we describe 
the stellar population models and the role of light-weighted ages on different photometric bands
in retrieving information on recent events of SFH in galaxies.
In Sect.~\ref{sec:classical} we describe the possibility to infer differences
in light-weighted ages on different photometric bands using information
from pairs of individual optical and ultraviolet indices.
In Sect.~\ref{sec:observations} we describe a rigorous  
method to build realistic observations, which closely 
mimic spectra that will be observed by WEAVE for StePS. 
In Sect.~\ref{sec:SP} we present our ability 
to measure spectral indices through the whole spectral range 
and the results of our Bayesian analysis.
In Sect.~\ref{sec:conclusions} we summarize our results 
and their implications in StePS and StePS-like analyses,
giving our conclusions.
We adopt $H_0=69.6$ km s$^{-1}$ Mpc$^{-1}$, $\Omega_M=0.286$, 
and $\Omega_{\Lambda}=0.714$ as cosmological parameters throughout this work \citep{Wright2006}.


\section{Stellar populations models \label{sec:models}}

In this work, spectral models of stellar populations play a two-fold crucial role.
On the one hand, they are used as the base ingredient to generate mock StePS-like observations, 
with a realistic wavelength coverage and noise behaviour together with 
a full variety of SFHs, chemical evolution histories, and
dust attenuation values, that well reproduce the complexity of observed galaxy spectra.
On the other hand, a comprehensive library of such models is required to derive the physical
parameters of interest from such mock observations via the Bayesian statistical tools described  
in Sect.~\ref{subsec:bayes} \citep[see also][]{Gallazzi2005,Zibetti2017}.
Note that the model libraries employed for these two purposes need not be the same. However,
we opt to have them stem from the same ``parent library'' in order to factor out all
possible systematic effects arising from the base Simple Stellar Population (SSP) model, and
the mathematical prescriptions for SFHs, chemical enrichment, and dust treatment.
These libraries are described in Sect.~\ref{subsec:models}.

We focus our analysis on mean stellar ages weighted on the light in different photometric bands,
in particular the SDSS $r$ and $u$ bands. We define the parameter \texttt{$\Delta$(age)} as the
difference between these two light-weighted ages and propose it as a simple tool to detect the 
presence of a younger stellar population coexisting with the bulk of an older one. 
We motivate this choice by using basic stellar physics arguments, 
as well as by showing how the overall general properties of the model SFHs 
correlate with \texttt{$\Delta$(age)}, irrespective  
of metallicity and dust content.
As we will see in more detail in Sect.~\ref{sec:SP}, marginalizing 
over metallicity and dust content broadens the probability density 
functions (PDFs) of the observed values of light-weighted ages and \texttt{$\Delta$(age)}, 
but without introducing major systematic offsets.
In this work, the model library used does not include emission lines.  
We focus on the stellar content of galaxies, which can be
accurately described once emission lines are properly modeled 
and subtracted from observed data using standard fitting algorithms,
such as Gas AND Absorption Line Fitting \citep[\texttt{GANDALF};][]{Sarzi2006}.

\subsection{Modeling the complex SFH of galaxies \label{subsec:models}}

Despite recent advances in our understanding of the star-formation histories of galaxies, 
thanks both to improved stellar population inference techniques and to more
detailed simulations of galaxy evolution, we still face a fundamental ignorance
about the detailed shape of a galaxy's star-formation and chemical-enrichment histories,
which stems from the huge diversity and stochastic nature of the physical mechanisms from which
they are affected. Therefore, in trying to model
the SFHs of galaxies, we pursue the goal of covering the observable parameter space
with as many as possible different forms and with different chemical-enrichment histories as well as
dust attenuation properties, in order to take parameter degeneracy
properly and fully into account. Our final aim, however, 
is not to extract from the (simulated) observations the full complexity 
that enters in the building of these models, but to focus on simple parameters 
and their relationship with basic but fundamental properties of galaxy SFHs.
In particular, we focus on light-weighted ages in different bands and their differences, 
and marginalize over the ``details'' on which they depend. 
The width of their posterior PDF will give back the actual ability 
of constraining these parameters with a realistic, StePS-like dataset.

We base our work on the comprehensive library of $500\,000$ models
introduced by \citet{Zibetti2017}, which forms our
``parent library''. 
Each model is characterized by a SFH, a chemical enrichment history, and a
two-component dust attenuation prescription.
The base models for our libraries are the \citet{Bruzual2003} 
SSP models, in the 2016 revised version, which adopt the \citet{Chabrier2003} 
initial mass function, updated evolutionary tracks \citep{Girardi2000, Marigo2013} and the MILES stellar spectral library 
\citep[][2.5 $\AA$ FWHM resolution]{SanchezBlazquez2006, FalconBarroso2011}, extended in the ultraviolet 
($911 \lesssim \lambda \lesssim 3500$ $\AA$) 
and the theoretical high-resolution ($\text{FWHM}=1$ \AA) models of \cite{Martins2005}. 

\begin{figure}[t]
\centering
\includegraphics[scale=0.35, trim=0cm 0cm 0cm 0cm, clip=true]{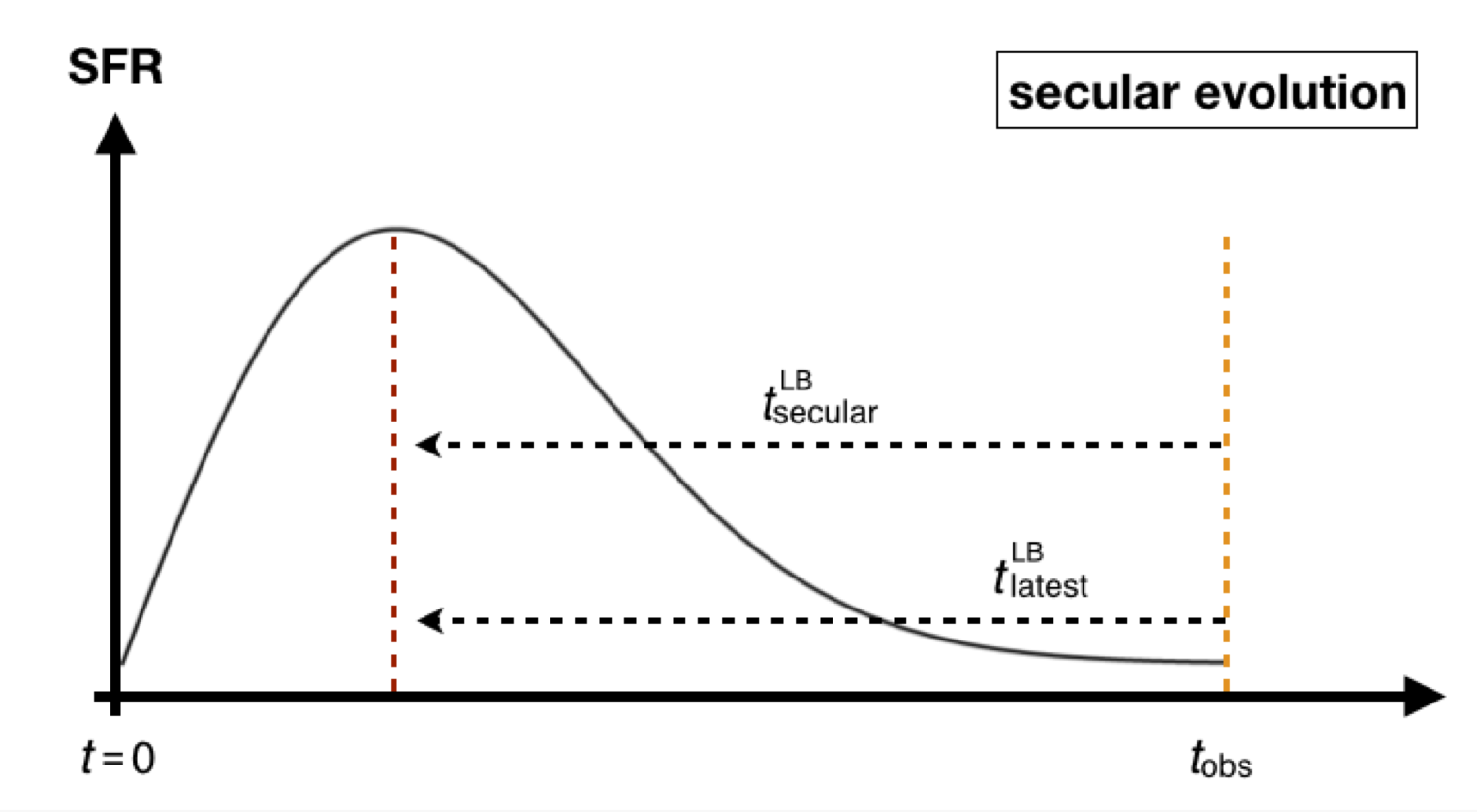}
\caption{Sketch of secular SFH of a galaxy that does not display 
a secondary burst event.
The orange dashed line corresponds to the time from the beginning of
the SFH ($t=0$) to the time at which the galaxy was observed ($t_{\rm obs}$), 
while the red dashed line corresponds to the peak of the secular SFH. 
Since the galaxy experienced no burst events, the look-back time to the peak of the secular SFH 
coincides with the look-back time to the latest peak of star formation 
($t^{\rm LB}_{\rm secular}$=$t^{\rm LB}_{\rm latest}$).}
\label{fig:SFH_sketch_noburst}
\end{figure}

The SFHs are modeled as a superposition of a continuous or ``secular'' component
and stochastic bursts. 
The secular component is described by a \cite{Sandage1986} law:
\begin{equation}
SFR_{\tau}(t) \propto \dfrac{t}{\tau} \exp{\left( -\dfrac{t^2}{2\tau^2} \right)} \, ,
\label{eq:SFR}
\end{equation}
allowing for both an increasing and decaying phase for the star-formation rate (SFR; see Fig.~\ref{fig:SFH_sketch_noburst}), 
whose delay time and steepness are regulated by $\tau$ in the time interval from $t=0$ to $t_\text{obs}$, which
is randomly generated from a uniform logarithmic distribution
from $5 \times 10^8\,\text{yr}$ up to $\sim 1.7 \times 10^{10}\,\text{yr}$.
The value of $\tau$ is generated as a
random number multiplied by $t_\text{obs}$, so that $\tau$ can vary between $1/50 \times t_\text{obs}$
(almost an instantaneous burst) and $2 \times t_\text{obs}$ (resulting in a still monotonically increasing
SFR at $t_{\rm obs}$). Overall, the distribution of the mean age 
of stars integrated between $0$ and $t_\text{obs}$ 
is almost uniform in logarithm between $3 \times 10^8\,\text{yr}$ and $8 \times 10^{9}\,\text{yr}$ and
drops to zero at $\sim 1.5 \times 10^8\,\text{yr}$ and $\sim 1.7 \times 10^{10}\,\text{yr}$. 
Our choice of the classical Sandage law for describing SFHs is justified by observational 
evidence that the majority of galaxies display first a rising and then declining phase 
in their SFH, irrespective of their specific parametrization, as suggested by the 
integrated view of galaxy evolution and more recent simulations 
\citep{Behroozi2013, Gladders2013, Abramson2015, Pacifici2016, Diemer2017, LopezFernandez2018}. 
 
In two thirds of the models random bursts are superimposed onto the secular SFH, 
in order to include also short-lived features in SFHs. Up to 6 bursts can
occur, with a total mass formed in these episodes ranging between $1/1000$ and $2$ times the total 
stellar mass formed in the secular component. The age of each burst (i.e., the look-back time at
which they occur) is randomly generated from a log-uniform distribution in the range 
$10^5\,\text{yr}$ to $t_\text{obs}$. 
For the youngest bursts ($\text{age}<10^{7}\,\text{yr}$), we allow
only a reduced mass fraction to be formed, in order to avoid models whose light is completely
dominated by the burst.  The presence of random bursts of star formation greatly alleviates the relative rigidity 
of the simple parametric form chosen to reproduce the secular component, enabling us to model the presence of sharp transitions in SFHs.


\begin{figure}[t]
\centering
\includegraphics[scale=0.35, trim=0cm 0cm 0cm 0cm, clip=true]{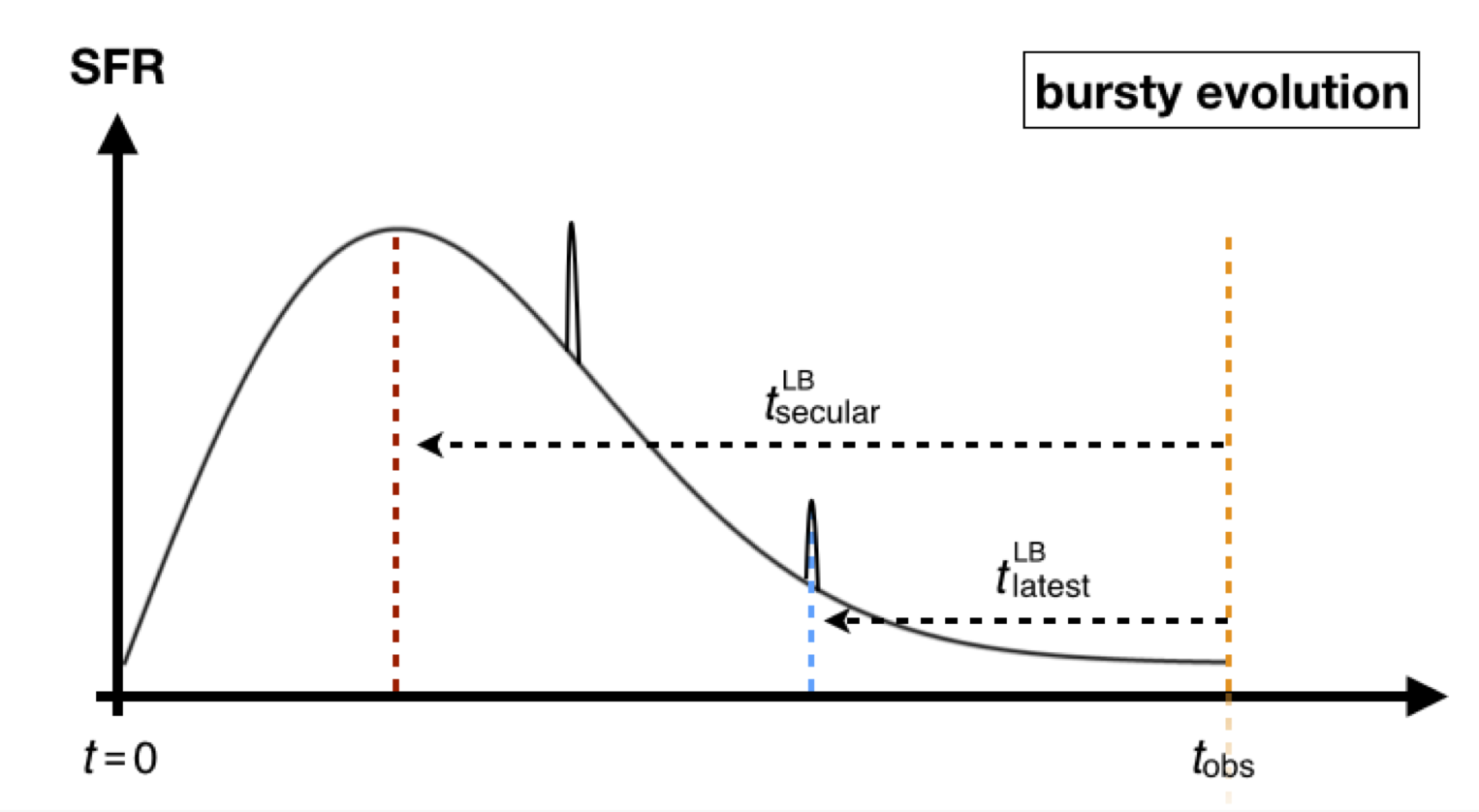}
\caption{Sketch of bursty SFH of a galaxy that displays 
additional burst events after the peak of the secular SFH.
The orange dashed line corresponds to the time from the beginning of
the SFH ($t=0$) to the time at which the galaxy was observed ($t_{\rm obs}$), 
the blue dashed line stands for the latest additional event of star formation superimposed to the secular SFH, 
while the red dashed line corresponds to the peak of the secular SFH. 
The look-back time to the peak of the secular SFH ($t^{\rm LB}_{\rm secular}$) in this case differs from 
the look-back time to the latest peak of star formation ($t^{\rm LB}_{\rm latest}$).}
\label{fig:SFH_sketch_burst}
\end{figure}

Figs.~\ref{fig:SFH_sketch_noburst} and \ref{fig:SFH_sketch_burst} present two sketches of example
SFHs, that is, $SFR$ as a function of time. We label $t^{\rm LB}_{\rm secular}$ the look-back time 
from $t_\text{obs}$ to the peak of the secular SFH and $t^{\rm LB}_{\rm latest}$ 
the look-back time from $t_\text{obs}$ to the latest peak of 
star formation. In general $t^{\rm LB}_{\rm latest}$ < $t^{\rm LB}_{\rm secular}$, 
but if a galaxy either experienced no secondary events of star formation
or such secondary event happened before the peak of the secular star formation, then 
$t^{\rm LB}_{\rm secular} = t^{\rm LB}_{\rm latest}$.

In our models, we allow for chemical enrichment to occur along the SFH. We randomly
generate the initial and final metallicity as well as a parameter describing how quickly the 
enrichment occurs as a function of fraction of formed mass over the total, as detailed in 
\cite{Zibetti2017}. Finally, we also apply a two-component dust attenuation to the spectra,
following the formalism of \cite{Charlot2000}, with the random parameter distribution described
in \cite{Zibetti2017}.

In order to provide a homogeneous and fine sampling of the PDFs 
over the whole space of observable parameters, we equalize the library to have a roughly constant
density of models in the plane of H$\delta+$H$\gamma$ vs $\text{D}_\text{n}4000$. This is done by
building an initial library of 5 million objects and 
selecting a subsample of 500\,000 so as to obtain an equalized distribution. 

The various prescriptions described so far result in a roughly uniform distribution in the 
$r$-band light-weighted mean log age - Z
plane, over the range $8.5$ to $10.2$ in $\log\text{(age/yr)}$ and 0.02 and 2.5 times solar in metallicity.

This combination of ``secular'' SFH and bursts allows us to generate a complete coverage of the observed
parameter space of optical stellar absorption indices (such as $\text{D}_\text{n}4000$, the Balmer
Lick indices, and various metal-sensitive absorption indices), as we verified by comparison with
the observed distribution of spatially resolved regions in CALIFA and with integrated measurements 
in the SDSS. Notably, with our models we can reproduce and populate the region of 
$\text{D}_\text{n}4000\lesssim1.5$ and extremely low H$\delta$, which is characterized by the
``frosting'' of a tiny mass fraction of a young (age $<500\,$Myr) stellar population on top of a
mass-dominant older one (see Sect.~\ref{sec:classical}). 
This is particularly relevant in light of our attempt to identify
stellar populations with a large spread in age among its components.


From this rich ``parent library'', we randomly select a small but statistically 
significant chunk of 12\,500 models, which are used 
as the base to generate mock ``StePS-like'' observations (see Sect.~\ref{subsec:StepS-like}). 
In reality, observed spectra appear broadened with different line-of-sight velocity distributions 
(or, simplifying, velocity dispersion). To properly retrieve physical quantities from spectral indices analysis, 
we will convolve the original-resolution models in order to match the estimates of velocity dispersion object by object. 
For simplicity, we assume here the same velocity dispersion for our mock observations 
and we convolve both the mocks and the models with a fixed velocity dispersion of 150 km s$^{-1}$. 
We note that the results of the analysis are not affected by the particular choice of velocity dispersion 
since the sensitivity of the spectral indices to the physical parameters of interest does not change 
greatly in the range of velocity dispersions foreseen for the galaxy masses targeted 
by StePS (provided that the data and the model are compared consistently).

\subsection{Light-weighted ages \label{sec:age}}

Ideally, the full reconstruction of the SFH is the goal
of the stellar population analysis of a galaxy. Unfortunately, despite numerous attempts in the literature
(e.g., \texttt{ppxf}: \citealt{Cappellari2004}, \texttt{starlight}: \citealt{CidFernandes2005},
\texttt{steckmap}: \citealt{Ocvirk2006}, \texttt{vespa}: \citealt{Tojeiro2009}, 
\texttt{prospector}: \citealt{Leja2017}, \texttt{bagpipes}: \citealt{Carnall2018}) 
this goal has not yet been reached. One reason for this 
is the relatively low $SNR$ usually achievable in spectra
of typical surveys at intermediate redshift,
an issue typically addressed by the use of stacked spectra. Moreover,
the full reconstruction of a galaxy's SFH turns out to be an ill-posed inversion problem, 
hampered by a large number of degeneracies \citep[see e.g.,][]{Ocvirk2006, Cappellari2017}.
Vice versa, the lower moments of the age distribution of the stars in a galaxy, and the 
mean in particular, are more easily constrained by the observations and their intrinsic 
degeneracy-driven uncertainties can be robustly quantified by means of model libraries such as
the one presented in the previous section. As demonstrated and discussed, e.g., in 
\cite{Gallazzi2005} and \cite{Zibetti2017}, with optical rest-frame spectra having $SNR\gtrsim 10$
one can obtain uncertainties on $r-$band light-weighted mean stellar ages of the order of $\pm 0.1$--$0.2$\,dex
based on five indices ($\text{D4000}_\text{n}$, $\text{H}\beta$, $\text{H}\gamma + \text{H}\delta$,
$[\text{Mg}_2\,\text{Fe}]$, and $[\text{Mg}\,\text{Fe}]^\prime $), plus possibly broad-band optical SDSS $ugriz$ fluxes;
uncertainties on age vary little with galaxy spectral type, 
though with a tendency of smaller relative uncertainties for older galaxies.

Here we specifically refer to \emph{light-weighted} ages, defined as 
\begin{equation}
\texttt{age} = \dfrac{\int_{t=0}^{t_{\rm obs}} (t_{\rm obs} - t) \, {SFR}(t) \, L(t) \, dt}{\int_{t=0}^{t_{\rm obs}} {SFR}(t) \, L(t) \, dt} \, ,\label{eq:lwage}
\end{equation}
where SFR($t$) is the star formation rate as a function of time, and
$L(t)$ is the luminosity emitted in the considered band per unit formed stellar mass 
from the ensemble of single stellar populations of age ($t - t_{\rm obs}$) in the galaxy
(see Fig.~\ref{fig:SFH_sketch_noburst}).
Keeping in mind the spectral evolution of a simple stellar population as a function of time
\citep[e.g., Fig.~9 of][]{Bruzual2003}, it is easy to understand that $L(t)$ is a strong
function of the considered photometric band.
In particular, the young stellar populations represent
a relatively stronger component in the ultraviolet bands than in the optical/NIR ones and, 
conversely, the old stellar populations have a higher weight in the 
optical/NIR bands than in the ultraviolet ones.


\begin{figure}[t]
\centering
\includegraphics[scale=0.55, trim=0cm 0cm 0cm 1.4cm, clip=true]{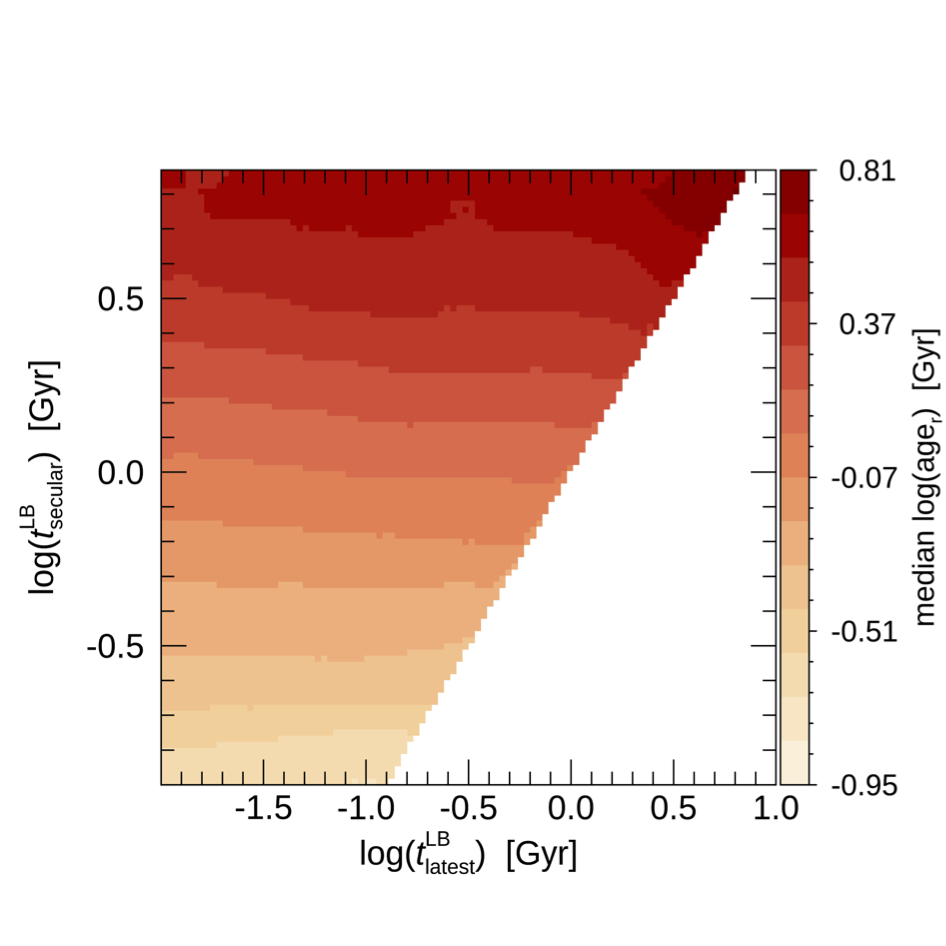}
\caption{Look-back time to the peak of the secular SFH ($t^{\rm LB}_{\rm secular}$) as a function of
look-back time to last episode of star formation ($t^{\rm LB}_{\rm latest}$),
color-coded according to $r$-band light-weighted age (\texttt{age$_{\rm r}$}) for models 
having ages smaller than the age of the Universe at redshift $z = 0.55$.}
\label{fig:SFH_age_r}
\end{figure}


As a consequence, for any composite stellar population (i.e., a superposition of multiple SSPs 
of different ages), the light-weighted ages in different bands are expected to differ, namely
in the sense of ages weighted in bluer bands being younger than those weighted in redder bands. 
The actual difference in the ages weighted in any two bands is determined by the characteristics of the 
SFH and specifically by the age spread by the different SSPs (see Sect.~\ref{subsec:delta_age}).
Such a spread can occur even in a simple $\tau$ model, provided that  $t^{\rm LB}_{\rm secular}$ is small enough, 
or $\tau$ is large enough relative to $t^{\rm LB}_{\rm secular}$ 
to ensure that the young and the old SSPs contribute a comparable amount of light. 
Similarly, when $t^{\rm LB}_{\rm secular}$ is large, and the galaxy is in the declining phase of its SFH, the presence of
a new burst of star formation may cause a significant difference in light-weighted ages in different photometric bands.

In this work, we focus on exploiting this effect to trace the coexistence of widely
diverse stellar populations and provide an essential but significant characterization of the SFH.
We elect as the reference ``red'' band the $r$ band, and as the reference ``blue'' band
the $u$ band.
Thus, we define $r$-band light-weighted ages (\texttt{age$_{\rm r}$}), $u$-band light-weighted ages
(\texttt{age$_{\rm u}$}), and their difference 
\texttt{$\Delta$(age)} = \texttt{age$_{\rm r}$} -\texttt{age$_{\rm u}$}, respectively.
The choice is justified by the fact that these bands are well covered 
in the rest-frame of intermediate redshift galaxies that will be
observed using new generation spectrographs, like WEAVE or 4MOST \citep{deJong2019}.
Note that, in principle, estimates of age weighted on a given
band can be obtained irrespective of that band being observed or not, because that age is a
model-derived quantity, which can be ``extrapolated''. 
However, it is clear that much better 
constrained values can be obtained if that band is covered by spectroscopic observations.

We also note that we could have used the mass-weighted age 
(i.e., the actual first moment of the SFH, defined by replacing $L(t)$ with $1$ in 
Eq.~(\ref{eq:lwage})), because it is maximally sensitive to the age of the bulk 
of the stars \citep[see also discussion in][]{Zibetti2017}. 
Yet its determination is much more model-dependent than for any light-weighted quantity, 
as the oldest stellar populations easily leave the spectrum unaffected, 
almost irrespective of their mass contribution. On the contrary, \texttt{age$_{\rm r}$} is, by definition, 
much better constrained by the observed spectrum, yet it is a reasonably good proxy to the 
mass-weighted age, as we demonstrate in the following sections.

\subsection{\texttt{age$_{\rm r}$} and \texttt{age$_{\rm u}$} diagnostics \label{sec:ager_ageu}}

In order to better illustrate the physical meaning of
\texttt{age$_{\rm r}$} and \texttt{age$_{\rm u}$} in the case of
realistic SFHs, we revert to the model formalism developed in the previous sections. We stress,
however, that the conclusions are not specific to this formalism but can be extended to more
complex or even completely stochastic SFHs.
We focus on redshift $z=0.55$, which corresponds to the median redshift where the 
bulk of StePS galaxies are going to be observed. We therefore select the subset of 
model galaxies having ages less than the age of the Universe at this redshift. 


\begin{figure}[t]
\centering
\includegraphics[scale=0.55, trim=0cm 0cm 0cm 1.4cm, clip=true]{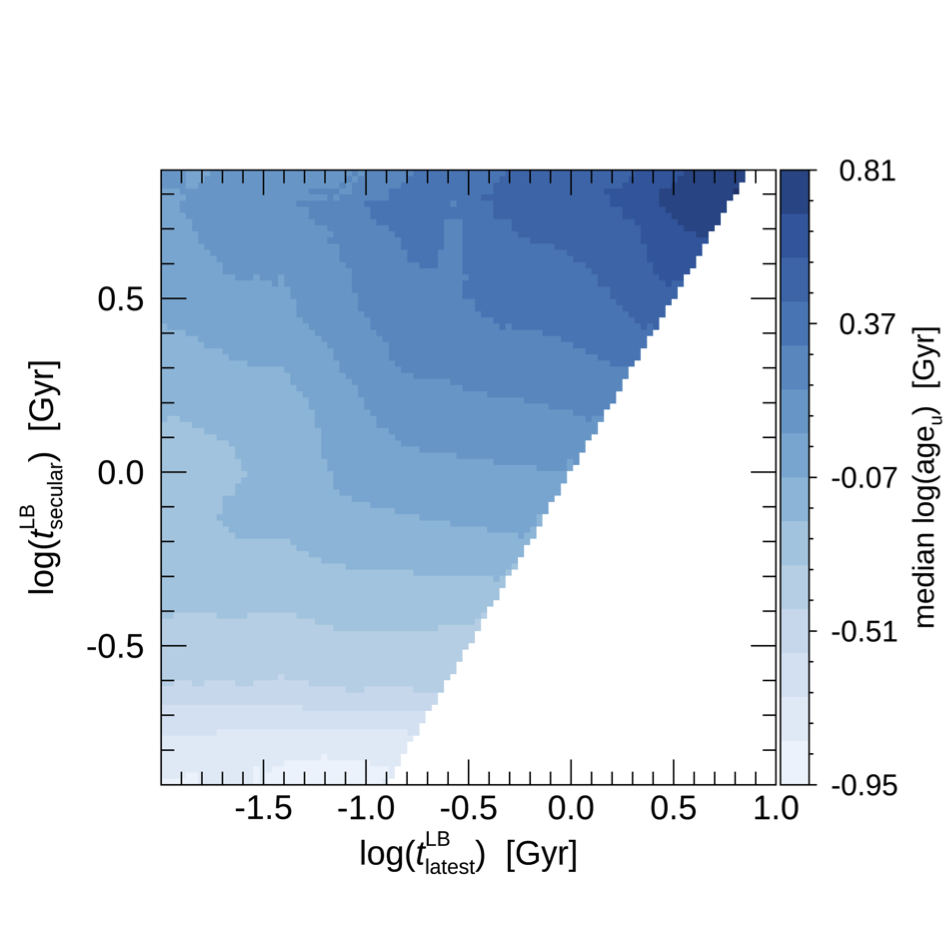}
\caption{Look-back time to the peak of the secular SFH ($t^{\rm LB}_{\rm secular}$) as a function of
look-back time to last episode of star formation ($t^{\rm LB}_{\rm latest}$),
color-coded according to $u$-band light-weighted age (\texttt{age$_{\rm u}$}) for models 
having ages smaller than the age of the Universe at redshift $z = 0.55$.}
\label{fig:SFH_age_u}
\end{figure}


In Fig.~\ref{fig:SFH_age_r} we highlight the role of \texttt{age$_{\rm r}$} in
tracing the evolution of the bulk of the stellar population
using the parameter plane of $t^{\rm LB}_{\rm secular}$ (the look-back time when the peak 
of the secular SFH occurs) and $t^{\rm LB}_{\rm latest}$ (the look-back time of the latest burst).
Galaxies evolving secularly according to Fig.~\ref{fig:SFH_sketch_noburst}
are located along the diagonal ($t^{\rm LB}_{\rm secular}$ = $t^{\rm LB}_{\rm latest}$)
and display a clear regular trend: \texttt{age$_{\rm r}$} smoothly increases from younger galaxies 
(smaller values of $t^{\rm LB}_{\rm secular}$) to older galaxies (higher $t^{\rm LB}_{\rm secular}$ values).
Galaxies that experience a secondary burst in their SFHs (see Fig.~\ref{fig:SFH_sketch_burst}) are offset from the 
diagonal position, but no significant gradient appears moving from smaller to higher $t^{\rm LB}_{\rm latest}$, 
making \texttt{age$_{\rm r}$} largely unaffected by the presence and timing of the most recent secondary burst. 
In other words, the value of \texttt{age$_{\rm r}$} retains the memory of the epoch of 
formation of the bulk of the stellar 
mass, being mostly insensitive to recently formed younger stellar populations.

The picture changes with \texttt{age$_{\rm u}$}, as shown
in Fig.~\ref{fig:SFH_age_u}, which highlights the different behaviour 
of \texttt{age$_{\rm u}$} relative to \texttt{age$_{\rm r}$} in the same plane.
For log($t^{\rm LB}_{\rm secular}$/Gyr) $\gtrsim -0.3$, 
\texttt{age$_{\rm u}$} appears to depend on both parameters and, in particular, 
to be very sensitive to the youngest bursts 
(log($t^{\rm LB}_{\rm latest}$/Gyr) $\lesssim -0.5$), a sensitivity that increases as $t^{\rm LB}_{\rm latest}$ decreases.

It is worth noting that even for secularly evolving galaxies, 
and more clearly for bursty SFHs, there is a difference
between $u$- and $r$-band
light-weighted ages at fixed $t^{\rm LB}_{\rm secular}$, with \texttt{age$_{\rm u}$} 
appearing to always indicate slightly younger ages than \texttt{age$_{\rm r}$}, as expected.

\subsection{\texttt{$\Delta$(age)} diagnostics \label{subsec:delta_age}}


\begin{figure}[t]
\centering
\includegraphics[scale=0.55, trim=0cm 0cm 0cm 1.55cm, clip=true]{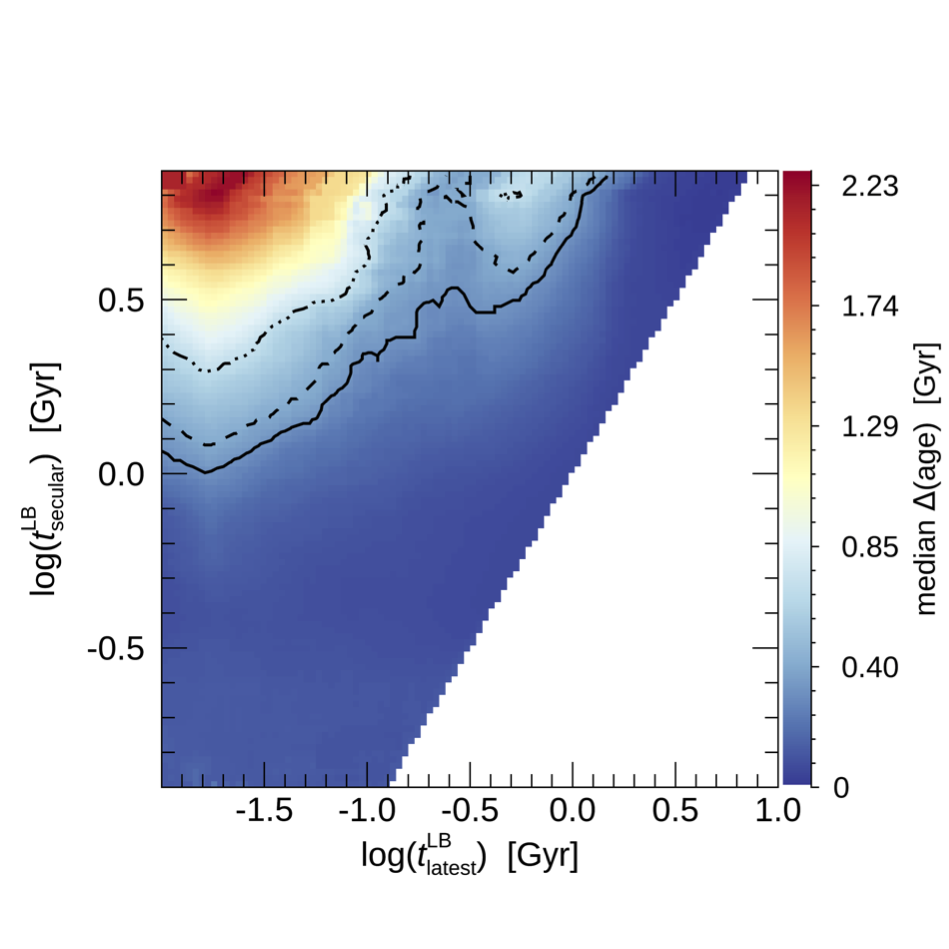}
\caption{Look-back time to the peak of the secular SFH ($t^{\rm LB}_{\rm secular}$) as a function of
look-back time to last episode of star formation ($t^{\rm LB}_{\rm latest}$),
color-coded according to \texttt{$\Delta$(age)} for models 
having ages smaller than the age of the Universe at redshift $z = 0.55$.
As discussed in Sect.~\ref{subsec:SFH_delta_age}, contour lines indicate the level for which we can reliably
assess a value of \texttt{$\Delta$(age)$\neq0$} at different $\mathit{SNR}_{\rm I, obs}$.
In particular, the threshold is \texttt{$\Delta$(age)}=0.7 at $\mathit{SNR}_{\rm I, obs}=10$ (black dashed-dotted line), 
\texttt{$\Delta$(age)}=0.4 at $\mathit{SNR}_{\rm I, obs}=20$ (black dashed line), 
and \texttt{$\Delta$(age)}=0.3 at $\mathit{SNR}_{\rm I, obs}=30$ (black solid line), respectively.}
\label{fig:SFH_delta_age}
\end{figure}


We compare the different behaviour of \texttt{age$_{\rm r}$} and \texttt{age$_{\rm u}$}
in the ($t^{\rm LB}_{\rm secular}$, $t^{\rm LB}_{\rm latest}$) plane, exploring in Fig.~\ref{fig:SFH_delta_age}
how \texttt{$\Delta$(age)} may be used to highlight the presence of a young stellar component 
superimposed on an older stellar population.
In this plane, the vertical gradient suggests that
a greater \texttt{$\Delta$(age)} value 
may arise only when the bulk of the galaxy stellar population is older.
More importantly, the horizontal trend implies that galaxies which experienced 
an additional burst event display a positive and non-negligible value of \texttt{$\Delta$(age)},
with more recent bursts associated with higher \texttt{$\Delta$(age)} values.
Thus, the larger the time separation between the peak of the secular star formation
and the latest peak of star formation, the larger \texttt{$\Delta$(age)} becomes.
When, on the contrary, either $t^{\rm LB}_{\rm latest}$ is quite large (greater than 1.5 Gyr) 
or the age of the peak of the secular SFH is quite small (lower than 1 Gyr), the value
of \texttt{$\Delta$(age)} becomes negligible. These are the cases
where the youngest population is not young enough to leave an imprint on the value 
of \texttt{$\Delta$(age)} or the bulk of the galaxy population is still young, 
so that no \texttt{$\Delta$(age)} is appreciable. 
It is worth noting that a non-zero value of \texttt{$\Delta$(age)} arises also if
the galaxy is still experiencing the peak of its SFH and $\tau$ is long enough
to allow the presence of a significant fraction of young stars within the galaxy,
as already mentioned in Sect.~\ref{sec:age}.


\begin{figure}[t]
\centering
\includegraphics[scale=0.55, trim=0cm 0cm 0cm 1.4cm, clip=true]{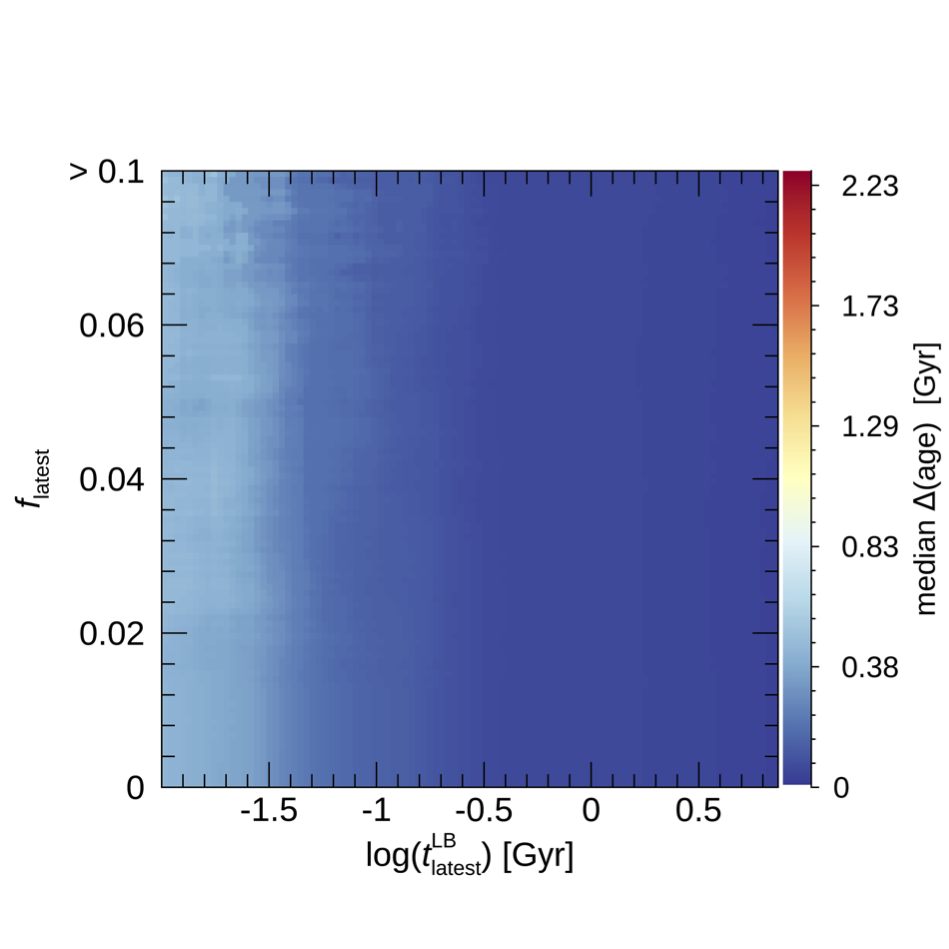}
\caption{Mass fraction ($f_{\rm latest}$) involved in last burst event as a function of
the look-back time to last episode of star formation ($t^{\rm LB}_{\rm latest}$),
color-coded according to \texttt{$\Delta$(age)} for models 
having ages smaller than the age of the Universe at redshift $z=0.55$.
$f_{\rm latest} > 0.1$ stands for $\sim$10\% of models having a large fraction of mass involved in the burst.}
\label{fig:SFH_mass_fraction}
\end{figure}


\begin{figure*}[t]
\centering
\includegraphics[scale=0.6, trim=1.cm 0cm 0cm 1.4cm, clip=true]{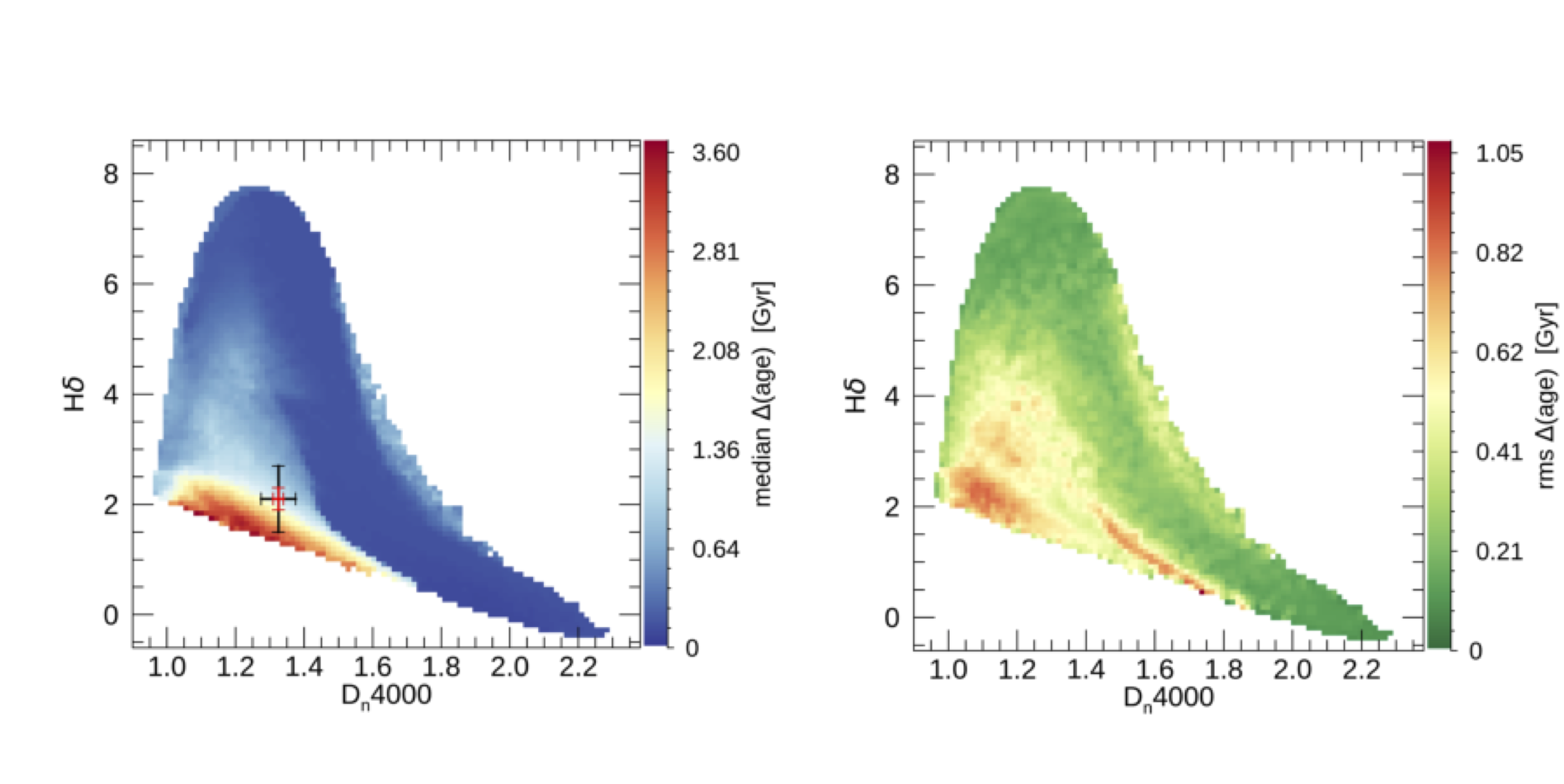}
\caption{(D$_{\rm n}$4000, H$\delta$) diagnostics color-coded according to the difference
between $r$- and $u$-band light-weighted ages for all models of the StePS-like spectral library
(left panel), and color-coded according to rms in \texttt{$\Delta$(age)} (right panel). 
We superimposed typical 1$\sigma$ average uncertainties in measuring spectral indices
in this plane at $SNR = 10$ (black error bars) and $SNR = 30$ (red error bars), respectively.}
\label{fig:dn4000_hdelta_delta_age}
\end{figure*}


\begin{figure*}[t]
\centering
\includegraphics[scale=0.6, trim=1.cm 0cm 0cm 1.4cm, clip=true]{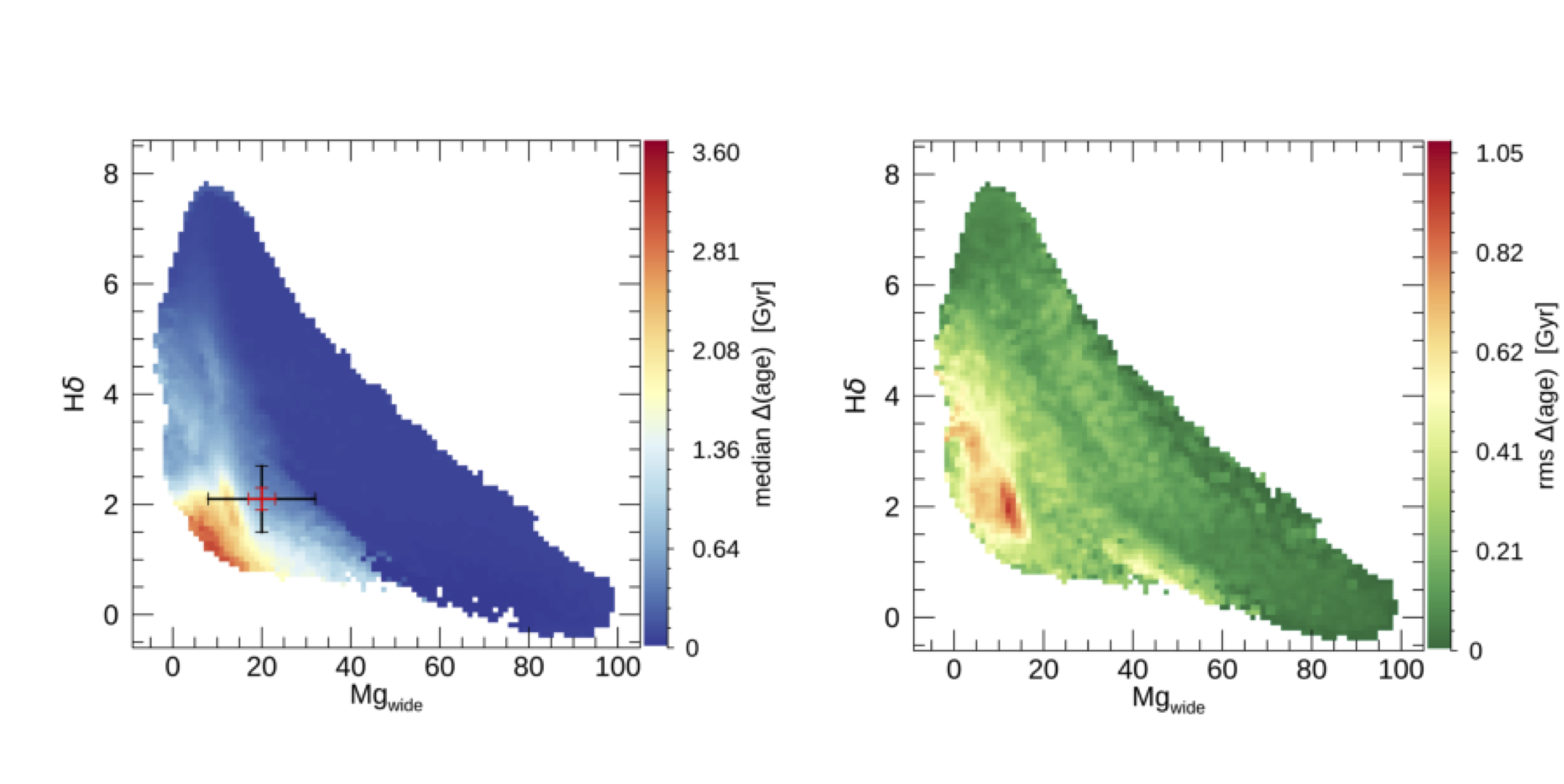}
\caption{(Mg$_{\rm wide}$, H$\delta$) diagnostics color-coded according to the difference
between $r$- and $u$-band light-weighted ages for all models of the StePS-like spectral library
(left panel), and color-coded according to rms in \texttt{$\Delta$(age)} (right panel). 
We superimposed typical 1$\sigma$ average uncertainties in measuring spectral indices
in this plane at $SNR = 10$ (black error bars) and $SNR = 30$ (red error bars), respectively.} 
\label{fig:hdelta_mg_delta_age}
\end{figure*}


One may expect that the mass fraction involved in the last burst event ($f_{\rm latest}$)
plays a significant role in affecting the value of \texttt{$\Delta$(age)}. 
This possibility is explored in Fig.~6, 
where the plane $f_{\rm latest}$ vs $t^{\rm LB}_{\rm latest}$ is color-coded according to the value of \texttt{$\Delta$(age)}.
While no trend appears in $f_{\rm latest}$, a gradient is seen
from higher to smaller values of $t^{\rm LB}_{\rm latest}$,
highlighting the fact that the time elapsed since the latest burst event is more important 
in affecting \texttt{$\Delta$(age)} values than 
the fraction of mass involved in it.
This trend also holds when considering higher mass fractions involved in the last burst event
since no systematic variation appears in the last bin (upper part of Fig.~6),
which corresponds to $\sim10\%$ of galaxies having  $0.1 < f_{\rm latest} < 0.5$.
Thus, we conclude that a significant value of \texttt{$\Delta$(age)} 
is present when a recent burst of star formation unsettles
the undisturbed secular evolution of galaxies, almost irrespective of the mass fraction involved.
This lack of a clear dependence on mass fraction can be understood by recalling that we are considering 
light-weighted ages and that the luminosity is a strong function of age itself. What is most relevant
in determining \texttt{$\Delta$(age)}, besides the difference in age of the two components, is 
the fraction of luminosity (not of mass) in the two components. If the luminosity in one of them is negligible
relative to the other, no age difference emerges. Moreover, for any given mass ratio, the relative 
luminosity difference in the two components is determined by their absolute age in addition to their age difference.
These three factors (mass ratio, absolute ages, and age difference) all play degenerate roles in determining 
\texttt{$\Delta$(age)}. For these
reasons we conclude that \texttt{$\Delta$(age)} is a powerful probe of age spread in stellar populations,
and specifically of young bursts in old galaxies, yet its amplitude cannot be directly used to estimate
the mass fraction and/or to pin down the precise age of the burst.

We further checked and confirmed the absence of any systematic dependence on metallicity of \texttt{$\Delta$(age)} values, 
splitting models in sub-solar, solar, and super-solar bins. 
This remarkable result enables the use of these new diagnostic, disregarding the well-known degeneracy
between age and metallicity, and marginalizing over the latter.

Finally, even if we are proposing an interpretation of \texttt{age$_{\rm r}$} and \texttt{$\Delta$(age)}
based on the specific SFH of our library models, we would like to stress that these
parameters per se are very general and of simple interpretation even when changing 
the specific parametric form chosen to define galaxy secular evolution. 
Their values may easily be used as simple heuristic information 
to be correlated with the variety of other intrinsic galaxy properties, such as morphology, mass,
size, and environment.


\section{Classical view from optical and ultraviolet spectral indices \label{sec:classical}}

In Sect.~\ref{subsec:delta_age}, we have shown that the value of \texttt{$\Delta$(age)}
is directly related to the presence of 
younger stellar components that coexist with older ones.
In the literature many attempts to describe and reconstruct the complexity of
the SFHs of galaxies on the basis of the combination of spectral indices 
can be found \citep{Thomas2005, Jorgensen2013, Lonoce2014}. 
In particular, the (D$_{\rm n}$4000, H$\delta$) diagnostic has been widely used to infer 
the recent star formation in galaxies both in the local Universe \citep{Kauffmann2003}
and at intermediate \citep{Wu2018} or high redshift \citep{Onodera2012}.

More recently, ultraviolet spectral features have been used to derive 
the physical properties of the stellar content of galaxies 
\citep{Fanelli1992, Daddi2005, Maraston2009, Vazdekis2016}.
In particular, the Mg$_{\rm wide}$ index has been demonstrated to be sensitive 
to the presence of stars older than 0.5 Gyr and it can be used as an
alternative to the D$_{\rm n}$4000 index as a signature of the ages of the 
old stellar components in galaxies \citep{Daddi2005}.

In the following, we thus explore the possibility of inferring 
\texttt{$\Delta$(age)} using both classical spectral indices in the optical wavelength range
(e.g., D$_{\rm n}$4000 and H$\delta$) and combining spectral indices
in the ultraviolet and optical wavelength range (e.g., Mg$_{\rm wide}$ and H$\delta$).

In Fig.~\ref{fig:dn4000_hdelta_delta_age} (left panel),
we present the classic (D$_{\rm n}$4000, H$\delta$) diagnostic as derived for the ``parent library'' models, 
color-coded according to the difference between the $r$- and $u$-band light-weighted ages. 
Galaxies that present a negligible amplitude of $\Delta$(age)$\sim$0 are
arranged in the bell-shaped sequence characteristic of the time evolution of 
SSPs \citep[e.g., Fig.~3 of][]{Kauffmann2003}, i.e.,
generations of stars with negligible spread in age. The sequence is defined
starting with the youngest SSPs at the lowest D$_{\rm n}$4000 and intermediate H$\delta$, 
then moving to the $\sim 1-2$\,Gyr-old SSPs 
peaking at the maximum values of H$\delta$ at intermediate/low D$_{\rm n}$4000, 
and finally to the oldest SSPs at the highest values of D$_{\rm n}$4000 and the lowest values of H$\delta$.
The region below the bell shape is characterized by increasing values of \texttt{$\Delta$(age)}, moving from the 
top to the bottom. This can be understood in terms of mixing old and young stellar populations, which are both
characterized by low H$\delta$ and whose high and low values of D$_{\rm n}$4000, respectively, average out
to an intermediate value for the break. In particular, we note the extreme region of 
H$\delta\lesssim 2.5$ and intermediate D$_{\rm n}$4000, which is characterized by 
\texttt{$\Delta$(age)}$\gtrsim 2$\,Gyr. Models with these properties are the result of a tiny fraction 
($\lesssim 0.1\%$) of young stars (age $\lesssim 10^8$\,yr) ``frosted'' on the top of a several-Gyr-old
stellar population.

However, in order to fully appreciate the usefulness of such a plot
as a tool to estimate \texttt{$\Delta$(age)}, we need to consider the intrinsic scatter
of \texttt{$\Delta$(age)} in the same plane (Fig.~\ref{fig:dn4000_hdelta_delta_age}, right panel), 
as well as the typical observational errors in measuring spectral indices (Fig.~\ref{fig:dn4000_hdelta_delta_age}, left panel).
The scatter reveals the intrinsic degeneracy between
the values of the spectral indices and the physical quantity \texttt{$\Delta$(age)}, and makes this diagnostic 
a less straightforward proxy for \texttt{$\Delta$(age)}, even if it ideally
provides a still good characterization of the difference of light-weighted ages. 
Considering typical uncertainties on spectral indices,
the overall picture inevitably blurs (see Fig.~\ref{fig:dn4000_hdelta_delta_age} and
Sect.~\ref{subsec:indices}). This makes almost impossible to infer
a reliable value of \texttt{$\Delta$(age)} using only the information
provided by two spectral indices, especially in the region of the plane
of higher \texttt{$\Delta$(age)} values.


\begin{figure*}[t]
\centering
\includegraphics[scale=0.38, trim=0.5cm 0cm 2cm 0cm, clip=true]{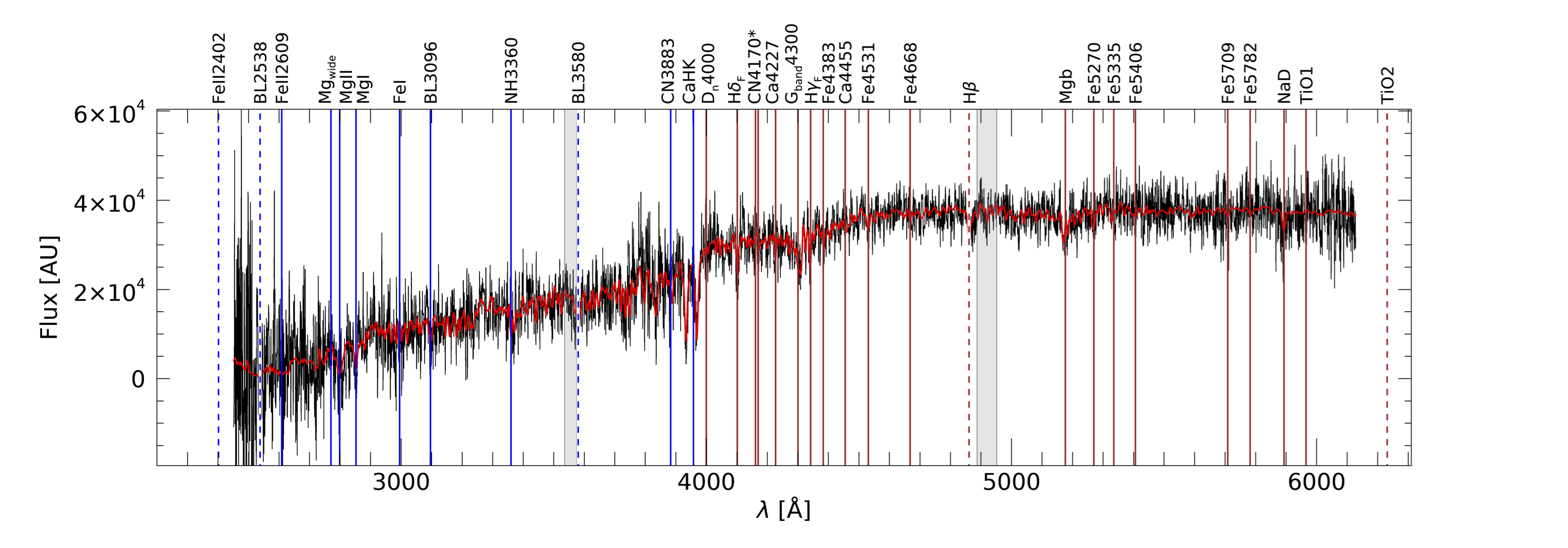}
\caption{Example of a rest-frame template (red curve) and observed spectrum (black curve) 
that mimic the StePS observation of a galaxy at $z$ = 0.55 and $\mathit{SNR}_{\rm I,  obs}=10$. 
The central rest-frame wavelengths of ultraviolet spectral indices are marked with blue solid lines,
while those of optical spectral indices are marked with red solid lines. 
CN4170 and CN$_2$ features share similar positions (*) in the plot.
Gray vertical bands represent gaps in the WEAVE spectrograph. It is worth noting that 
FeII2402, BL2538, and TiO2 indices are outside the considered spectral range, while
BL3580 and H$\beta$ features fall in the gap of the CCD at this particular redshift (dashed lines).}
\label{fig:template_SN10_z055}
\end{figure*}


Since the ultraviolet wavelength flux correlates well with the presence of young 
stellar populations,
one may expect that moving to ultraviolet spectral indices may help to
alleviate the intrinsic scatter shown in Fig.~\ref{fig:dn4000_hdelta_delta_age}.
Thus, we explore in Fig.~\ref{fig:hdelta_mg_delta_age} (left panel) the possibility 
of using the diagnostic plane (Mg$_{\rm wide}$, H$\delta$) 
that takes advantage of the wide magnesium absorption feature in the bluer part of the spectrum,
to trace the presence of recent events of star formation.
This feature becomes available for galaxies at redshift $z\gtrsim0.55$ in spectrographs
whose coverage in wavelength starts at 3800 $\AA$ (as foreseen for WEAVE and 4MOST; 
see also Sect.~\ref{subsec:indices}).
As in Fig.~\ref{fig:dn4000_hdelta_delta_age} (left panel), we can identify a continuum sequence followed by
coeval stellar populations, from high values of the Mg$_{\rm wide}$ index combined 
with low values of H$\delta$, up to low values of Mg$_{\rm wide}$ combined with 
high values of H$\delta$. As in the classical diagram, we can identify 
a region where the Mg$_{\rm wide}$ index assumes values lower than 50, 
which indicate stellar populations younger than a few Gyrs, and H$\delta$ values lower 
than $4-5$, which reveal the presence of very young components (i.e., ages lower than 0.1 Gyr).
The two indices in this region point toward the presence of stellar populations of different ages, 
and indeed they correspond to the region where $\Delta$(age) is larger than $0.4-0.5$ Gyr on average.
In our ``parent library'', it appears that 
the use of ultraviolet indices in the new (Mg$_{\rm wide}$, H$\delta$) diagnostic
reduces the degree of intrinsic uncertainty in the transition region of 
\texttt{$\Delta$(age)} $\sim$ 1 Gyr (see Fig.~\ref{fig:hdelta_mg_delta_age}, right panel).
On the other hand, as stated before and as shown in the following sections,
since we have to consider the non-zero (and usually non negligible, even at high $SNR$)
uncertainties in measuring spectral indices,
only by taking into account the whole spectral information
and dealing with a Bayesian analysis is it possible to properly
retrieve and constrain \texttt{$\Delta$(age)} of galaxies.


\section{From models to simulated observations \label{sec:observations}}

One of the main goals of this paper is to provide a realistic appraisal of 
scientific opportunities opened by data from new generation wide-field spectrographs 
at four-meter-class telescopes: large samples of galaxy spectra of high resolution ($R\sim5000$), 
moderate quality ($\mathit{SNR} \gtrsim 10$ \AA$^{-1}$), and wide wavelength coverage. 
 
In this context, StePS can be considered
the forerunner of the new observations that these spectrographs will provide. 
It uses the WEAVE spectrograph on the WHT in its so-called Low-Resolution MOS mode 
($R\sim5000$, over the wavelength range of $3660-9590$ \AA) to study galaxy evolution 
out to $z=0.7$ for a sample of $\sim$25\,000 galaxies. Its target sample is
selected using a simple magnitude criterion (I$_{\rm AB} \leq 20.5$ mag) 
coupled with a photometric (spectroscopic when available) redshift 
pre-selection ($z_{\rm phot} \geq 0.3$). The survey strategy foresees 
seven hours of observations for each target, split in 21 Observation Blocks 
of $\sim 20$ minutes each, to enable us to reach $SNR \gtrsim 10$ \AA$^{-1}$ 
in I band for most of the targets. 

We simulate mock observed galaxy spectra using a typical StePS 
observational set-up and the known parameters of the WEAVE spectrograph on the WHT 
(see also Jin et al. \emph{in prep.} for a set of comprehensive simulations of the forthcoming WEAVE data), 
and use this sample to explore the physical information that can be extracted from it. 

Even though we use StePS as the blueprint for our simulations, 
the flexibility of our method and analysis is such that it can be easily 
extended/modified to closely reproduce data from other upcoming facilities 
\citep[e.g., 4MOST at VISTA or PFS at SUBARU, to mention similar projects in advanced stages of completion;][]{Takada2014}.

\subsection{StePS-like simulated observations: the ingredients\label{subsec:StepS-like}}

As already anticipated in Sect.~\ref{subsec:models}, we make use of a 
representative, randomly chosen sub-sample of $12\,500$ rest-frame templates, stemming from the 
full ``parent library'' of $500\,000$ model spectra, to produce mock StePS-like 
observations, that is, spectra that closely mimic forthcoming WEAVE-StePS data.  
For each considered redshift ($z=[0.3,0.55,0.7]$),
we make a selection of models from the chunk, which are required to have
(i) $t_\text{obs}$ not exceeding 
the age of the Universe at that redshift,
(ii) $g$-band effective attenuation  
$A_g\equiv M_{g,\text{emerging}}-M_{g,\text{emitted}}<2\,\text{mag}$, 
(iii) $r$-band light-weighted mean metallicity $0.1 < Z/Z_{\sun} < 2$. Both metallicity and attenuation values are  
appropriate for the typical mass range of galaxies observed in surveys with bright selection magnitude limit 
\citep[I$_{\rm AB} < 20.5$ mag, see][]{Zahid2014}.
With this selection, we obtain 4848 models at $z=0.3$, 4300 models at $z=0.55$,
and 3967 models at $z=0.7$, respectively.
It is worth noting that this sub-sample library is fully 
representative of the comparison library,
ensuring no bias in any physical property 
(in particular in SFHs), and sufficient in size to enable 
an exploration of the whole parameter 
space of galaxy physical properties with statistically significant numbers.

Two ingredients are needed in order to reproduce mock StePS-like observations 
starting from these model spectra: 
\begin{itemize}
\item[(a)] $\mathit{RF_{\lambda, \rm obs}}$: the throughput of the combined atmospheric transmission, optics of the
WHT and WEAVE spectrograph, including also the gaps due to the presence 
of inter-CCD spacing in both blue and red spectrograph arms;
\item[(b)] $\mathit{N_{\lambda, \rm obs}}$: the contribution to the noise for all our mock StePS-like spectra due to 
the expected Poisson noise from sky continuum background and the detector noise of WEAVE CCDs.  
\end{itemize} 

The value of $\mathit{RF_{\lambda, \rm obs}}$ is obtained from \citet{Dalton2016} for the typical 
atmospheric transmission values and for combined WEAVE+WHT throughput.
The sky Poisson noise is computed assuming a typical sky continuum spectrum as provided in 
ESO Exposure Time Calculators and a dark-sky surface brightness magnitude of $V \sim 22.0$ mag arcsec$^{-2}$, 
that can be considered typical for sky brightness of dark nights in La Palma \citep{BennEllison1998}. 
Using the function $\mathit{RF_{\lambda, \rm obs}}$ described in (a) and the WEAVE 
fibers' footprint on sky (1.3 arcsec as defined by diameter of fiber core), we obtain the 
expected sky background Poisson noise in counts px$^{-1}$ on the CCDs (gain = 1). 
Notice that in our estimate of sky Poisson noise we neglect the contribution 
of sky emission lines, as the relatively high resolution 
used in StePS ($R\sim 5000$) is such that the pixels affected 
by the presence of such lines can be easily masked in the analysis.  

The WEAVE detectors' contribution to noise is mainly due to read-out noise, that 
from WEAVE specifications \citep{Dalton2016} we assume to amount to $\sim$ 2.5 e$^-$ px$^{-1}$. We estimate its 
actual value in counts for our spectra by considering an exposure time of 
$21\times20$ mins $\sim 7$ hours and a typical spatial diameter of a fiber on the CCDs of $\sim$4.5 px, 
corresponding to an expected integration window for profile weighted extraction of approx 1.5$\times$FWHM px. 
We neglected the contribution of thermal noise due to dark current, as the exposure time for single OBs 
in StePS will be short ($\sim20$ min) and Dark Current for WEAVE detectors is quite low ($\leq 0.1$ e$^-$/hour). 

While the first ingredient, $\mathit{RF_{\lambda, \rm obs}}$, is used to convert all 
our model spectra fluxes to expected counts on the CCDs, thus defining (in arbitrary units) 
the shape of the spectra as observed, the second one, $\mathit{N_{\lambda, \rm obs}}$, 
provides a realistic wavelength-dependent noise contribution in counts that is equal for 
all our sources. This noise is added in quadrature to the specific Poisson noise for 
each template to generate mock StePS-like observed spectra. Its value is used to anchor, 
depending on the desired $SNR$ in I-band, the actual value of counts (and therefore the corresponding 
Poisson noise contribution) of the observed mock StePS-like spectra. 
These two ingredients are obviously specific to WEAVE+WHT, 
but can easily be changed to reproduce  
different spectrographs and telescopes characteristics. 


\begin{figure}[t]
\centering
\includegraphics[scale=0.41, trim=0.3cm 0.5cm 0.8cm 0cm, clip=true]{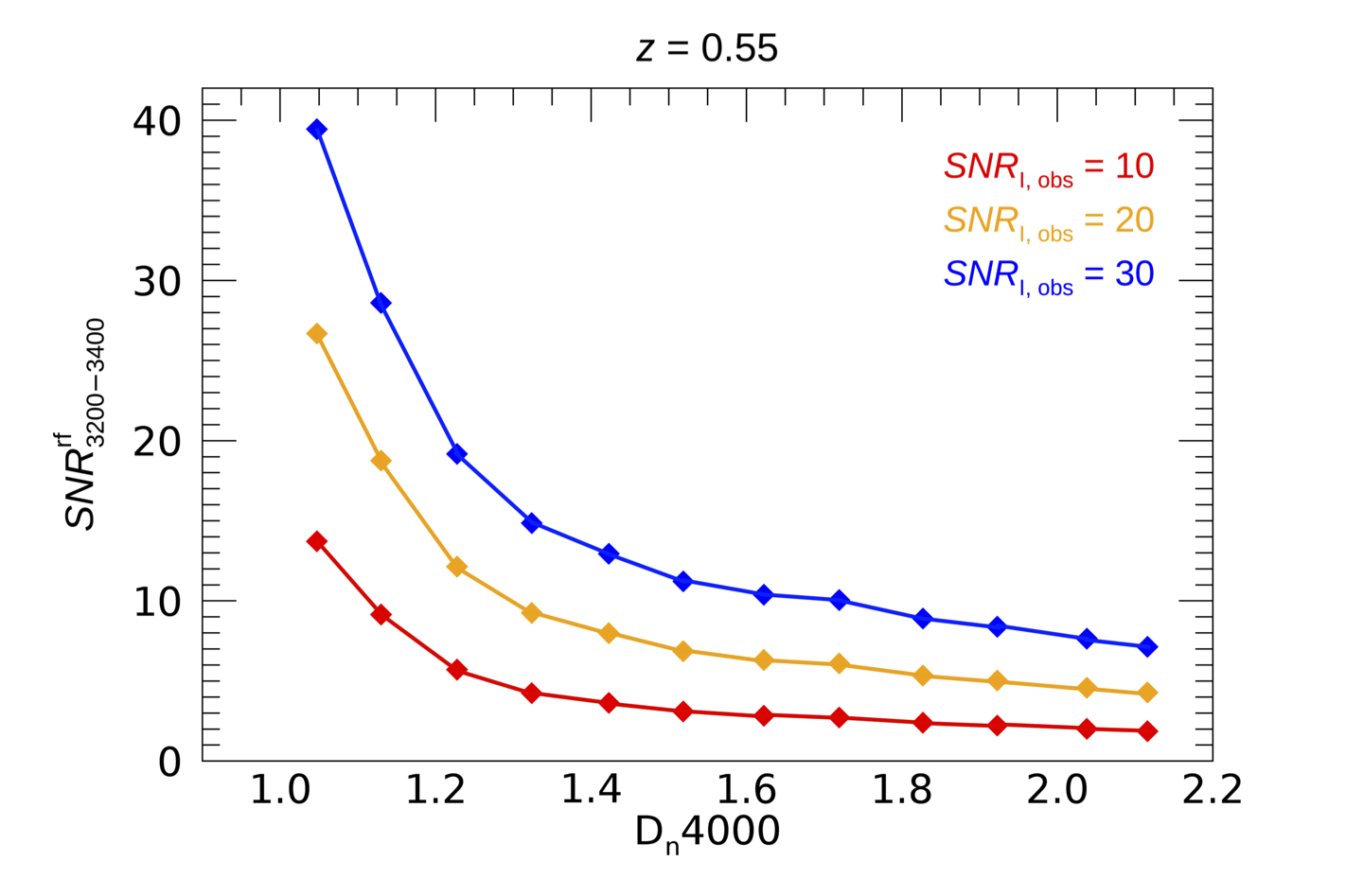}
\caption{Median value of $SNR^{\rm rf}_{3200-3400}$ as a function of D$_{\rm n}$4000 at fixed $z$ = 0.55.
Different colors correspond to $\mathit{SNR}_{\rm I, obs}=10$ (red diamonds), 
$\mathit{SNR}_{\rm I, obs}=20$ (orange diamonds), 
and $\mathit{SNR}_{\rm I, obs}=30$ (blue diamonds).}
\label{fig:SN_blue_I}
\end{figure}


\subsection{StePS-like simulated observations: the recipe \label{subsec:observed_spectra}}

In this section we detail the procedure to move from our representative sub-sample 
of rest-frame templates to mock StePS-like observed 
spectra at three different redshifts $z=[0.3,0.55,0.7]$, and
three $\mathit{SNR}_{\rm I, obs} = [10,20,30]$ \AA$^{-1}$.
\newline

In sequence, we perform the following operations: 

\begin{itemize}
\item[(1)] each model template $\mathcal{T}_{\lambda}$ is redshifted to each of the three 
desired redshift values $z=[0.3,0.55,0.7]$, rebinned to the observed pixel size equal to 1 \AA, 
and finally trimmed to fit the WEAVE spectral range, becoming $\mathcal{T}_{\lambda, {\rm obs}}$;
\item[(2)] each template $\mathcal{T}_{\lambda, {\rm obs}}$ is then multiplied by the  
function $\mathit{RF_{\lambda, \rm obs}}$, in order to retrieve the correct shape, in arbitrary units, 
of observed counts on the CCDs, becoming \texttt{T}$_{\lambda, {\rm obs}}$ [counts $\AA^{-1}$];
\item[(3)] for each template, we then compute the value $S_{\rm I, obs}$, that is, the mean number 
of counts per $\AA^{-1}$ needed to get the desired I-band  $\mathit{SNR}_{\rm I, obs} = [10,20,30]$. 
We use the function $N_{\lambda, \rm obs}$ to derive $N_{\rm I, obs}$, that is, the expected 
sky+CCD noise contribution in counts $\AA^{-1}$ in I-band. Then, 
for each $\mathit{SNR}_{\rm I, obs}$ defined above, $S_{\rm I, obs}$
can be simply obtained using the formula: 
\begin{equation}
\mathit{SNR}_{\rm I, obs} = \dfrac{S_{\rm I,obs}}{\sqrt{S_{\rm I,obs} + N^2_{\rm I,obs}}}\, ;
\end{equation} 
\item[(4)] the value of $S_{\rm I, obs}$ is then used to normalize each template 
\texttt{T}$_{\lambda, {\rm obs}}$ [counts $\AA^{-1}$], to obtain spectra of $\mathit{SNR}_{\rm I, obs} = [10,20,30]$;
\item[(5)] at each wavelength, we calculate the Poisson noise contribution in counts for each 
normalized template and add it in quadrature to the sky+CCD noise $\mathit{N_{\lambda, \rm obs}}$ 
contribution, to obtain the total noise budget $\mathit{N^{\rm tot}_{\lambda, \rm obs}}$;
\item[(6)] each normalized template \texttt{T}$_{\lambda, {\rm obs}}$ is perturbed with a noise contribution 
randomly generated from Gaussians of width equal to $\mathit{N^{\rm tot}_{\lambda, \rm obs}}$, 
obtaining observed spectra \texttt{S}$_{\lambda, \rm obs}$; 
\item[(7)] we apply back the sensitivity function to all spectra \texttt{S}$_{\lambda, \rm obs}$, 
obtaining the final spectra $\mathcal{S}_{\lambda, {\rm obs}}$ back in flux units.
\end{itemize}

At the end of this procedure we then have, for each model template $\mathcal{T}_{\lambda}$ , 
realistic mock StePS-like observed spectra $\mathcal{S}_{\lambda, {\rm obs}}$ 
that mimic StePS observations at the three redshifts for each of the three $\mathit{SNR}_{\rm I, obs}$ 
chosen (see Fig.~\ref{fig:template_SN10_z055}, as an example). 

The realism introduced by the wavelength dependence of $\mathit{SNR}_{\rm \lambda, obs}$ 
enables us to take into account in our simulations the difficulties of dealing with 
the variety of spectral types (including redder galaxies) and the reduced efficiency 
of the WEAVE+WHT system in going to bluer wavelengths. 
Our library of mock StePS-like observed spectra can assess, in a reliable fashion, 
what can be obtained from spectra of different $\mathit{SNR}_{\rm I, obs}$ values
and spectral types at different redshifts.
However, we note that with our simulations we only reproduce the random 
uncertainties due to the noise in the spectra, while we do not include any subdominant systematic
errors that might be present in the observed data (e.g., sky subtraction, flux calibration). 


\begin{table}
\caption{Ultraviolet and optical spectral indices.}
\centering
\begin{tabular}{cccccc}
\hline
UV index			&	$z$         	&	ref.        	&   opt. index          & 	$z$         	&	ref.            \\
(1) 			    &	(2)         	&	(3)            	&   (1)                 & 	(2)         	&	(3)             \\
\hline
FeII2402	   		&	> 0.66			&	 a				&	D$_{\rm n}$4000		&    all       		&    c				\\
BL2538              &   > 0.56   		&    a           	&   CN4170		        &	 all       		&    d           	\\
FeII2609	        &   > 0.48          &    a           	&   H$\delta_{\rm F}$   &	 all       		&    d           	\\
MgII       	      	&  	> 0.38      	&    a          	&   Ca4227	            &	 all       		&    d           	\\
MgI	           	    &   > 0.35          &    a              &   H$\gamma_{\rm F}$	&	 all      		&    d           	\\
Mg$_{\rm wide}$     &   > 0.54       	&    a           	&   G$_{\rm band}$4300	&	 all       		&    d           	\\
FeI	           		&   > 0.31       	&    a           	&   Fe4383              &    all       		&    d           	\\
BL3096	            &   > 0.25       	&    a           	&   Ca4455	            &    all       		&    d           	\\
NH3360	            &   > 0.14       	&    a           	&   Fe4531	            &    all       		&    d           	\\
BL3580	            &   > 0.09     		&    a           	&   Fe4668	            &    all       		&    d           	\\
CN3883	            &   > 0.01     		&    a           	&   H$\beta$	        &    all       		&    d           	\\
CaHK	            &   all        		&    b           	&   Fe5015	            &    all       		&    d           	\\
	            	&		       		&               	&   Mgb	                &    all       		&    d           	\\
                	&           		&                	&   Fe5270	            &    < 0.79   		&    d           	\\
					&   	       		&               	&   Fe5335	            &    < 0.77    		&    d           	\\
    	            &         	    	&               	&   Fe5406	            &    < 0.75  		&    d           	\\
    	            &          	    	&               	&   Fe5709	            &    < 0.66  		&    d           	\\
            	    &         	    	&               	&   Fe5782	            &    < 0.63       	&    d           	\\
             	    &              		&                  	&   NaD	                &    < 0.60      	&    d           	\\
                    &              		&               	&   TiO1	            &  	 < 0.56      	&    d           	\\
                    &              		&               	&   TiO2	            &  	 < 0.48   		&    d           	\\
\hline  
\end{tabular}
\tablefoot{(1) index name; (2) redshift range in which the index
is within the spectral range of the WEAVE spectrograph ($z_{\rm max} = 0.8$); (3) reference for 
indices definition: (a) \citet{Fanelli1992}, (b) \citet{Serven2005}, 
(c) \citet{Balogh1999}, (d) \citet{Worthey1994}.
}
\label{tab:indices}
\end{table}


To quantify our capability to retrieve ultraviolet indices in observed spectra, 
in Fig.~\ref{fig:SN_blue_I} we analyze $SNR^{\rm rf}_{3200-3400}$, that is the trend of $SNR$ as directly measured from mock observed spectra in the rest-frame 
window $3200-3400$ $\AA$ as a function of different D$_{\rm n}$4000 values 
for three values of  $\mathit{SNR}_{\rm I, obs} = [10,20,30]$ at fixed redshift $z=0.55$. 
Our ability of measuring ultraviolet indices strongly depends not only on the 
$\mathit{SNR}_{\rm I, obs}$ value, but also on 
the spectral type of the galaxies as identified by D$_{\rm n}$4000 values.
In particular, for D$_{\rm n}4000<1.5$, the $SNR$ in the blue region of 
the spectra increases significantly, thus increasing the reliability of 
the analysis based on ultraviolet indices. 
This trend does not change when moving to redshift $z=0.3$ and $z=0.7$.


\section{Stellar population analysis \label{sec:SP}}

As we discussed in Sect.~\ref{sec:classical}, a simple index vs index
plane, even when including ultraviolet indices, cannot fully capture the
complexity that is beyond the median value of \texttt{$\Delta$(age)}
corresponding to a specific position in such a plane. The large
scatter of possible \texttt{$\Delta$(age)} values and the typical
uncertainties associated with measured spectral indices
in the positions where the presence of differences in stellar ages 
becomes appreciable 
make it very difficult to apply such diagnostics tools in
individual galaxies (see Figs.~\ref{fig:dn4000_hdelta_delta_age} 
and \ref{fig:hdelta_mg_delta_age}). 

In this section we show how the ability to resolve the
key absorption-line indices both in the ultraviolet and optical
wavelength range enables us, when adopting a full Bayesian approach, 
to reliably obtain the value of \texttt{age$_{\rm r}$}, \texttt{age$_{\rm u}$}, and 
\texttt{$\Delta$(age)}.
It is worth noting that the power of a full Bayesian analysis
relies on the ability to retrieve for each individual galaxy the full PDF
of any physical parameter chosen, while providing a good representation of the overall uncertainties.
The analysis is performed at all redshifts and $\mathit{SNR}_{\rm I, obs}$ defined in the previous section.
However, we provide representative results obtained at $z=0.55$, the median redshift for StePS.

\subsection{Spectral indices \label{subsec:indices}}

We measure all spectral indices that can be observed in the 
WEAVE spectral range corresponding to each chosen redshift $z = [0.3, 0.55, 0.7]$, 
as listed in Table \ref{tab:indices}. 
Indices defined for band-pass wavelength greater than \mbox{4000 $\AA$} are 
considered optical indices, while those 
defined for band-pass wavelengths lower than 4000 $\AA$ are considered ultraviolet ones.
An example of measured indices for an observed spectrum at $z=0.55$ 
is shown in Fig.~\ref{fig:template_SN10_z055}.


\begin{figure}[t]
\centering
\includegraphics[scale=0.52, trim=0.2cm 0cm 1.5cm 0cm, clip=true]{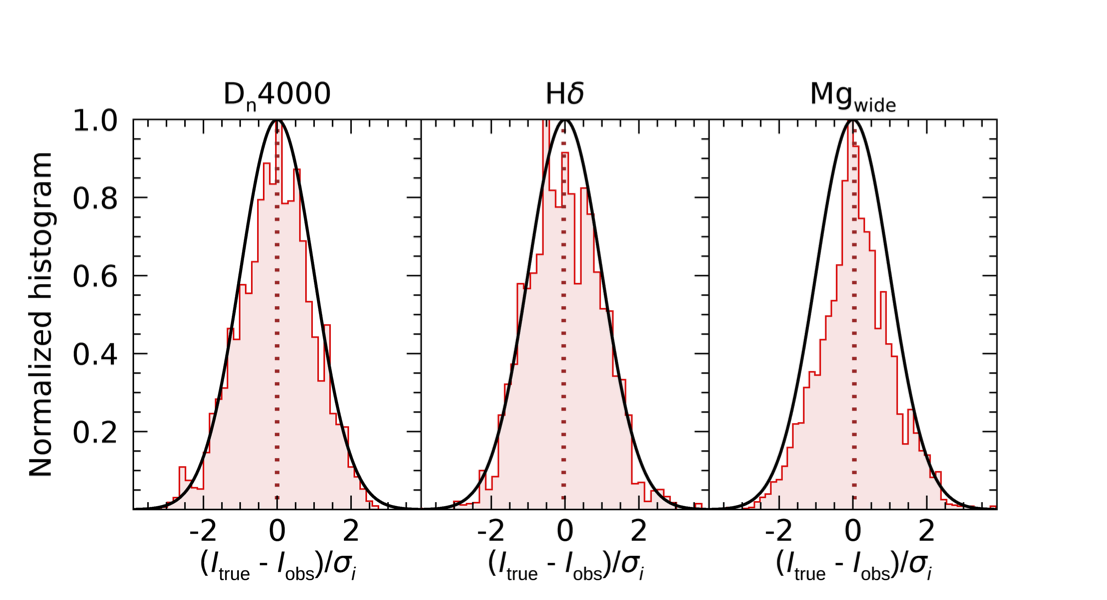}
\caption{Distribution of systematic errors at  $z=0.55$ and $\mathit{SNR}_{\rm I, obs}=10$
for D$_{\rm n}$4000, H$\delta$, and Mg$_{\rm wide}$, respectively.
The red dotted line corresponds to the median value of each distribution.
For reference, a solid black line shows a Gaussian distribution with unit standard deviation.}
\label{fig:systematics}
\end{figure}


In order to estimate observational errors associated with each index, we generate 
1000 random realizations of noise contribution for each template \texttt{T}$_{\lambda, {\rm obs}}$, 
at each considered $\mathit{SNR}_{\rm I, obs}$ and redshift.
Then, we measure all spectral indices for each realization and 
evaluate the systematics $\sigma_{\rm syst}$ and statistical $\sigma_{\rm stat}$ observational 
errors as the mean and standard deviation of the relative difference between
true and measured values. 
To realistically estimate uncertainties on the D$_{\rm n}$4000 spectral index,
whose window of estimate is quite wide,
we included a generous extra term ($\sim 5\%$) to the error budget to account for the 
uncertainty expected on the spectrophotometric calibration.
In Fig.~\ref{fig:systematics} we show the systematic deviation between true and observed values
of three indices, namely, Mg$_{\rm wide}$, D$_{\rm n}$4000, and H$\delta$, 
for observed spectra with $\mathit{SNR}_{\rm I, obs}=10$: 
even in the lowest bin of $\mathit{SNR}_{\rm I, obs}$, there are no systematics in retrieving 
the true value of measured indices, confirming a good accuracy of our estimates.

\begin{figure}[t]
\centering
\includegraphics[scale=0.52, trim=0.2cm 0cm 1.5cm 0cm, clip=true]{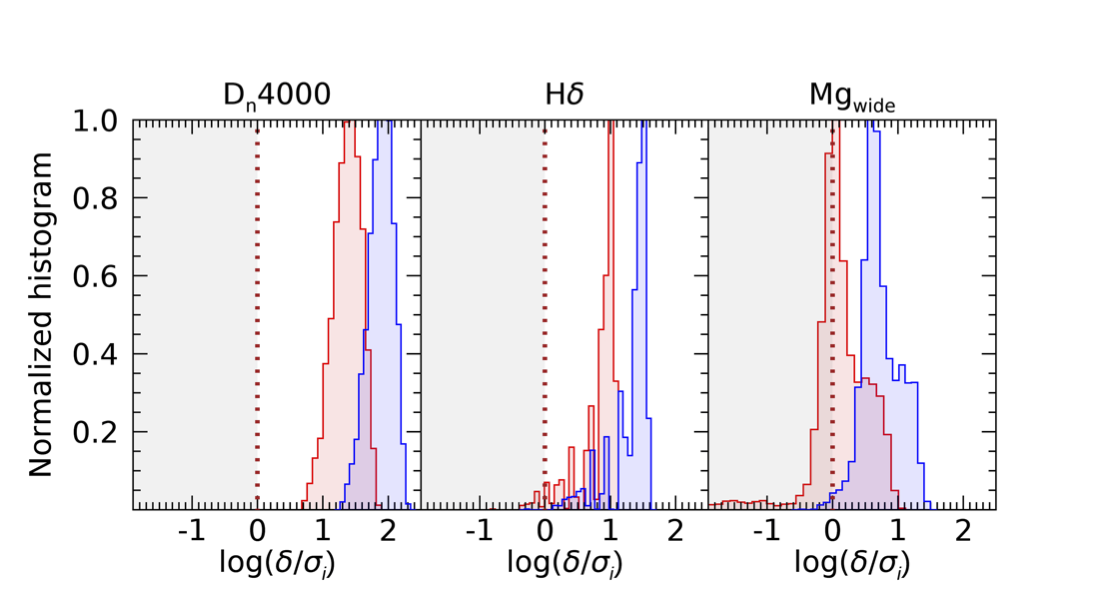}
\caption{Distribution of the resolving power log($\delta_i/\sigma_i$) at $z=0.55$
for D$_{\rm n}$4000, H$\delta$, and Mg$_{\rm wide}$, respectively.
Red histograms correspond to $\mathit{SNR}_{\rm I, obs}=10$, 
while blue histograms stand for $\mathit{SNR}_{\rm I, obs}=30$.
The red dotted line separates the region where log($\delta_i/\sigma_i$) < 0 (in gray) from the one where 
log($\delta_i/\sigma_i$) > 0 (in white).}
\label{fig:resolving_power}
\end{figure}

In order to evaluate the actual contribution of
each index in retrieving light-weighted ages in different bands, 
we calculate their resolving power log($\delta_i$/$\sigma_{i}$), as defined in \citet{Gallazzi2005}.
In this definition, $\sigma_{i}$ is the statistical observational uncertainty associated to
a particular index ($\sigma_{i} = \sigma_{\rm stat}$, where $i$ runs on different indices), 
while $\delta_i$ (with $i$ running on different indices) 
is the dynamical range of each index, which corresponds to the $5-95$\% 
percentile range of the distribution of index strengths for all observed galaxies at a given redshift. 
A resolving power of $0$ or less indicates that the measurement error is such that the 
index is broadly consistent with the full range of models, so that little or no information is provided. 
On the contrary, $\log(\delta_i/\sigma_i) >> 0$ implies that we are able to locate 
the value of the index very precisely within its allowed range, thus we can obtain significant information.  
To ensure a more refined characterization of the resolving power of each spectral index measurement, 
for indices other than D$_{\rm n}$4000,
we consider the dynamical range $\delta_i$ at fixed value of D$_{\rm n}$4000 equal to the one of the galaxy, 
whereby D$_{\rm n}$4000 is used as a proxy of galaxy spectral type.
Thus, we calculate a resolving power log($\delta_i$/$\sigma_{i}$) that depends on the spectral 
type of observed galaxies. 
In Fig.~\ref{fig:resolving_power} we show the resolving power 
of three indices, namely, Mg$_{\rm wide}$, D$_{\rm n}$4000, and H$\delta$, 
for observed spectra with $\mathit{SNR}_{\rm I, obs}=10$ and  $\mathit{SNR}_{\rm I, obs}=30$, while
an example of the actual dynamical range of 
H$\delta$ ($\delta_{H\delta}$) as a function of D$_{\rm n}$4000 can 
be obtained from Fig.~\ref{fig:dn4000_hdelta_delta_age}, where clearly 
the dynamical range for H$\delta$ is large ($\delta_{\rm H\delta} \sim [2,8]$) 
at lower values of D$_{\rm n}$4000, and smaller ($\delta_{\rm H\delta} \sim [0,1]$) 
at higher values of D$_{\rm n}$4000.


As expected, the resolving power of different spectral indices has an important dependence on 
the galaxy spectral type. 
Moreover, since red (blue) galaxies have a lower (higher)
$SNR^{\rm rf}_{3200-3400}$, taking this dependence into account is the only way to 
robustly estimate the actual contribution  
of each measured index to our Bayesian analysis, particularly in the bluer region of the spectrum. 

The expected performances of StePS are remarkable, as 
even at the lowest $\mathit{SNR}_{\rm I, obs}$ the bluer indices are retrieved  
with a precision that delivers a resolving power $\log(\delta_i/\sigma_i) > 0$ for most galaxies. In Fig.~\ref{fig:resolving_power} 
this is shown for the index Mg$_{\rm wide}$, where 
log($\delta_{\rm Mg_{\rm wide}}$/$\sigma_{\rm Mg_{\rm wide}}$) > 0 in $\sim70$\% of the realizations even at 
$\mathit{SNR}_{\rm I, obs}\sim 10$. This
percentage increases to almost 100\% at higher values of $\mathit{SNR}_{\rm I, obs}$.
The same trend roughly holds for most of the ultraviolet indices explored.

\subsection{Bayesian inference \label{subsec:bayes}}

Bayesian statistics provides a powerful framework for inferring the intrinsic properties of galaxies. 
As a consequence of the Bayes' theorem, 
the probability that a certain physical parameter $\Theta$ is described by observed data $\theta$
is given by the posterior PDF:
\begin{equation}
{\rm p}(\Theta | \theta)  = \dfrac{{\rm p}(\Theta) \, {\rm p}(\theta | \Theta)}{{\rm p}(\theta)} \, ,
\end{equation}
where p($\Theta$) is the prior distribution on $\Theta$, p($\theta | \Theta$) is the 
likelihood function, and p($\theta$) is a normalization factor independent of the assumed model.
Our application of the Bayesian approach relies on the comparison
between a set of \emph{observable quantities}, consistently measured in the data (real or mock) and in the
models of the reference library (see Sect.~\ref{subsec:models}). 
In order to keep our mock data and the reference library
fully independent, the latter is selected from the complement to the chunk in the ``parent library''.
We exclude from the reference library only models that have $t_\text{obs}$ exceeding
the age of the Universe at the considered redshift by more than $1\,\text{Gyr}$.
We allow this buffer of unphysical ages (older than the age of the Universe at the redshift
considered) in order to avoid skewing the PDF of age values by cutting the old wing.


\begin{figure}[t]
\centering
\includegraphics[scale=0.37, trim=1cm 0cm 1cm 0cm, clip=true]{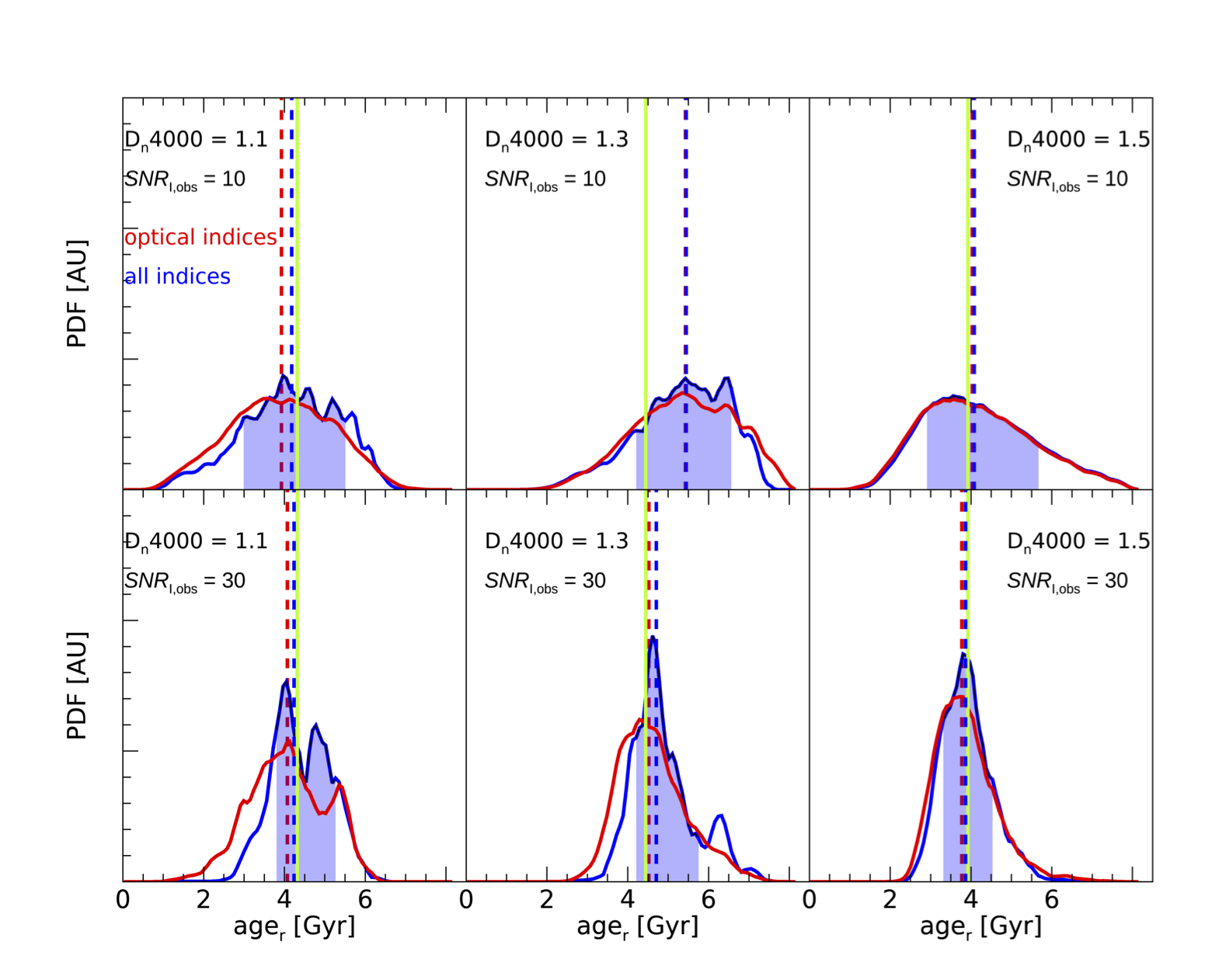}
\caption{Examples of PDFs of \texttt{age$_{\rm r}$} of observed galaxies at $z=0.55$; different 
values of D$_{\rm n}4000=[1.1,1.3,1.5]$ (from left to right), and $\mathit{SNR}_{\rm I, obs}=10$ (upper panels) 
and $\mathit{SNR}_{\rm I, obs}=30$ (bottom panels) are shown.  
PDFs are retrieved using only optical indices (red distributions)
and both ultraviolet and optical indices (blue distributions).
The blue shaded region represents the confidence interval corresponding to the 16-84 per 
cent percentile range of the PDF. The solid green line corresponds to the true value of \texttt{age$_{\rm r}$} = 4.31 Gyr, 
\texttt{age$_{\rm r}$} = 4.44 Gyr, and \texttt{age$_{\rm r}$} = 3.94 Gyr, respectively;
the dashed vertical red (blue) line indicates the median value of the corresponding distribution.}
\label{fig:PDF_example_age_r}
\end{figure}


Assuming errors normally distributed with a known correlation matrix, 
the likelihood is proportional to the chi-square statistic
\begin{equation}
\chi^2(\theta) = \sum_{i=1}^{N} \left( \dfrac{I_i^{\rm obs} - I_i^{\rm mod}}{\sigma_{i}}  \right)^2 \, ,
\label{eq:chi}
\end{equation}
where $i$ runs over all observed spectral indices, $I_i^{\rm obs}$ and $I_i^{\rm mod}$ are
the values of $i$-th spectral index in the observed and model spectra, respectively, 
and $\sigma_{i}$ is the observational error of $i$-th spectral index, computed as described in Sect.~\ref{subsec:indices}.
Thus, the likelihood of a given set of observed spectral indices $\theta$ being described
by a spectrum with intrinsic properties $\Theta$ is given by
\begin{equation}
{\rm p}(\theta | \Theta) \propto e^{-\chi(\theta)/2} \, .
\end{equation}

The Bayesian approach allows us to marginalize over different parameters,
focusing on relevant physical quantities instead of the complex SFH of each galaxy
\citep{Kauffmann2003, Gallazzi2005, Zibetti2017}.
The PDF of each parameter is obtained by the
distribution of the likelihood of all the models in the reference spectral library, in that parameter space. 
We focus on the PDFs of three physical quantities:
\begin{itemize}
\item[$\bullet$] \texttt{age$_{\rm r}$}, i.e., $r$-band light-weighted age;
\item[$\bullet$] \texttt{age$_{\rm u}$}, i.e., $u$-band light-weighted age;
\item[$\bullet$] \texttt{$\Delta$(age)} =  \texttt{age$_{\rm r}$} -  \texttt{age$_{\rm u}$}.
\end{itemize}
The fiducial value of each parameter is estimated as the marginalized median value of the PDF,
while errors $\sigma_{\rm obs}$ are evaluated as half of 
68 per cent confidence interval, that corresponds to the 16-84 per cent
percentile range of the PDF, being equivalent to $\pm$ 1$\sigma$ for a Gaussian distribution.


\begin{figure}[t]
\centering
\includegraphics[scale=0.37, trim=1cm 0cm 1cm 0cm, clip=true]{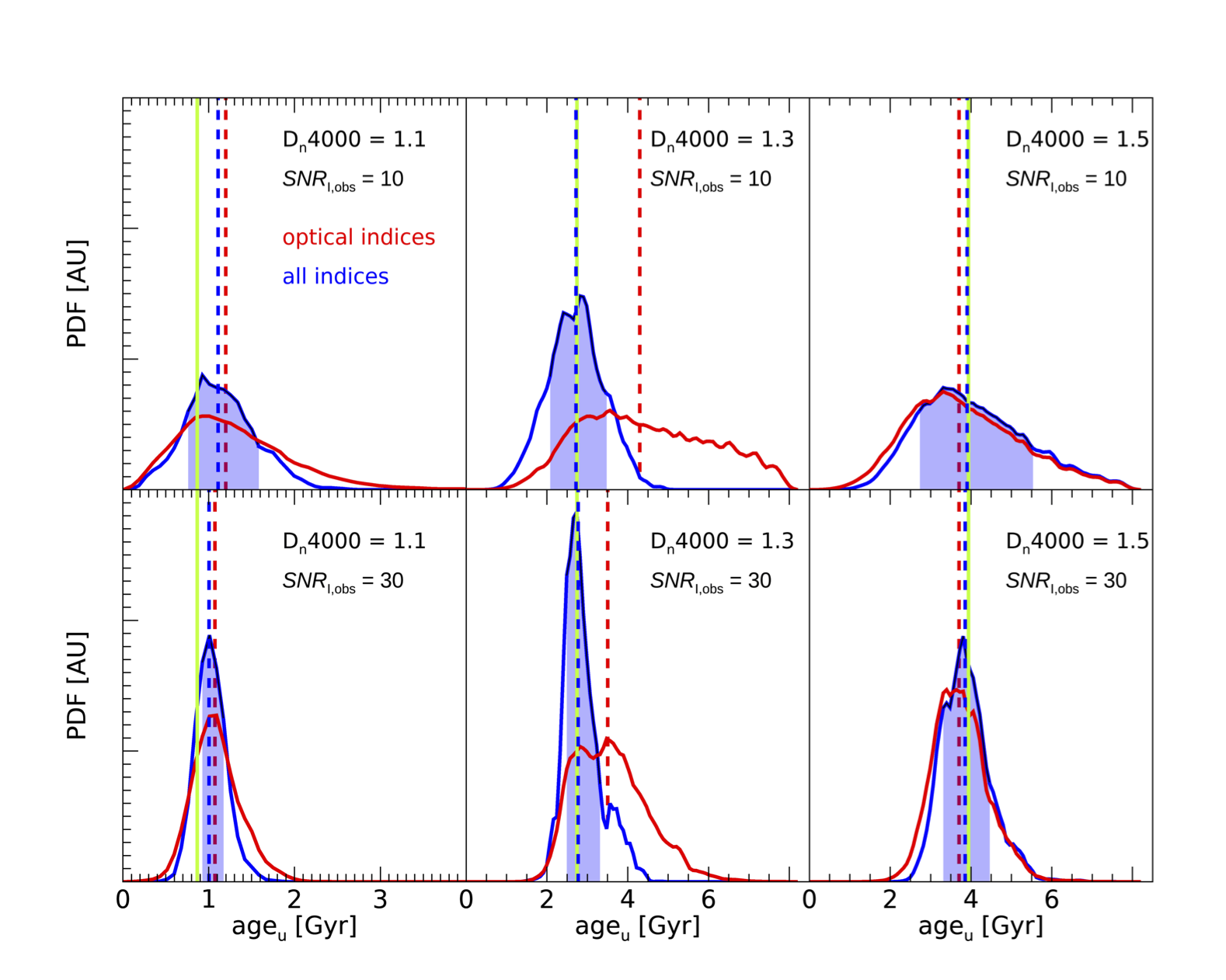}
\caption{As in Fig.~\ref{fig:PDF_example_age_r}, but for \texttt{age$_u$}.
The solid green line corresponds to the true value of \texttt{age$_{\rm u}$} = 0.87 Gyr, 
\texttt{age$_{\rm u}$} = 2.74 Gyr, and \texttt{age$_{\rm u}$} = 3.92 Gyr, respectively;
the dashed vertical red (blue) line indicates to the median value of the corresponding distribution.}
\label{fig:PDF_example_age_u}
\end{figure}


\begin{figure*}[t!]
\centering
\includegraphics[scale=0.57, trim=0.7cm 0.5cm 1cm 0cm, clip=true]{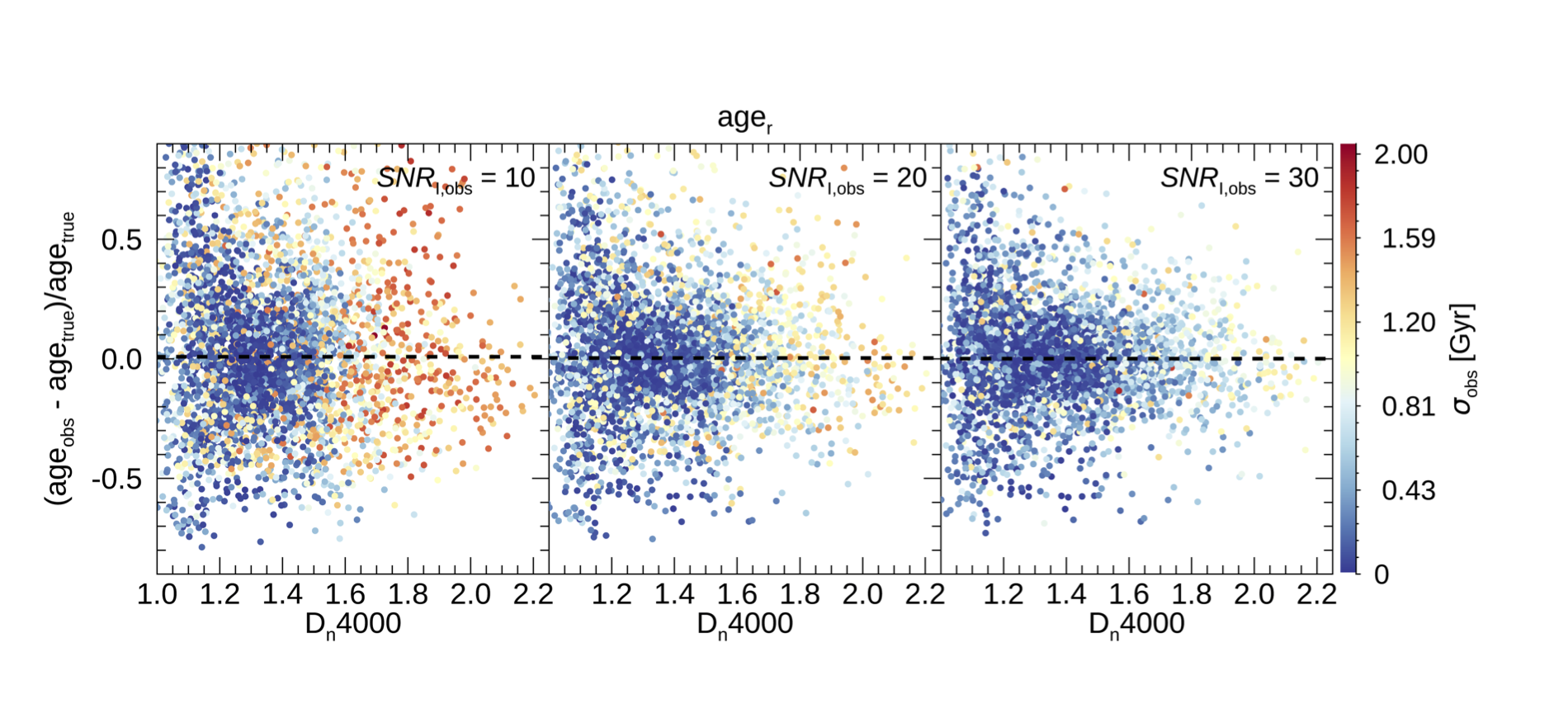}
\caption{Difference between observed and true value of \texttt{age$_{\rm r}$} (weighted on true values)
as a function of D$_{\rm n}4000$, color-coded according to the observed errors retrieved from PDFs.
Galaxies are observed at $z=0.55$ and $\mathit{SNR}_{\rm I, obs}$ = [10, 20, 30] (from left to right).
The black horizontal dashed line corresponds to median value of the distribution.}
\label{fig:Dn4000_age_r}
\end{figure*}


\begin{figure*}[t!]
\centering
\includegraphics[scale=0.57, trim=0.7cm 0.5cm 1cm 0cm, clip=true]{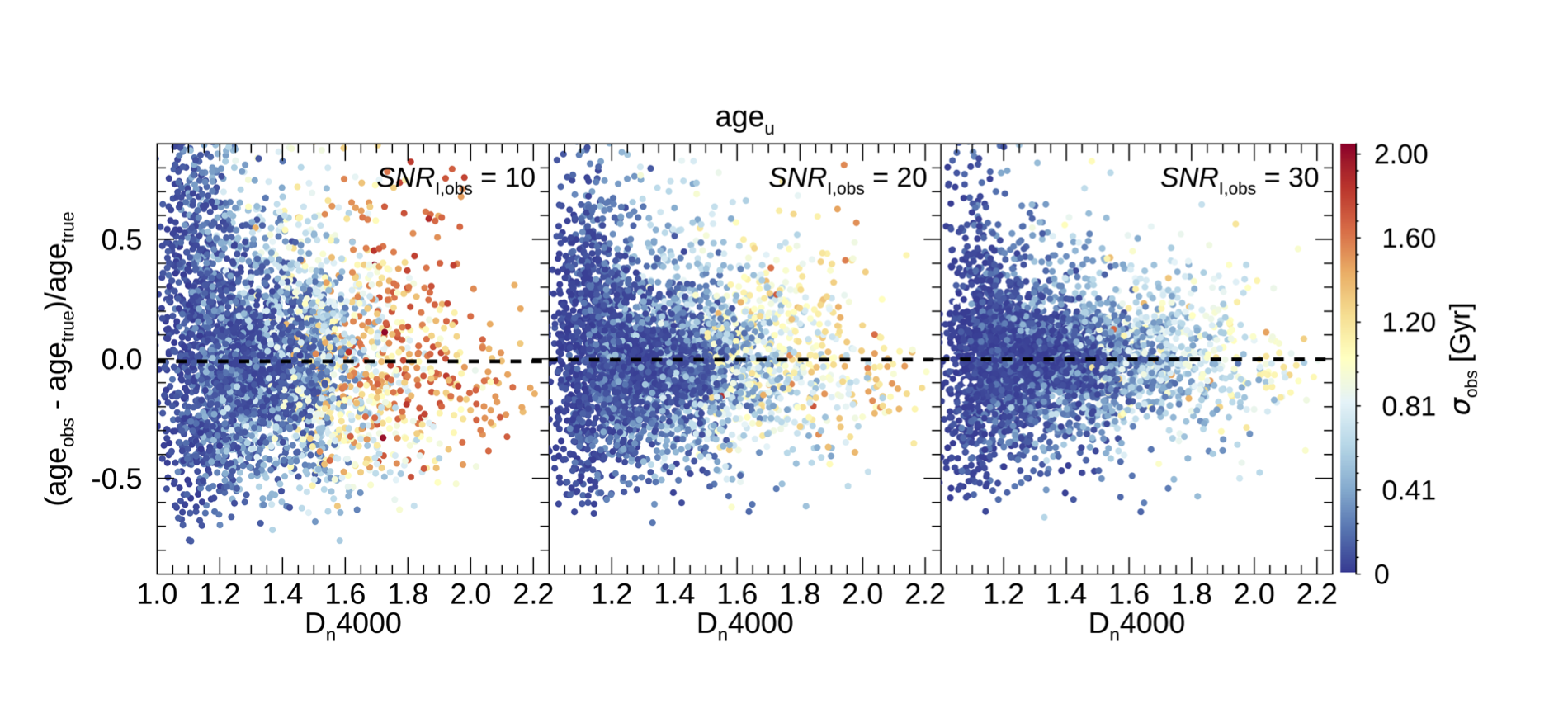}
\caption{As in Fig.~\ref{fig:Dn4000_age_r}, but for \texttt{age$_{\rm u}$}.}
\label{fig:Dn4000_age_u}
\end{figure*}


\subsubsection{Results: estimates of \texttt{age$_{\rm r}$} and \texttt{age$_{\rm u}$} \label{subsec:SFH_age}}

We use the method described in Sect.~\ref{subsec:bayes} to retrieve the PDF 
of \texttt{age$_{\rm r}$} and \texttt{age$_{\rm u}$} of our mock observed galaxies 
and explore the role of ultraviolet indices in helping to constrain
light-weighted ages at different values of $\mathit{SNR}_{\rm I, obs}$. 


\begin{figure*}[t!]
\centering
\includegraphics[scale=0.55, trim=1cm 0cm 1cm 0cm, clip=true]{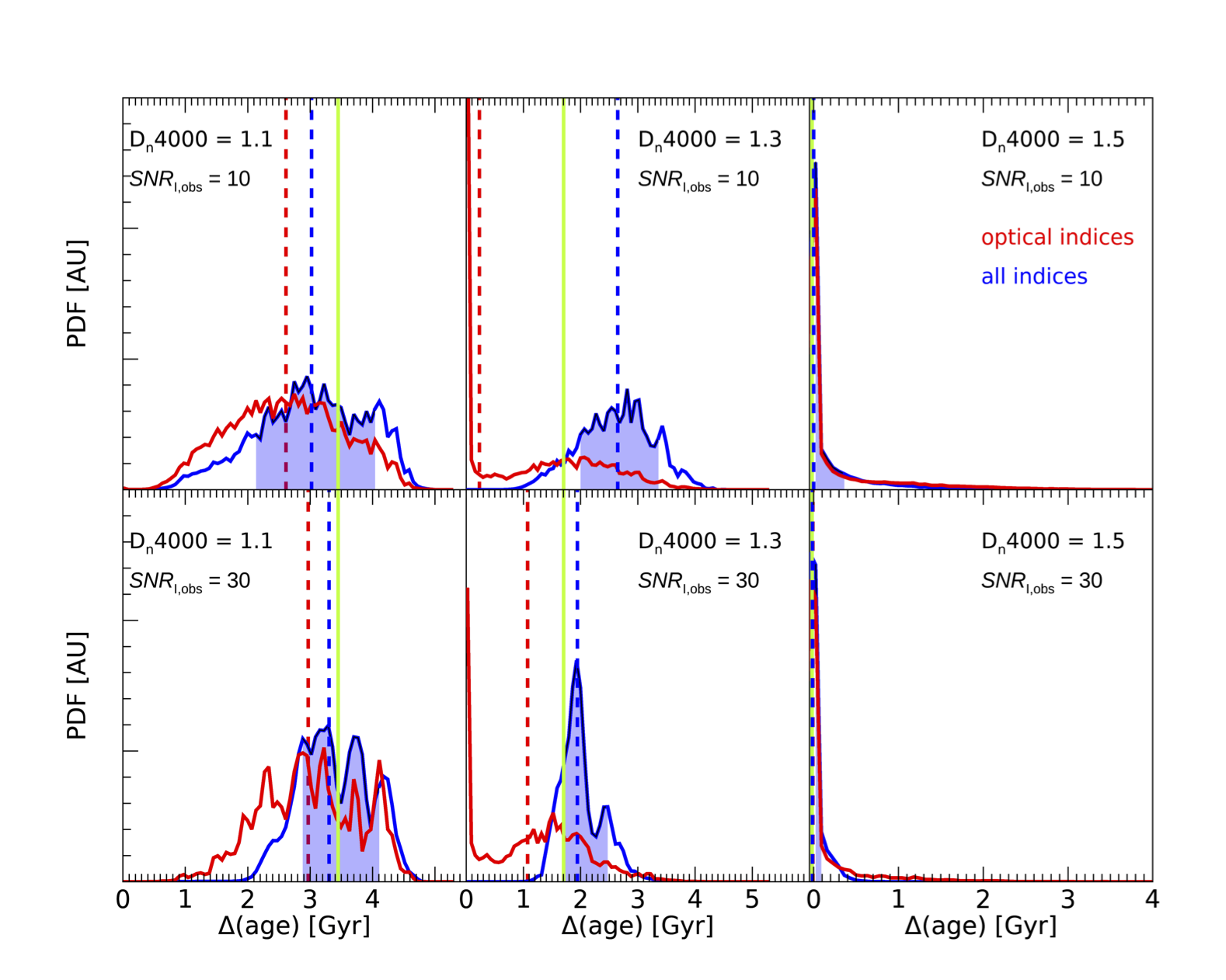}
\caption{As in Fig.~\ref{fig:PDF_example_age_r}, but for \texttt{$\Delta$(age)}.
The solid green line corresponds to the true value of \texttt{$\Delta$(age)} = 3.45 Gyr, 
\texttt{$\Delta$(age)} = 1.70 Gyr, and \texttt{$\Delta$(age)} = 0.02 Gyr, respectively.}
\label{fig:PDF_example_delta_age}
\end{figure*}


For illustration, we show in Figs.~\ref{fig:PDF_example_age_r} and \ref{fig:PDF_example_age_u} 
how the PDF of \texttt{age$_{\rm r}$} and \texttt{age$_{\rm u}$} vary using either 
only optical or all available indices. We choose as examples three galaxies with 
different spectral types as parametrized by their value of 
D$_{\rm n}4000\simeq[1.1,1.3,1.5]$ and two $\mathit{SNR}_{\rm I, obs}$ = [10, 30]. 
There is a clear improvement in constraining \texttt{age$_{\rm u}$} using 
ultraviolet indices, while their roles are almost negligible in 
constraining \texttt{age$_{\rm r}$}. 

Obviously the value of $\mathit{SNR}_{\rm I, obs}$ always plays a role,
with an expected narrowing of the PDF width around the true value as $\mathit{SNR}_{\rm I, obs}$ increases. 
However, and more interestingly, there is a noticeable difference as a function galaxy spectral type. 
For values of D$_{\rm n}4000 \gtrsim 1.5$ there is no improvement in constraining 
\texttt{age$_{\rm u}$} or \texttt{age$_{\rm r}$} by adding the information that comes from the bluer 
region of the spectrum. This result holds at both $\mathit{SNR}_{\rm I, obs}$ considered,
hinting at the physical interpretation that when a galaxy is truly old 
the information provided by the ultraviolet and optical indices coincide.  
On the contrary, for values of D$_{\rm n}4000 \lesssim 1.5$ and even 
at $\mathit{SNR}_{\rm I, obs}=10$, the role of ultraviolet indices in increasing our
ability of constraining \texttt{age$_{\rm u}$} is quite clear. Their usage helps in retrieving the true value 
and in reducing the uncertainty in the measurement. 
Even at $\mathit{SNR}_{\rm I, obs}=30$,
the use of optical indices alone cannot compete with the gain of information
introduced by the bluer part of the spectrum 
(see middle panels in Fig.~\ref{fig:PDF_example_age_u}, as an example).

Moving to the general picture, in Figs.~\ref{fig:Dn4000_age_r} and \ref{fig:Dn4000_age_u} we 
show the systematic relative deviations between true and measured values 
of \texttt{age$_{\rm r}$} and \texttt{age$_{\rm u}$} (using all available spectral indices)
as a function of D$_{\rm n}4000$ values, respectively. Each point on these plots is
color-coded according to the errors $\sigma_{\rm obs}$ as obtained from the PDF distribution.
We find no significant systematic deviations between true and measured values, 
as the median value of the distribution is consistent with zero.

We observe a clear trend in increasing $\sigma_{\rm obs}$ as a function of increasing  D$_{\rm n}4000$ values, both for \texttt{age$_{\rm r}$} and \texttt{age$_{\rm u}$} values.
The trend is more obvious at $\mathit{SNR}_{\rm I, obs}=10$, but persists as  
$\mathit{SNR}_{\rm I, obs}$ increases, even as the actual typical values reached by $\sigma_{\rm obs}$ decrease.
This is not surprising as the spectral evolution proceeds more slowly as a stellar population ages (i.e. for larger
values of D$_{\rm n}4000$), so that at older ages it takes larger absolute age differences to produce the same variation in observable quantities than at younger ages. In general, the relative (fractional) precision of age determination stays roughly constant with age. In fact, 
we find that $\sigma_{\rm obs}/$\texttt{age$_{\rm r,u}$} is
$\sim$20\%, $\sim$15\%, and $\sim$10\% for $\mathit{SNR}_{\rm I, obs}=[10, 20, 30]$ over the full age range, respectively.

\subsubsection{Results: \texttt{$\Delta$(age)} \label{subsec:SFH_delta_age}}

In this section we move to estimate the quantity 
\texttt{$\Delta$(age)} for our observed galaxies 
using the method described in Sect.~\ref{subsec:bayes}. 
The value of this physical parameter is directly computed
for each model in the library and its full PDF is estimated from the Bayesian 
analysis\footnote{That is, we do not estimate \texttt{$\Delta$(age)} from the difference of 
the fiducial estimates of \texttt{age$_{\rm r}$} and \texttt{age$_{\rm u}$}, 
but we treat \texttt{$\Delta$(age)} as a proper physical parameter in the models.}.

In Fig.~\ref{fig:PDF_example_delta_age} we plot the PDF of \texttt{$\Delta$(age)},
calculated by means of  either only optical or all available
indices 
at different $\mathit{SNR}_{\rm I, obs} = [10,30]$ and various values of 
D$_{\rm n}4000\simeq[1.1,1.3,1.5]$.
Similarly to \texttt{age$_{\rm u}$}, the role of ultraviolet indices in improving the retrieval
of the true value of \texttt{$\Delta$(age)} is significant, both in the accuracy and the precision of the estimate. 
At higher D$_{\rm n}4000$ values, \texttt{$\Delta$(age)} is intrinsically lower,
and equally constrained using only optical or all indices available;
however, the complexity of SFHs at lower D$_{\rm n}4000$ values results
in differences between $u$- and $r$-band light-weighted ages,
which are actually more accurately defined using bluer indices.
More important, the use of ultraviolet indices reduces the ambiguity of the peak
at \texttt{$\Delta$(age)} $\simeq 0$, which is present when only optical indices are used
(see middle panels in Fig.~\ref{fig:PDF_example_delta_age}, as an example).
Indeed, the information present in the bluer part of the spectrum allows us to 
prove the coexistence within the galaxy of a younger and older stellar population, 
information that would be irretrievable when looking only at redder wavelengths
\citep{Vazdekis2016}.


\begin{figure*}[t]
\centering
\includegraphics[scale=0.57, trim=0.7cm 0.5cm 1cm 0cm, clip=true]{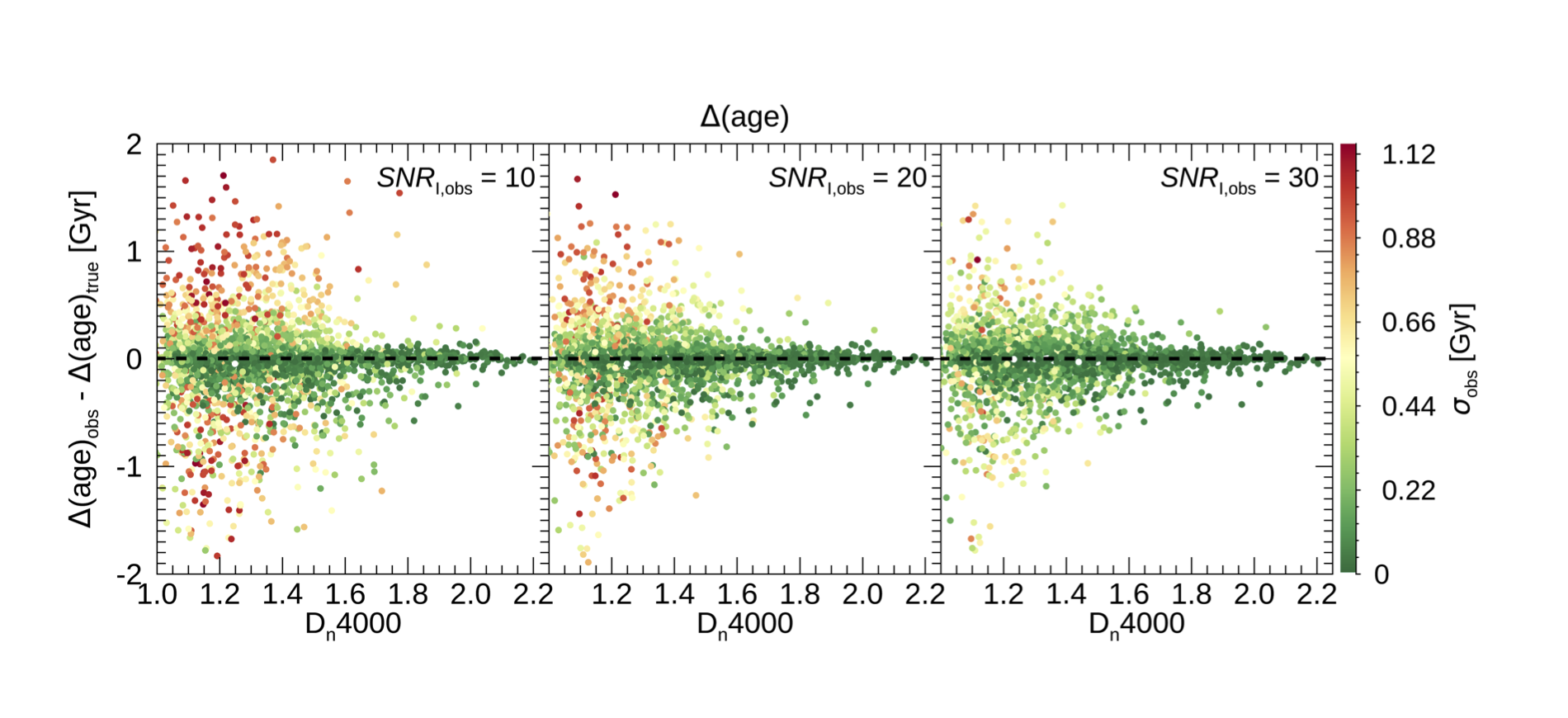}
\caption{Difference between observed and true value of \texttt{$\Delta$(age)} as a function of
D$_{\rm n}4000$, color-coded according to the observed errors retrieved from PDFs.
Galaxies are observed at $z=0.55$ and $\mathit{SNR}_{\rm I, obs} = [10,20,30]$ (from left to right).
The black horizontal dashed line corresponds to median value of the distribution.}
\label{fig:Dn4000_delta_age}
\end{figure*}


\begin{figure*}[t]
\centering
\includegraphics[scale=0.57, trim=0.7cm 0.5cm 1cm 0cm, clip=true]{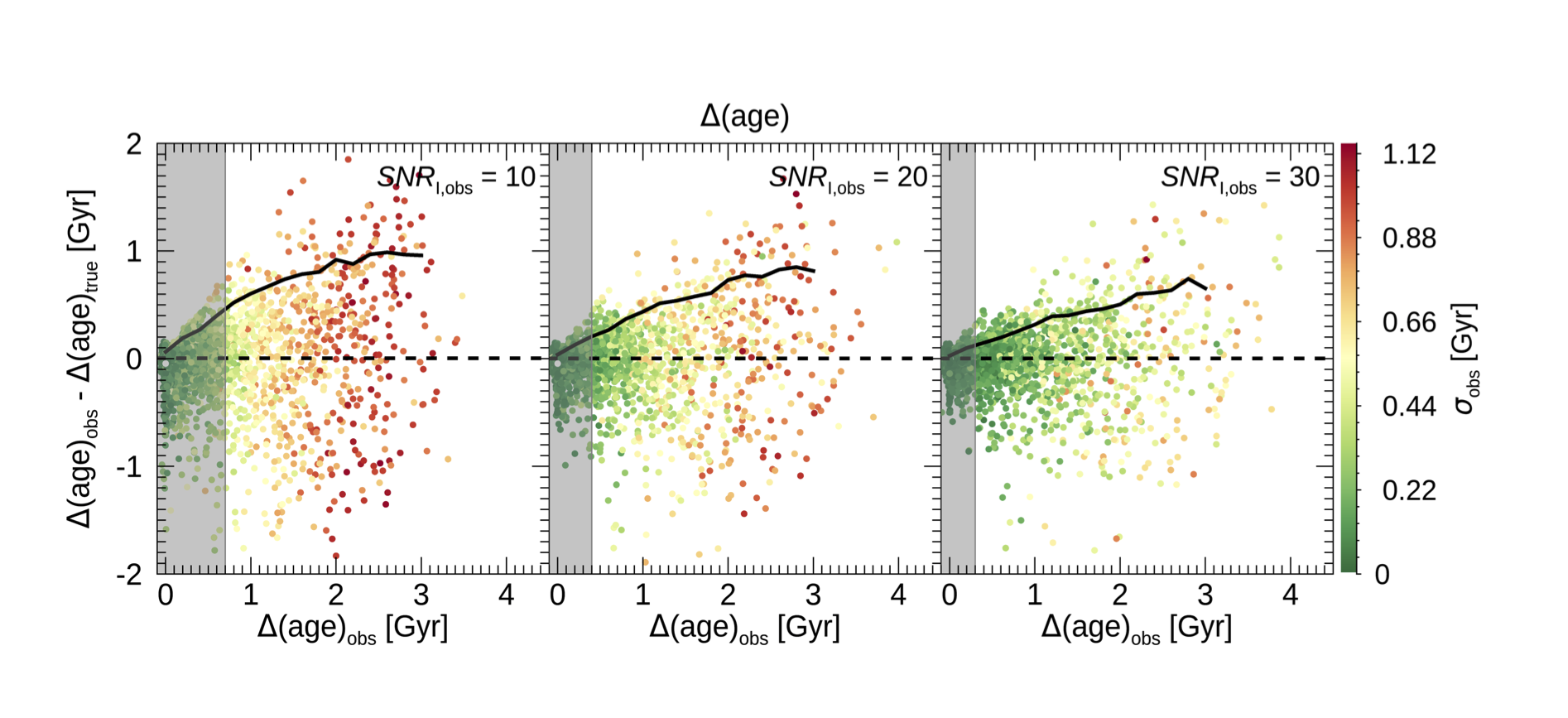}
\caption{Difference between observed and true values of \texttt{$\Delta$(age)} as a function of
\texttt{$\Delta$(age)$_{\rm obs}$}, color-coded according to the observed errors retrieved from PDFs.
Galaxies are observed at $z=0.55$ and $\mathit{SNR}_{\rm I, obs} = [10,20,30]$ (from left to right).
The black horizontal dashed line corresponds to the median value of
\texttt{$\Delta$(age)$_{\rm obs}$} - \texttt{$\Delta$(age)$_{\rm true}$},
while black solid line represents the trend of observed errors at various values of \texttt{$\Delta$(age)$_{\rm obs}$}.
The gray shaded area represents the region compatible with \texttt{$\Delta$(age)$_{\rm obs}=0$}.}
\label{fig:delta_age_delta_age}
\end{figure*}


In Fig.~\ref{fig:Dn4000_delta_age} we show our ability to recover the
true value of \texttt{$\Delta$(age)} as a function of D$_{\rm n}4000$
at $\mathit{SNR}_{\rm I, obs} = [10,20,30]$.
Each point on this plot is
color-coded according to $\sigma_{\rm obs}$ as obtained from the PDF.
The systematic error in recovered \texttt{$\Delta$(age)}
values is negligible even at the lower $\mathit{SNR}_{\rm I, obs}$. 
Furthermore, the scatter around the true value of \texttt{$\Delta$(age)}
is higher at lower D$_{\rm n}4000$ values and decreases at higher $\mathit{SNR}_{\rm I, obs}$, 
A similar trend holds for the value of $\sigma_{\rm obs}$.
It is worth noting that, if D$_{\rm n}4000$ > 1.5, where the expected values of \texttt{$\Delta$(age)} 
are intrinsically low, the ability of our methodology in retrieving the true value with small uncertainties is remarkable.
On the other hand, the higher observational uncertainties $\sigma_{\rm obs}$
at D$_{\rm n}4000$ < 1.5 are associated with the fact that this spectral feature is not an unambiguous indicator for \texttt{$\Delta$(age)}.
In the region where D$_{\rm n}4000$ < 1.5, there are galaxies whose intrinsic \texttt{$\Delta$(age)}
values may be quite large. The presence of 
higher scatter and higher observational errors is therefore the consequence of having
galaxies whose PDFs are centered on intrinsically higher values of \texttt{$\Delta$(age)} with broader PDFs.
This is confirmed by plotting in  Fig.~\ref{fig:delta_age_delta_age} the absolute difference between
observed and true values of \texttt{$\Delta$(age)} as a function of observed value of
\texttt{$\Delta$(age)}, again color-coded as in Fig.~\ref{fig:Dn4000_delta_age}.
In this plot higher values of \texttt{$\Delta$(age)} 
have higher observational errors (i.e., broader PDFs). When $\mathit{SNR}_{\rm I, obs}$
increases, the precision in the determination of \texttt{$\Delta$(age)} 
increases (i.e., lower observational errors), 
as well as its accuracy (i.e., lower scatter around zero).
However, even at lower $\mathit{SNR}_{\rm I, obs}$ the information provided
by the Bayesian analysis outperforms the coarse inference that classical diagnostics 
as those discussed in Sect.~\ref{sec:classical} can provide.

Fig.~\ref{fig:delta_age_delta_age} clearly shows that, irrespective of D$_{\rm n}4000$ values,
if the expected values of \texttt{$\Delta$(age)} are intrinsically low, 
the ability to retrieve the true value with small uncertainties is remarkable ($\sigma_{\rm obs} \sim 0.1$ Gyr).
This means that, even at $\mathit{SNR}_{\rm I, obs} = 10$, 
we are able not only to detect significant differences of \texttt{$\Delta$(age)} values,
but also to assert with reliable accuracy when \texttt{$\Delta$(age)} is consistent with zero.

The importance of \texttt{$\Delta$(age)} 
as a proxy to infer the presence of a significant spread in age between different stellar populations in individual 
galaxies relies on our ability to significantly determine a cutoff
for defining a non-zero value of this parameter.
Taking advantage of the Bayesian analysis and the full information provided by the PDF
allows us to identify the values of \texttt{$\Delta$(age)} that are robustly
different from zero at different $\mathit{SNR}_{\rm I, obs}$.
Thus, we divide the observed galaxies into bins of 0.1 Gyr in true \texttt{$\Delta$(age)} values
and evaluate for each galaxy the area of the PDF that is below \texttt{$\Delta$(age)} < 0.1 Gyr,
a value that we define to be the threshold under which the measurement is compatible with \texttt{$\Delta$(age)} = 0 Gyr.
In each bin of \texttt{$\Delta$(age)}, we then calculate the average value of 
the probability of being compatible with zero
considering each single galaxy in that bin.
We define as a reliable cutoff the bin at which the average probability is less than 10\%,
finding \texttt{$\Delta$(age)} = 0.7 Gyr at $\mathit{SNR}_{\rm I, obs}=10$,
\texttt{$\Delta$(age)} = 0.4 Gyr at $\mathit{SNR}_{\rm I, obs}=20$, and
\texttt{$\Delta$(age)} = 0.3 Gyr at $\mathit{SNR}_{\rm I, obs}=30$.
This is the gray shaded area shown in Fig.~\ref{fig:delta_age_delta_age}.

Combining this information with that in Fig.~\ref{fig:SFH_delta_age},
at the highest $SNR$ this analysis allows us to be sensitive to secondary episodes 
of star formation up to an age of $\sim 0.1$ Gyr for stellar populations older
$\sim 1.5$ Gyr, pushing up to an age of $\sim 1$ Gyr
for stellar populations older than $\sim 5$ Gyr.

By means of Bayesian analysis and the availability of ultraviolet indices, we have been able to drastically reduce the uncertainties 
on the differences of $u$- and $r$-band light-weighted ages,
making this parameter suitable for inferring physical properties in individual galaxies.
Moreover, the overall information contained in the PDF of \texttt{$\Delta$(age)}
allows us to properly distinguish between galaxies that display a coexistence of 
widely different stellar populations from those that present a more homogeneous, nearly coeval stellar population. 


\section{Summary and conclusions \label{sec:conclusions}}

The extragalactic community is on the eve of a new era in which large-field, high-multiplexing 
spectrographs with a remarkable wavelength span, mounted on four to eight meter class telescopes, 
will allow the collection of moderately-high $SNR$ spectra for tens of thousands 
of galaxies at intermediate to high redshift, encompassing both their rest-frame ultraviolet and optical range. 
While the ultimate science goal of our efforts in analyzing galaxy stellar continuum is to 
reconstruct the galaxy's star formation and chemical enrichment history, we must be aware that 
both theoretical limitations (intrinsic degeneracies) and technical limitations 
(chiefly the limited $SNR$ that can be attained in these surveys) 
will force us to focus on an essential but significant characterization.


\begin{figure}[t]
\centering
\includegraphics[scale=0.40, trim=0.5cm 0cm 1cm 0cm, clip=true]{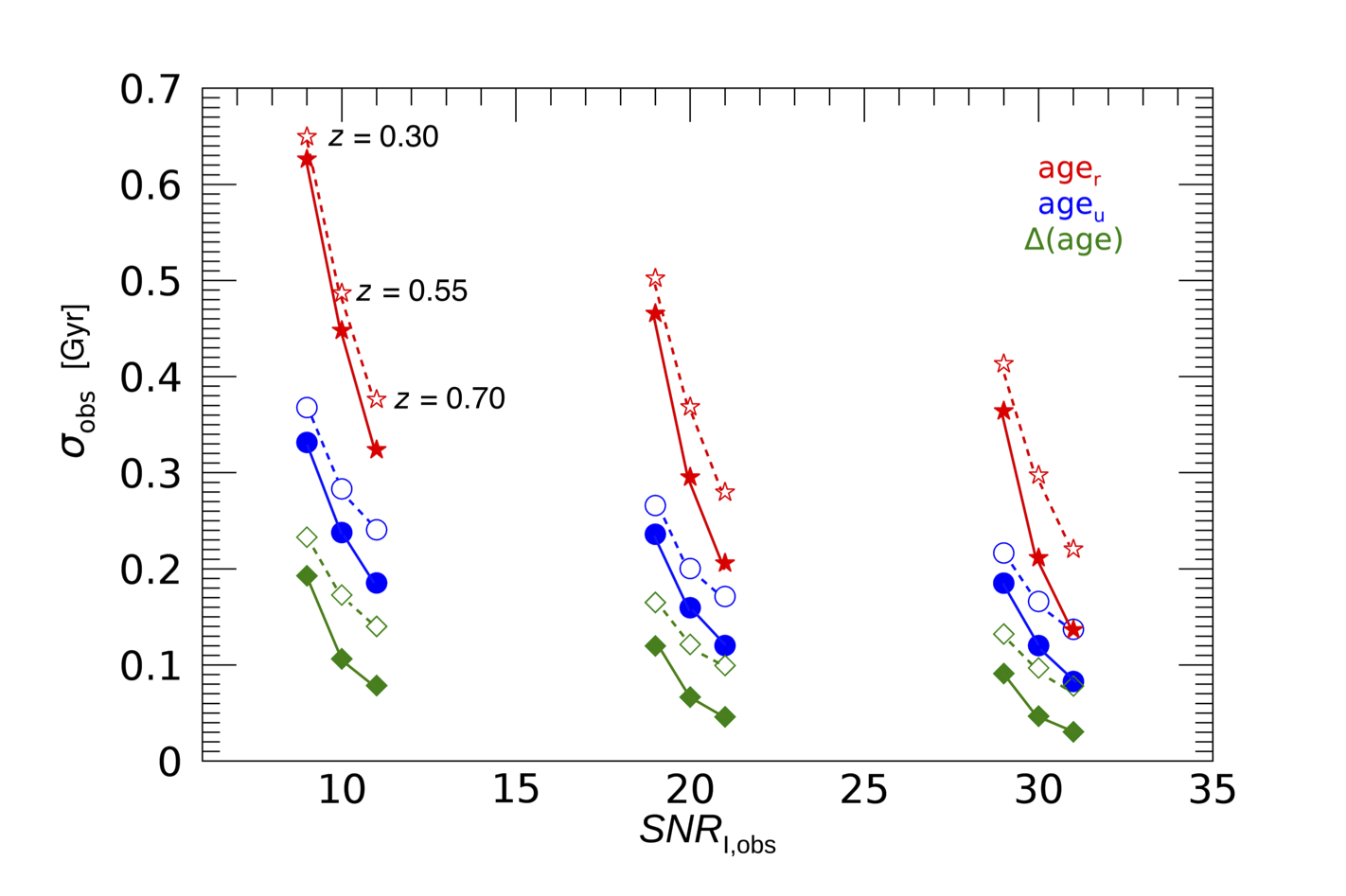}
\caption{Median value of observational errors on \texttt{age$_{\rm r}$} (red stars), 
\texttt{age$_{\rm u}$} (blue circles), and \texttt{$\Delta$(age)} (green diamonds)
as a function of $\mathit{SNR}_{\rm I, obs}$. Filled symbols correspond to Bayesian
analysis where all measured indices were used, 
while empty symbols correspond to PDF retrieved using only optical indices.}
\label{fig:sigma_pdf_SN_z}
\end{figure}


In this work we concentrate in particular on the potential given by the simultaneous 
ultraviolet and visible rest-frame coverage, despite a moderate $SNR$ ($\sim10-30$). 
Starting from the consideration that stellar populations of different ages affect 
the spectrum at different wavelengths differently, we investigate how the 
difference \texttt{$\Delta$(age)} between the age light-weighted in the rest-frame 
$u$- and $r$-band can be used as a simple tool to 
discern the complexity of star-formation histories in galaxies.
\texttt{$\Delta$(age)} is essentially a diagnostic of coexistence of old 
(ultraviolet-faint) stellar populations and young (ultraviolet-bright) stellar populations. 
Using a vast spectral library of SFHs, including both smooth and bursty ones, 
we have demonstrated that a large \texttt{$\Delta$(age)} is a symptom of either a 
smooth but very extended SFH or of a SFH with widely spaced peaks 
(e.g., an old "secular" peak and a recent burst). Although \texttt{$\Delta$(age)} 
cannot discriminate between these two scenarios by itself, it has the great 
advantage of being robust against degeneracies and largely independent from 
the actual implementation of SFHs in models, provided that enough complexity is included.

Based on realistic simulations of WEAVE-StePS spectra, we have assessed our future 
ability to measure \texttt{$\Delta$(age)} at different redshifts and $SNR$s, 
and its consequent diagnostic power. More specifically, we have simulated spectra at three 
different redshifts $z=[0.3,0.55,0.7]$ and three different
$SNR$ in the observed-frame I-band, $\mathit{SNR}_{\rm I, obs}=[10,20,30]$. 
We have taken into account the typical
transmission, efficiency, and noise derived from the characteristic of the observing site, 
telescope, fibers, spectrograph, and CCDs for the low-resolution configuration adopted 
by StePS for the WEAVE spectrograph on the WHT in La Palma. 

We have adopted the Bayesian approach developed by \citet{Gallazzi2005} and 
confronted the spectral absorption indices as measured in the simulated spectra with those in our vast library of models, 
to obtain the marginalized PDF for \texttt{$\Delta$(age)}, as well as for \texttt{age$_{\rm u}$} and \texttt{age$_{\rm r}$}. 
Thanks to the Bayesian approach, we have obtained remarkable accuracy and precision in our estimates, in particular 
when indices in both the ultraviolet and visible range are combined. 
In Fig.~\ref{fig:sigma_pdf_SN_z} we show the trend of the median uncertainty in  
\texttt{age$_{\rm r}$}, \texttt{age$_{\rm u}$}, and \texttt{$\Delta$(age)} at different $\mathit{SNR}_{\rm I, obs}$ as a 
function of redshift $z=[0.3,0.55,0.7]$. 
Errors on every parameter decrease as both $\mathit{SNR}_{\rm I, obs}$ and redshift increase.
While the first behaviour is expected, the improvement at higher redshift can
be ascribed to the increasing number of ultraviolet spectral features that become available,
and the better $SNR$ at which they are measured due to the shift of the spectrum 
toward the I-band, where $\mathit{SNR}$ is fixed (see Sect.~\ref{subsec:observed_spectra}).
Under the assumption of the existence of an old underlying stellar population (as is the case for the
massive galaxies we are mainly going to target in StePS), these errors translate into the ability to
identify secondary episodes
of star formation younger than $\sim 1$ Gyr for stellar populations older than
$\sim 5$ Gyr, or younger than $\sim 0.1$ Gyr for stellar populations older than
$\sim 1.5$ Gyr. In this sense, \texttt{$\Delta$(age)} and \texttt{age$_{\rm r}$} together present
a powerful diagnostic of rejuvenation episodes in old galaxies, or, conversely, allow us to check for
the truly passive evolution of old stellar populations with a much larger sensitivity than \texttt{age$_{\rm r}$} alone. 

Finally, it is worth stressing that the analysis presented in this paper can be easily adapted for
similar instruments/surveys and can be tested with different stellar population libraries, thus allowing
for the benchmarking of both instrumental effects and theoretical models and assumptions.

\begin{acknowledgements} 
We thank the anonymous referee for carefully reading the manuscript
and for the suggestions that helped us to improve the way we presented our results.
We wish to thank M.~Irwin 
for very useful discussions and valuable suggestions.
L.~C.~wish to acknowledge financial support from Premiale 2015 MITiC, program 1.05.06.10  
and Comunidad de Madrid under Atracci\'on de Talento grant 2018-T2/TIC-11612.
A.~I., S.~Z., M.~L., A.~G., A.~M., and C.~T.~acknowledge the financial support from the INAF PRIN-SKA 2017
program 1.05.01.88.04.ESKAPE-HI.
I.~L.~acknowledges financial support from INAF-WEAVE funds, program 1.05.03.04.05. 
R.~G.~B.~acknowledges support from the Spanish Ministry of Economy and Competitiveness 
through grant 205 AYA2016-77846-P and the State Agency for Research of 
the Spanish MCIU through the ``Center of Excellence Severo Ochoa'' award to 2017 
the Instituto de Astrof\'isica de Andaluc\'ia (SEV-2017-0709).
A.~F.~M.~has received financial support through the Postdoctoral Junior 
Leader Fellowship Programme from ``La Caixa'' Banking Foundation (LCF/BQ/LI18/11630007).
B.~V. acknowledges the financial support from INAF Main Stream 2018 (P.~I.:~B.~Vulcani).
\end{acknowledgements}


\bibliographystyle{aa}
\bibliography{Costantin}

\begin{thebibliography}{82}
\expandafter\ifx\csname natexlab\endcsname\relax\def\natexlab#1{#1}\fi

\bibitem[{{Abramson} {et~al.}(2015){Abramson}, {Gladders}, {Dressler},
  {Oemler}, {Poggianti}, \& {Vulcani}}]{Abramson2015}
{Abramson}, L.~E., {Gladders}, M.~D., {Dressler}, A., {et~al.} 2015, \apjl,
  801, L12

\bibitem[{{Baldry} {et~al.}(2006){Baldry}, {Balogh}, {Bower}, {Glazebrook},
  {Nichol}, {Bamford}, \& {Budavari}}]{Baldry2006}
{Baldry}, I.~K., {Balogh}, M.~L., {Bower}, R.~G., {et~al.} 2006, \mnras, 373,
  469

\bibitem[{{Baldry} {et~al.}(2004){Baldry}, {Glazebrook}, {Brinkmann},
  {Ivezi{\'c}}, {Lupton}, {Nichol}, \& {Szalay}}]{Baldry2004}
{Baldry}, I.~K., {Glazebrook}, K., {Brinkmann}, J., {et~al.} 2004, \apj, 600,
  681

\bibitem[{{Balogh} {et~al.}(1999){Balogh}, {Morris}, {Yee}, {Carlberg}, \&
  {Ellingson}}]{Balogh1999}
{Balogh}, M.~L., {Morris}, S.~L., {Yee}, H.~K.~C., {Carlberg}, R.~G., \&
  {Ellingson}, E. 1999, \apj, 527, 54

\bibitem[{{Bamford} {et~al.}(2009){Bamford}, {Nichol}, {Baldry}, {Land},
  {Lintott}, {Schawinski}, {Slosar}, {Szalay}, {Thomas}, {Torki}, {Andreescu},
  {Edmondson}, {Miller}, {Murray}, {Raddick}, \& {Vandenberg}}]{Bamford2009}
{Bamford}, S.~P., {Nichol}, R.~C., {Baldry}, I.~K., {et~al.} 2009, \mnras, 393,
  1324

\bibitem[{{Behroozi} {et~al.}(2013){Behroozi}, {Wechsler}, \&
  {Conroy}}]{Behroozi2013}
{Behroozi}, P.~S., {Wechsler}, R.~H., \& {Conroy}, C. 2013, \apj, 770, 57

\bibitem[{{Benn} \& {Ellison}(1998)}]{BennEllison1998}
{Benn}, C.~R. \& {Ellison}, S.~L. 1998, \nar, 42, 503

\bibitem[{{Blanton} {et~al.}(2003){Blanton}, {Hogg}, {Bahcall}, {Baldry},
  {Brinkmann}, {Csabai}, {Eisenstein}, {Fukugita}, {Gunn}, {Ivezi{\'c}},
  {Lamb}, {Lupton}, {Loveday}, {Munn}, {Nichol}, {Okamura}, {Schlegel},
  {Shimasaku}, {Strauss}, {Vogeley}, \& {Weinberg}}]{Blanton2003}
{Blanton}, M.~R., {Hogg}, D.~W., {Bahcall}, N.~A., {et~al.} 2003, \apj, 594,
  186

\bibitem[{{Blanton} \& {Moustakas}(2009)}]{BlantonMoustakas2009}
{Blanton}, M.~R. \& {Moustakas}, J. 2009, \araa, 47, 159

\bibitem[{{Bruzual} \& {Charlot}(2003)}]{Bruzual2003}
{Bruzual}, G. \& {Charlot}, S. 2003, \mnras, 344, 1000

\bibitem[{{Cappellari}(2017)}]{Cappellari2017}
{Cappellari}, M. 2017, \mnras, 466, 798

\bibitem[{{Cappellari} \& {Emsellem}(2004)}]{Cappellari2004}
{Cappellari}, M. \& {Emsellem}, E. 2004, \pasp, 116, 138

\bibitem[{{Carnall} {et~al.}(2018){Carnall}, {McLure}, {Dunlop}, \&
  {Dav{\'e}}}]{Carnall2018}
{Carnall}, A.~C., {McLure}, R.~J., {Dunlop}, J.~S., \& {Dav{\'e}}, R. 2018,
  \mnras, 480, 4379

\bibitem[{{Chabrier}(2003)}]{Chabrier2003}
{Chabrier}, G. 2003, \apjl, 586, L133

\bibitem[{{Charlot} \& {Fall}(2000)}]{Charlot2000}
{Charlot}, S. \& {Fall}, S.~M. 2000, \apj, 539, 718

\bibitem[{{Chauke} {et~al.}(2019){Chauke}, {van der Wel}, {Pacifici},
  {Bezanson}, {Wu}, {Gallazzi}, {Straatman}, {Franx}, {Bari{\v s}i{\'c}},
  {Bell}, {van Houdt}, {Maseda}, {Muzzin}, {Sobral}, \& {Spilker}}]{Chauke2019}
{Chauke}, P., {van der Wel}, A., {Pacifici}, C., {et~al.} 2019, \apj, 877, 48

\bibitem[{{Cid Fernandes} {et~al.}(2005){Cid Fernandes}, {Mateus}, {Sodr{\'e}},
  {Stasi{\'n}ska}, \& {Gomes}}]{CidFernandes2005}
{Cid Fernandes}, R., {Mateus}, A., {Sodr{\'e}}, L., {Stasi{\'n}ska}, G., \&
  {Gomes}, J.~M. 2005, \mnras, 358, 363

\bibitem[{{Costantin} {et~al.}(2018){Costantin}, {M{\'e}ndez-Abreu}, {Corsini},
  {Eliche-Moral}, {Tapia}, {Morelli}, {Dalla Bont{\`a}}, \&
  {Pizzella}}]{Costantin2018}
{Costantin}, L., {M{\'e}ndez-Abreu}, J., {Corsini}, E.~M., {et~al.} 2018, \aap,
  609, A132

\bibitem[{{Costantin} {et~al.}(2017){Costantin}, {M{\'e}ndez-Abreu}, {Corsini},
  {Morelli}, {Aguerri}, {Dalla Bont{\`a}}, \& {Pizzella}}]{Costantin2017}
{Costantin}, L., {M{\'e}ndez-Abreu}, J., {Corsini}, E.~M., {et~al.} 2017, \aap,
  601, A84

\bibitem[{{Daddi} {et~al.}(2005){Daddi}, {Renzini}, {Pirzkal}, {Cimatti},
  {Malhotra}, {Stiavelli}, {Xu}, {Pasquali}, {Rhoads}, {Brusa}, {di Serego
  Alighieri}, {Ferguson}, {Koekemoer}, {Moustakas}, {Panagia}, \&
  {Windhorst}}]{Daddi2005}
{Daddi}, E., {Renzini}, A., {Pirzkal}, N., {et~al.} 2005, \apj, 626, 680

\bibitem[{{Dalton} {et~al.}(2016){Dalton}, {Ham}, {Trager}, {Abrams},
  {Bonifacio}, {Aguerri}, {Middleton}, {Benn}, {Rogers}, {Stuik}, {Carrasco},
  {Vallenari}, {Jin}, \& {Lewis}}]{Dalton2016}
{Dalton}, G., {Ham}, S.~J., {Trager}, S., {et~al.} 2016, in Society of
  Photo-Optical Instrumentation Engineers (SPIE) Conference Series, Vol. 9913,
  \procspie, 99132X

\bibitem[{{Dalton} {et~al.}(2012){Dalton}, {Trager}, {Abrams}, {Carter},
  {Bonifacio}, {Aguerri}, {MacIntosh}, {Evans}, {Lewis}, {Navarro}, {Agocs},
  {Dee}, {Rousset}, {Tosh}, {Middleton}, {Pragt}, {Terrett}, {Brock}, {Benn},
  {Verheijen}, {Cano Infantes}, {Bevil}, {Steele}, {Mottram}, {Bates},
  {Gribbin}, {Rey}, {Rodriguez}, {Delgado}, {Guinouard}, {Walton}, {Irwin},
  {Jagourel}, {Stuik}, {Gerlofsma}, {Roelfsma}, {Skillen}, {Ridings},
  {Balcells}, {Daban}, {Gouvret}, {Venema}, \& {Girard}}]{Dalton2012}
{Dalton}, G., {Trager}, S.~C., {Abrams}, D.~C., {et~al.} 2012, in Society of
  Photo-Optical Instrumentation Engineers (SPIE) Conference Series, Vol. 8446,
  Ground-based and Airborne Instrumentation for Astronomy IV, 84460P

\bibitem[{{Davidzon} {et~al.}(2017){Davidzon}, {Ilbert}, {Laigle}, {Coupon},
  {McCracken}, {Delvecchio}, {Masters}, {Capak}, {Hsieh}, {Le F{\`e}vre},
  {Tresse}, {Bethermin}, {Chang}, {Faisst}, {Le Floc'h}, {Steinhardt}, {Toft},
  {Aussel}, {Dubois}, {Hasinger}, {Salvato}, {Sanders}, {Scoville}, \&
  {Silverman}}]{Davidzon2017}
{Davidzon}, I., {Ilbert}, O., {Laigle}, C., {et~al.} 2017, \aap, 605, A70

\bibitem[{{de Jong} {et~al.}(2019){de Jong}, {Agertz}, {Berbel}, {Aird},
  {Alexander}, {Amarsi}, {Anders}, {Andrae}, {Ansarinejad}, {Ansorge},
  {Antilogus}, {Anwand -Heerwart}, {Arentsen}, {Arnadottir}, {Asplund},
  {Auger}, {Azais}, {Baade}, {Baker}, {Baker}, {Balbinot}, {Baldry}, {Banerji},
  {Barden}, {Barklem}, {Barth{\'e}l{\'e}my-Mazot}, {Battistini}, {Bauer},
  {Bell}, {Bellido-Tirado}, {Bellstedt}, {Belokurov}, {Bensby}, {Bergemann},
  {Bestenlehner}, {Bielby}, {Bilicki}, {Blake}, {Bland-Hawthorn}, {Boeche},
  {Boland}, {Boller}, {Bongard}, {Bongiorno}, {Bonifacio}, {Boudon}, {Brooks},
  {Brown}, {Brown}, {Br{\"u}ggen}, {Brynnel}, {Brzeski}, {Buchert},
  {Buschkamp}, {Caffau}, {Caillier}, {Carrick}, {Casagrande}, {Case}, {Casey},
  {Cesarini}, {Cescutti}, {Chapuis}, {Chiappini}, {Childress}, {Christlieb},
  {Church}, {Cioni}, {Cluver}, {Colless}, {Collett}, {Comparat}, {Cooper},
  {Couch}, {Courbin}, {Croom}, {Croton}, {Daguis{\'e}}, {Dalton}, {Davies},
  {Davis}, {de Laverny}, {Deason}, {Dionies}, {Disseau}, {Doel}, {D{\"o}scher},
  {Driver}, {Dwelly}, {Eckert}, {Edge}, {Edvardsson}, {Youssoufi}, {Elhaddad},
  {Enke}, {Erfanianfar}, {Farrell}, {Fechner}, {Feiz}, {Feltzing}, {Ferreras},
  {Feuerstein}, {Feuillet}, {Finoguenov}, {Ford}, {Fotopoulou}, {Fouesneau},
  {Frenk}, {Frey}, {Gaessler}, {Geier}, {Fusillo}, {Gerhard}, {Giannantonio},
  {Giannone}, {Gibson}, {Gillingham}, {Gonz{\'a}lez-Fern{\'a}ndez},
  {Gonzalez-Solares}, {Gottloeber}, {Gould}, {Grebel}, {Gueguen}, {Guiglion},
  {Haehnelt}, {Hahn}, {Hansen}, {Hartman}, {Hauptner}, {Hawkins}, {Haynes},
  {Haynes}, {Heiter}, {Helmi}, {Aguayo}, {Hewett}, {Hinton}, {Hobbs}, {Hoenig},
  {Hofman}, {Hook}, {Hopgood}, {Hopkins}, {Hourihane}, {Howes}, {Howlett},
  {Huet}, {Irwin}, {Iwert}, {Jablonka}, {Jahn}, {Jahnke}, {Jarno}, {Jin},
  {Jofre}, {Johl}, {Jones}, {J{\"o}nsson}, {Jordan}, {Karovicova}, {Khalatyan},
  {Kelz}, {Kennicutt}, {King}, {Kitaura}, {Klar}, {Klauser}, {Kneib}, {Koch},
  {Koposov}, {Kordopatis}, {Korn}, {Kosmalski}, {Kotak}, {Kovalev}, {Kreckel},
  {Kripak}, {Krumpe}, {Kuijken}, {Kunder}, {Kushniruk}, {Lam}, {Lamer},
  {Laurent}, {Lawrence}, {Lehmitz}, {Lemasle}, {Lewis}, {Li}, {Lidman}, {Lind},
  {Liske}, {Lizon}, {Loveday}, {Ludwig}, {McDermid}, {Maguire}, {Mainieri},
  {Mali}, {Mandel}, {Mandel}, {Mannering}, {Martell}, {Martinez Delgado},
  {Matijevic}, {McGregor}, {McMahon}, {McMillan}, {Mena}, {Merloni}, {Meyer},
  {Michel}, {Micheva}, {Migniau}, {Minchev}, {Monari}, {Muller}, {Murphy},
  {Muthukrishna}, {Nandra}, {Navarro}, {Ness}, {Nichani}, {Nichol}, {Nicklas},
  {Niederhofer}, {Norberg}, {Obreschkow}, {Oliver}, {Owers}, {Pai},
  {Pankratow}, {Parkinson}, {Paschke}, {Paterson}, {Pecontal}, {Parry},
  {Phillips}, {Pillepich}, {Pinard}, {Pirard}, {Piskunov}, {Plank},
  {Pl{\"u}schke}, {Pons}, {Popesso}, {Power}, {Pragt}, {Pramskiy}, {Pryer},
  {Quattri}, {Queiroz}, {Quirrenbach}, {Rahurkar}, {Raichoor}, {Ramstedt},
  {Rau}, {Recio-Blanco}, {Reiss}, {Renaud}, {Revaz}, {Rhode}, {Richard},
  {Richter}, {Rix}, {Robotham}, {Roelfsema}, {Romaniello}, {Rosario},
  {Rothmaier}, {Roukema}, {Ruchti}, {Rupprecht}, {Rybizki}, {Ryde}, {Saar},
  {Sadler}, {Sahl{\'e}n}, {Salvato}, {Sassolas}, {Saunders}, {Saviauk},
  {Sbordone}, {Schmidt}, {Schnurr}, {Scholz}, {Schwope}, {Seifert}, {Shanks},
  {Sheinis}, {Sivov}, {Sk{\'u}lad{\'o}ttir}, {Smartt}, {Smedley}, {Smith},
  {Smith}, {Sorce}, {Spitler}, {Starkenburg}, {Steinmetz}, {Stilz}, {Storm},
  {Sullivan}, {Sutherland}, {Swann}, {Tamone}, {Taylor}, {Teillon}, {Tempel},
  {ter Horst}, {Thi}, {Tolstoy}, {Trager}, {Traven}, {Tremblay}, {Tresse},
  {Valentini}, {van de Weygaert}, {van den Ancker}, {Veljanoski}, {Venkatesan},
  {Wagner}, {Wagner}, {Walcher}, {Waller}, {Walton}, {Wang}, {Winkler},
  {Wisotzki}, {Worley}, {Worseck}, {Xiang}, {Xu}, {Yong}, {Zhao}, {Zheng},
  {Zscheyge}, \& {Zucker}}]{deJong2019}
{de Jong}, R.~S., {Agertz}, O., {Berbel}, A.~A., {et~al.} 2019, The Messenger,
  175, 3

\bibitem[{{de Lorenzo-C{\'a}ceres} {et~al.}(2013){de Lorenzo-C{\'a}ceres},
  {Falc{\'o}n-Barroso}, \& {Vazdekis}}]{deLorenzoCaceres2013}
{de Lorenzo-C{\'a}ceres}, A., {Falc{\'o}n-Barroso}, J., \& {Vazdekis}, A. 2013,
  \mnras, 431, 2397

\bibitem[{{Diemer} {et~al.}(2017){Diemer}, {Sparre}, {Abramson}, \&
  {Torrey}}]{Diemer2017}
{Diemer}, B., {Sparre}, M., {Abramson}, L.~E., \& {Torrey}, P. 2017, \apj, 839,
  26

\bibitem[{{Falc{\'o}n-Barroso} {et~al.}(2011){Falc{\'o}n-Barroso},
  {S{\'a}nchez-Bl{\'a}zquez}, {Vazdekis}, {Ricciardelli}, {Cardiel}, {Cenarro},
  {Gorgas}, \& {Peletier}}]{FalconBarroso2011}
{Falc{\'o}n-Barroso}, J., {S{\'a}nchez-Bl{\'a}zquez}, P., {Vazdekis}, A.,
  {et~al.} 2011, \aap, 532, A95

\bibitem[{{Fanelli} {et~al.}(1992){Fanelli}, {O'Connell}, {Burstein}, \&
  {Wu}}]{Fanelli1992}
{Fanelli}, M.~N., {O'Connell}, R.~W., {Burstein}, D., \& {Wu}, C.-C. 1992,
  \apjs, 82, 197

\bibitem[{{Ferr{\'e}-Mateu} {et~al.}(2014){Ferr{\'e}-Mateu},
  {S{\'a}nchez-Bl{\'a}zquez}, {Vazdekis}, \& {de la Rosa}}]{FerreMateu2014}
{Ferr{\'e}-Mateu}, A., {S{\'a}nchez-Bl{\'a}zquez}, P., {Vazdekis}, A., \& {de
  la Rosa}, I.~G. 2014, \apj, 797, 136

\bibitem[{{Gallazzi} {et~al.}(2014){Gallazzi}, {Bell}, {Zibetti}, {Brinchmann},
  \& {Kelson}}]{Gallazzi2014}
{Gallazzi}, A., {Bell}, E.~F., {Zibetti}, S., {Brinchmann}, J., \& {Kelson},
  D.~D. 2014, \apj, 788, 72

\bibitem[{{Gallazzi} {et~al.}(2005){Gallazzi}, {Charlot}, {Brinchmann},
  {White}, \& {Tremonti}}]{Gallazzi2005}
{Gallazzi}, A., {Charlot}, S., {Brinchmann}, J., {White}, S.~D.~M., \&
  {Tremonti}, C.~A. 2005, \mnras, 362, 41

\bibitem[{{Gargiulo} {et~al.}(2017){Gargiulo}, {Bolzonella}, {Scodeggio},
  {Krywult}, {De Lucia}, {Guzzo}, {Garilli}, {Granett}, {de la Torre}, {Abbas},
  {Adami}, {Arnouts}, {Bottini}, {Cappi}, {Cucciati}, {Davidzon}, {Franzetti},
  {Fritz}, {Haines}, {Hawken}, {Iovino}, {Le Brun}, {Le F{\`e}vre}, {Maccagni},
  {Ma{\l}ek}, {Marulli}, {Moutard}, {Polletta}, {Pollo}, {Tasca}, {Tojeiro},
  {Vergani}, {Zanichelli}, {Zamorani}, {Bel}, {Branchini}, {Coupon}, {Ilbert},
  {Moscardini}, \& {Peacock}}]{Gargiulo2017}
{Gargiulo}, A., {Bolzonella}, M., {Scodeggio}, M., {et~al.} 2017, \aap, 606,
  A113

\bibitem[{{Girardi} {et~al.}(2000){Girardi}, {Bressan}, {Bertelli}, \&
  {Chiosi}}]{Girardi2000}
{Girardi}, L., {Bressan}, A., {Bertelli}, G., \& {Chiosi}, C. 2000, Astronomy
  and Astrophysics Supplement Series, 141, 371

\bibitem[{{Gladders} {et~al.}(2013){Gladders}, {Oemler}, {Dressler},
  {Poggianti}, {Vulcani}, \& {Abramson}}]{Gladders2013}
{Gladders}, M.~D., {Oemler}, A., {Dressler}, A., {et~al.} 2013, \apj, 770, 64

\bibitem[{{Guglielmo} {et~al.}(2019){Guglielmo}, {Poggianti}, {Vulcani},
  {Maurogordato}, {Fritz}, {Bolzonella}, {Fotopoulou}, {Adami}, \&
  {Pierre}}]{Guglielmo2019}
{Guglielmo}, V., {Poggianti}, B.~M., {Vulcani}, B., {et~al.} 2019, \aap, 625,
  A112

\bibitem[{{Hahn} {et~al.}(2015){Hahn}, {Blanton}, {Moustakas}, {Coil}, {Cool},
  {Eisenstein}, {Skibba}, {Wong}, \& {Zhu}}]{Hahn2015}
{Hahn}, C., {Blanton}, M.~R., {Moustakas}, J., {et~al.} 2015, \apj, 806, 162

\bibitem[{{Haines} {et~al.}(2017){Haines}, {Iovino}, {Krywult}, {Guzzo},
  {Davidzon}, {Bolzonella}, {Garilli}, {Scodeggio}, {Granett}, {de la Torre},
  {De Lucia}, {Abbas}, {Adami}, {Arnouts}, {Bottini}, {Cappi}, {Cucciati},
  {Franzetti}, {Fritz}, {Gargiulo}, {Le Brun}, {Le F{\`e}vre}, {Maccagni},
  {Ma{\l}ek}, {Marulli}, {Moutard}, {Polletta}, {Pollo}, {Tasca}, {Tojeiro},
  {Vergani}, {Zanichelli}, {Zamorani}, {Bel}, {Branchini}, {Coupon}, {Ilbert},
  {Moscardini}, {Peacock}, \& {Siudek}}]{Haines2017}
{Haines}, C.~P., {Iovino}, A., {Krywult}, J., {et~al.} 2017, \aap, 605, A4

\bibitem[{{Iovino} {et~al.}(2010){Iovino}, {Cucciati}, {Scodeggio}, {Knobel},
  {Kova{\v{c}}}, {Lilly}, {Bolzonella}, {Tasca}, {Zamorani}, {Zucca}, {Caputi},
  {Pozzetti}, {Oesch}, {Lamareille}, {Halliday}, {Bardelli}, {Finoguenov},
  {Guzzo}, {Kampczyk}, {Maier}, {Tanaka}, {Vergani}, {Carollo}, {Contini},
  {Kneib}, {Le F{\`e}vre}, {Mainieri}, {Renzini}, {Bongiorno}, {Coppa}, {de la
  Torre}, {de Ravel}, {Franzetti}, {Garilli}, {Le Borgne}, {Le Brun},
  {Mignoli}, {Pell{\`o}}, {Peng}, {Perez-Montero}, {Ricciardelli}, {Silverman},
  {Tresse}, {Abbas}, {Bottini}, {Cappi}, {Cassata}, {Cimatti}, {Koekemoer},
  {Leauthaud}, {Maccagni}, {Marinoni}, {McCracken}, {Memeo}, {Meneux},
  {Porciani}, {Scaramella}, {Schiminovich}, \& {Scoville}}]{Iovino2010}
{Iovino}, A., {Cucciati}, O., {Scodeggio}, M., {et~al.} 2010, \aap, 509, A40

\bibitem[{{J{\o}rgensen} \& {Chiboucas}(2013)}]{Jorgensen2013}
{J{\o}rgensen}, I. \& {Chiboucas}, K. 2013, \aj, 145, 77

\bibitem[{{Kauffmann} {et~al.}(2003){Kauffmann}, {Heckman}, {White}, {Charlot},
  {Tremonti}, {Brinchmann}, {Bruzual}, {Peng}, {Seibert}, {Bernardi},
  {Blanton}, {Brinkmann}, {Castander}, {Cs{\'a}bai}, {Fukugita}, {Ivezic},
  {Munn}, {Nichol}, {Padmanabhan}, {Thakar}, {Weinberg}, \&
  {York}}]{Kauffmann2003}
{Kauffmann}, G., {Heckman}, T.~M., {White}, S.~D.~M., {et~al.} 2003, \mnras,
  341, 33

\bibitem[{{Kauffmann} {et~al.}(2004){Kauffmann}, {White}, {Heckman},
  {M{\'e}nard}, {Brinchmann}, {Charlot}, {Tremonti}, \&
  {Brinkmann}}]{Kauffmann2004}
{Kauffmann}, G., {White}, S. D.~M., {Heckman}, T.~M., {et~al.} 2004, \mnras,
  353, 713

\bibitem[{{Kaviraj} {et~al.}(2009){Kaviraj}, {Peirani}, {Khochfar}, {Silk}, \&
  {Kay}}]{Kaviraj2009}
{Kaviraj}, S., {Peirani}, S., {Khochfar}, S., {Silk}, J., \& {Kay}, S. 2009,
  \mnras, 394, 1713

\bibitem[{{Knobel} {et~al.}(2009){Knobel}, {Lilly}, {Iovino}, {Porciani},
  {Kova{\v{c}}}, {Cucciati}, {Finoguenov}, {Kitzbichler}, {Carollo}, {Contini},
  {Kneib}, {Le F{\`e}vre}, {Mainieri}, {Renzini}, {Scodeggio}, {Zamorani},
  {Bardelli}, {Bolzonella}, {Bongiorno}, {Caputi}, {Coppa}, {de la Torre}, {de
  Ravel}, {Franzetti}, {Garilli}, {Kampczyk}, {Lamareille}, {Le Borgne}, {Le
  Brun}, {Maier}, {Mignoli}, {Pello}, {Peng}, {Perez Montero}, {Ricciardelli},
  {Silverman}, {Tanaka}, {Tasca}, {Tresse}, {Vergani}, {Zucca}, {Abbas},
  {Bottini}, {Cappi}, {Cassata}, {Cimatti}, {Fumana}, {Guzzo}, {Koekemoer},
  {Leauthaud}, {Maccagni}, {Marinoni}, {McCracken}, {Memeo}, {Meneux}, {Oesch},
  {Pozzetti}, \& {Scaramella}}]{Knobel2009}
{Knobel}, C., {Lilly}, S.~J., {Iovino}, A., {et~al.} 2009, \apj, 697, 1842

\bibitem[{{Kova{\v{c}}} {et~al.}(2010){Kova{\v{c}}}, {Lilly}, {Knobel},
  {Bolzonella}, {Iovino}, {Carollo}, {Scarlata}, {Sargent}, {Cucciati},
  {Zamorani}, {Pozzetti}, {Tasca}, {Scodeggio}, {Kampczyk}, {Peng}, {Oesch},
  {Zucca}, {Finoguenov}, {Contini}, {Kneib}, {Le F{\`e}vre}, {Mainieri},
  {Renzini}, {Bardelli}, {Bongiorno}, {Caputi}, {Coppa}, {de la Torre}, {de
  Ravel}, {Franzetti}, {Garilli}, {Lamareille}, {Le Borgne}, {Le Brun},
  {Maier}, {Mignoli}, {Pello}, {Perez Montero}, {Ricciardelli}, {Silverman},
  {Tanaka}, {Tresse}, {Vergani}, {Abbas}, {Bottini}, {Cappi}, {Cassata},
  {Cimatti}, {Fumana}, {Guzzo}, {Koekemoer}, {Leauthaud}, {Maccagni},
  {Marinoni}, {McCracken}, {Memeo}, {Meneux}, {Porciani}, {Scaramella}, \&
  {Scoville}}]{Peng2010}
{Kova{\v{c}}}, K., {Lilly}, S.~J., {Knobel}, C., {et~al.} 2010, \apj, 718, 86

\bibitem[{{Krywult} {et~al.}(2017){Krywult}, {Tasca}, {Pollo}, {Vergani},
  {Bolzonella}, {Davidzon}, {Iovino}, {Gargiulo}, {Haines}, {Scodeggio},
  {Guzzo}, {Zamorani}, {Garilli}, {Granett}, {de la Torre}, {Abbas}, {Adami},
  {Bottini}, {Cappi}, {Cucciati}, {Franzetti}, {Fritz}, {Le Brun}, {Le
  F{\`e}vre}, {Maccagni}, {Ma{\l}ek}, {Marulli}, {Polletta}, {Tojeiro},
  {Zanichelli}, {Arnouts}, {Bel}, {Branchini}, {Coupon}, {De Lucia}, {Ilbert},
  {McCracken}, {Moscardini}, \& {Takeuchi}}]{Krywult2017}
{Krywult}, J., {Tasca}, L.~A.~M., {Pollo}, A., {et~al.} 2017, \aap, 598, A120

\bibitem[{{La Barbera} {et~al.}(2014){La Barbera}, {Pasquali}, {Ferreras},
  {Gallazzi}, {de Carvalho}, \& {de la Rosa}}]{LaBarbera2014}
{La Barbera}, F., {Pasquali}, A., {Ferreras}, I., {et~al.} 2014, \mnras, 445,
  1977

\bibitem[{{Leja} {et~al.}(2017){Leja}, {Johnson}, {Conroy}, {van Dokkum}, \&
  {Byler}}]{Leja2017}
{Leja}, J., {Johnson}, B.~D., {Conroy}, C., {van Dokkum}, P.~G., \& {Byler}, N.
  2017, \apj, 837, 170

\bibitem[{{Lonoce} {et~al.}(2014){Lonoce}, {Longhetti}, {Saracco}, {Gargiulo},
  \& {Tamburri}}]{Lonoce2014}
{Lonoce}, I., {Longhetti}, M., {Saracco}, P., {Gargiulo}, A., \& {Tamburri}, S.
  2014, \mnras, 444, 2048

\bibitem[{{L{\'o}pez Fern{\'a}ndez} {et~al.}(2018){L{\'o}pez Fern{\'a}ndez},
  {Gonz{\'a}lez Delgado}, {P{\'e}rez}, {Garc{\'\i}a-Benito}, {Cid Fernandes},
  {Schoenell}, {S{\'a}nchez}, {Gallazzi}, {S{\'a}nchez-Bl{\'a}zquez}, {Vale
  Asari}, \& {Walcher}}]{LopezFernandez2018}
{L{\'o}pez Fern{\'a}ndez}, R., {Gonz{\'a}lez Delgado}, R.~M., {P{\'e}rez}, E.,
  {et~al.} 2018, \aap, 615, A27

\bibitem[{{Maraston} {et~al.}(2009){Maraston}, {Nieves Colmen{\'a}rez},
  {Bender}, \& {Thomas}}]{Maraston2009}
{Maraston}, C., {Nieves Colmen{\'a}rez}, L., {Bender}, R., \& {Thomas}, D.
  2009, \aap, 493, 425

\bibitem[{{Marigo} {et~al.}(2013){Marigo}, {Bressan}, {Nanni}, {Girardi}, \&
  {Pumo}}]{Marigo2013}
{Marigo}, P., {Bressan}, A., {Nanni}, A., {Girardi}, L., \& {Pumo}, M.~L. 2013,
  \mnras, 434, 488

\bibitem[{{Martins} {et~al.}(2005){Martins}, {Gonz{\'a}lez Delgado},
  {Leitherer}, {Cervi{\~n}o}, \& {Hauschildt}}]{Martins2005}
{Martins}, L.~P., {Gonz{\'a}lez Delgado}, R.~M., {Leitherer}, C.,
  {Cervi{\~n}o}, M., \& {Hauschildt}, P. 2005, \mnras, 358, 49

\bibitem[{{M{\'e}ndez-Abreu} {et~al.}(2018){M{\'e}ndez-Abreu}, {Costantin},
  {Aguerri}, {de Lorenzo-C{\'a}ceres}, \& {Corsini}}]{MendezAbreu2018}
{M{\'e}ndez-Abreu}, J., {Costantin}, L., {Aguerri}, J.~A.~L., {de
  Lorenzo-C{\'a}ceres}, A., \& {Corsini}, E.~M. 2018, \mnras, 479, 4172

\bibitem[{{M{\'e}ndez-Abreu} {et~al.}(2012){M{\'e}ndez-Abreu},
  {S{\'a}nchez-Janssen}, {Aguerri}, {Corsini}, \&
  {Zarattini}}]{MendezAbreu2012}
{M{\'e}ndez-Abreu}, J., {S{\'a}nchez-Janssen}, R., {Aguerri}, J.~A.~L.,
  {Corsini}, E.~M., \& {Zarattini}, S. 2012, \apjl, 761, L6

\bibitem[{{Morelli} {et~al.}(2015){Morelli}, {Corsini}, {Pizzella}, {Dalla
  Bont{\`a}}, {Coccato}, \& {M{\'e}ndez-Abreu}}]{Morelli2015}
{Morelli}, L., {Corsini}, E.~M., {Pizzella}, A., {et~al.} 2015, \mnras, 452,
  1128

\bibitem[{{Nelson} {et~al.}(2015){Nelson}, {Pillepich}, {Genel},
  {Vogelsberger}, {Springel}, {Torrey}, {Rodriguez-Gomez}, {Sijacki}, {Snyder},
  {Griffen}, {Marinacci}, {Blecha}, {Sales}, {Xu}, \& {Hernquist}}]{Nelson2015}
{Nelson}, D., {Pillepich}, A., {Genel}, S., {et~al.} 2015, Astronomy and
  Computing, 13, 12

\bibitem[{{Ocvirk} {et~al.}(2006){Ocvirk}, {Pichon}, {Lan{\c c}on}, \&
  {Thi{\'e}baut}}]{Ocvirk2006}
{Ocvirk}, P., {Pichon}, C., {Lan{\c c}on}, A., \& {Thi{\'e}baut}, E. 2006,
  \mnras, 365, 74

\bibitem[{{Onodera} {et~al.}(2012){Onodera}, {Renzini}, {Carollo},
  {Cappellari}, {Mancini}, {Strazzullo}, {Daddi}, {Arimoto}, {Gobat}, {Yamada},
  {McCracken}, {Ilbert}, {Capak}, {Cimatti}, {Giavalisco}, {Koekemoer}, {Kong},
  {Lilly}, {Motohara}, {Ohta}, {Sanders}, {Scoville}, {Tamura}, \&
  {Taniguchi}}]{Onodera2012}
{Onodera}, M., {Renzini}, A., {Carollo}, M., {et~al.} 2012, \apj, 755, 26

\bibitem[{{Pacifici} {et~al.}(2016){Pacifici}, {Kassin}, {Weiner}, {Holden},
  {Gardner}, {Faber}, {Ferguson}, {Koo}, {Primack}, {Bell}, {Dekel}, {Gawiser},
  {Giavalisco}, {Rafelski}, {Simons}, {Barro}, {Croton}, {Dav{\'e}}, {Fontana},
  {Grogin}, {Koekemoer}, {Lee}, {Salmon}, {Somerville}, \&
  {Behroozi}}]{Pacifici2016}
{Pacifici}, C., {Kassin}, S.~A., {Weiner}, B.~J., {et~al.} 2016, \apj, 832, 79

\bibitem[{{Pandya} {et~al.}(2017){Pandya}, {Brennan}, {Somerville}, {Choi},
  {Barro}, {Wuyts}, {Taylor}, {Behroozi}, {Kirkpatrick}, {Faber}, {Primack},
  {Koo}, {McIntosh}, {Kocevski}, {Bell}, {Dekel}, {Fang}, {Ferguson}, {Grogin},
  {Koekemoer}, {Lu}, {Mantha}, {Mobasher}, {Newman}, {Pacifici}, {Papovich},
  {van der Wel}, \& {Yesuf}}]{Pandya2017}
{Pandya}, V., {Brennan}, R., {Somerville}, R.~S., {et~al.} 2017, \mnras, 472,
  2054

\bibitem[{{P{\'e}rez-Gonz{\'a}lez} {et~al.}(2008){P{\'e}rez-Gonz{\'a}lez},
  {Rieke}, {Villar}, {Barro}, {Blaylock}, {Egami}, {Gallego}, {Gil de Paz},
  {Pascual}, {Zamorano}, \& {Donley}}]{PerezGonzalez2008}
{P{\'e}rez-Gonz{\'a}lez}, P.~G., {Rieke}, G.~H., {Villar}, V., {et~al.} 2008,
  \apj, 675, 234

\bibitem[{{Poggianti} {et~al.}(2009){Poggianti}, {Fasano}, {Bettoni}, {Cava},
  {Dressler}, {Vanzella}, {Varela}, {Couch}, {D'Onofrio}, {Fritz},
  {Kjaergaard}, {Moles}, \& {Valentinuzzi}}]{Poggianti2009}
{Poggianti}, B.~M., {Fasano}, G., {Bettoni}, D., {et~al.} 2009, \apjl, 697,
  L137

\bibitem[{{S{\'a}nchez-Bl{\'a}zquez} {et~al.}(2006){S{\'a}nchez-Bl{\'a}zquez},
  {Peletier}, {Jim{\'e}nez-Vicente}, {Cardiel}, {Cenarro},
  {Falc{\'o}n-Barroso}, {Gorgas}, {Selam}, \& {Vazdekis}}]{SanchezBlazquez2006}
{S{\'a}nchez-Bl{\'a}zquez}, P., {Peletier}, R.~F., {Jim{\'e}nez-Vicente}, J.,
  {et~al.} 2006, \mnras, 371, 703

\bibitem[{{Sandage}(1986)}]{Sandage1986}
{Sandage}, A. 1986, \aap, 161, 89

\bibitem[{{Sarzi} {et~al.}(2006){Sarzi}, {Falc{\'o}n-Barroso}, {Davies},
  {Bacon}, {Bureau}, {Cappellari}, {de Zeeuw}, {Emsellem}, {Fathi},
  {Krajnovi{\'c}}, {Kuntschner}, {McDermid}, \& {Peletier}}]{Sarzi2006}
{Sarzi}, M., {Falc{\'o}n-Barroso}, J., {Davies}, R.~L., {et~al.} 2006, \mnras,
  366, 1151

\bibitem[{{Schaye} {et~al.}(2015){Schaye}, {Crain}, {Bower}, {Furlong},
  {Schaller}, {Theuns}, {Dalla Vecchia}, {Frenk}, {McCarthy}, {Helly},
  {Jenkins}, {Rosas-Guevara}, {White}, {Baes}, {Booth}, {Camps}, {Navarro},
  {Qu}, {Rahmati}, {Sawala}, {Thomas}, \& {Trayford}}]{Schaye2015}
{Schaye}, J., {Crain}, R.~A., {Bower}, R.~G., {et~al.} 2015, \mnras, 446, 521

\bibitem[{{Schiavon} {et~al.}(2006){Schiavon}, {Faber}, {Konidaris}, {Graves},
  {Willmer}, {Weiner}, {Coil}, {Cooper}, {Davis}, {Harker}, {Koo}, {Newman}, \&
  {Yan}}]{Schiavon2006}
{Schiavon}, R.~P., {Faber}, S.~M., {Konidaris}, N., {et~al.} 2006, \apjl, 651,
  L93

\bibitem[{{Serven} {et~al.}(2005){Serven}, {Worthey}, \& {Briley}}]{Serven2005}
{Serven}, J., {Worthey}, G., \& {Briley}, M.~M. 2005, \apj, 627, 754

\bibitem[{{Spilker} {et~al.}(2018){Spilker}, {Bezanson}, {Bari{\v s}i{\'c}},
  {Bell}, {Lagos}, {Maseda}, {Muzzin}, {Pacifici}, {Sobral}, {Straatman}, {van
  der Wel}, {van Dokkum}, {Weiner}, {Whitaker}, {Williams}, \&
  {Wu}}]{Spilker2018}
{Spilker}, J., {Bezanson}, R., {Bari{\v s}i{\'c}}, I., {et~al.} 2018, \apj,
  860, 103

\bibitem[{{Takada} {et~al.}(2014){Takada}, {Ellis}, {Chiba}, {Greene},
  {Aihara}, {Arimoto}, {Bundy}, {Cohen}, {Dor{\'e}}, {Graves}, {Gunn},
  {Heckman}, {Hirata}, {Ho}, {Kneib}, {Le F{\`e}vre}, {Lin}, {More},
  {Murayama}, {Nagao}, {Ouchi}, {Seiffert}, {Silverman}, {Sodr{\'e}},
  {Spergel}, {Strauss}, {Sugai}, {Suto}, {Takami}, \& {Wyse}}]{Takada2014}
{Takada}, M., {Ellis}, R.~S., {Chiba}, M., {et~al.} 2014, \pasj, 66, R1

\bibitem[{{Thomas} {et~al.}(2005){Thomas}, {Maraston}, {Bender}, \& {Mendes de
  Oliveira}}]{Thomas2005}
{Thomas}, D., {Maraston}, C., {Bender}, R., \& {Mendes de Oliveira}, C. 2005,
  \apj, 621, 673

\bibitem[{{Thomas} {et~al.}(2010){Thomas}, {Maraston}, {Schawinski}, {Sarzi},
  \& {Silk}}]{Thomas2010}
{Thomas}, D., {Maraston}, C., {Schawinski}, K., {Sarzi}, M., \& {Silk}, J.
  2010, \mnras, 404, 1775

\bibitem[{{Tojeiro} {et~al.}(2009){Tojeiro}, {Wilkins}, {Heavens}, {Panter}, \&
  {Jimenez}}]{Tojeiro2009}
{Tojeiro}, R., {Wilkins}, S., {Heavens}, A.~F., {Panter}, B., \& {Jimenez}, R.
  2009, \apjs, 185, 1

\bibitem[{{Trager} {et~al.}(2000){Trager}, {Faber}, {Worthey}, \&
  {Gonz{\'a}lez}}]{Trager2000}
{Trager}, S.~C., {Faber}, S.~M., {Worthey}, G., \& {Gonz{\'a}lez}, J.~J. 2000,
  \aj, 120, 165

\bibitem[{{van der Wel} {et~al.}(2016){van der Wel}, {Noeske}, {Bezanson},
  {Pacifici}, {Gallazzi}, {Franx}, {Mu{\~n}oz-Mateos}, {Bell}, {Brammer},
  {Charlot}, {Chauk{\'e}}, {Labb{\'e}}, {Maseda}, {Muzzin}, {Rix}, {Sobral},
  {van de Sande}, {van Dokkum}, {Wild}, \& {Wolf}}]{vanderWel2016}
{van der Wel}, A., {Noeske}, K., {Bezanson}, R., {et~al.} 2016, \apjs, 223, 29

\bibitem[{{Vazdekis} {et~al.}(2016){Vazdekis}, {Koleva}, {Ricciardelli},
  {R{\"o}ck}, \& {Falc{\'o}n-Barroso}}]{Vazdekis2016}
{Vazdekis}, A., {Koleva}, M., {Ricciardelli}, E., {R{\"o}ck}, B., \&
  {Falc{\'o}n-Barroso}, J. 2016, \mnras, 463, 3409

\bibitem[{{Worthey} {et~al.}(1994){Worthey}, {Faber}, {Gonzalez}, \&
  {Burstein}}]{Worthey1994}
{Worthey}, G., {Faber}, S.~M., {Gonzalez}, J.~J., \& {Burstein}, D. 1994,
  \apjs, 94, 687

\bibitem[{{Wright}(2006)}]{Wright2006}
{Wright}, E.~L. 2006, \pasp, 118, 1711

\bibitem[{{Wu} {et~al.}(2018){Wu}, {van der Wel}, {Gallazzi}, {Bezanson},
  {Pacifici}, {Straatman}, {Franx}, {Bari{\v s}i{\'c}}, {Bell}, {Brammer},
  {Calhau}, {Chauke}, {van Houdt}, {Maseda}, {Muzzin}, {Rix}, {Sobral},
  {Spilker}, {van de Sande}, {van Dokkum}, \& {Wild}}]{Wu2018}
{Wu}, P.-F., {van der Wel}, A., {Gallazzi}, A., {et~al.} 2018, \apj, 855, 85

\bibitem[{{York} {et~al.}(2000){York}, {Adelman}, {Anderson}, {Anderson},
  {Annis}, {Bahcall}, {Bakken}, {Barkhouser}, {Bastian}, {Berman}, {Boroski},
  {Bracker}, {Briegel}, {Briggs}, {Brinkmann}, {Brunner}, {Burles}, {Carey},
  {Carr}, {Castander}, {Chen}, {Colestock}, {Connolly}, {Crocker}, {Csabai},
  {Czarapata}, {Davis}, {Doi}, {Dombeck}, {Eisenstein}, {Ellman}, {Elms},
  {Evans}, {Fan}, {Federwitz}, {Fiscelli}, {Friedman}, {Frieman}, {Fukugita},
  {Gillespie}, {Gunn}, {Gurbani}, {de Haas}, {Haldeman}, {Harris}, {Hayes},
  {Heckman}, {Hennessy}, {Hindsley}, {Holm}, {Holmgren}, {Huang}, {Hull},
  {Husby}, {Ichikawa}, {Ichikawa}, {Ivezi{\'c}}, {Kent}, {Kim}, {Kinney},
  {Klaene}, {Kleinman}, {Kleinman}, {Knapp}, {Korienek}, {Kron}, {Kunszt},
  {Lamb}, {Lee}, {Leger}, {Limmongkol}, {Lindenmeyer}, {Long}, {Loomis},
  {Loveday}, {Lucinio}, {Lupton}, {MacKinnon}, {Mannery}, {Mantsch}, {Margon},
  {McGehee}, {McKay}, {Meiksin}, {Merelli}, {Monet}, {Munn}, {Narayanan},
  {Nash}, {Neilsen}, {Neswold}, {Newberg}, {Nichol}, {Nicinski}, {Nonino},
  {Okada}, {Okamura}, {Ostriker}, {Owen}, {Pauls}, {Peoples}, {Peterson},
  {Petravick}, {Pier}, {Pope}, {Pordes}, {Prosapio}, {Rechenmacher}, {Quinn},
  {Richards}, {Richmond}, {Rivetta}, {Rockosi}, {Ruthmansdorfer}, {Sandford},
  {Schlegel}, {Schneider}, {Sekiguchi}, {Sergey}, {Shimasaku}, {Siegmund},
  {Smee}, {Smith}, {Snedden}, {Stone}, {Stoughton}, {Strauss}, {Stubbs},
  {SubbaRao}, {Szalay}, {Szapudi}, {Szokoly}, {Thakar}, {Tremonti}, {Tucker},
  {Uomoto}, {Vanden Berk}, {Vogeley}, {Waddell}, {Wang}, {Watanabe},
  {Weinberg}, {Yanny}, {Yasuda}, \& {SDSS Collaboration}}]{York2010}
{York}, D.~G., {Adelman}, J., {Anderson}, John~E., J., {et~al.} 2000, \aj, 120,
  1579

\bibitem[{{Zahid} {et~al.}(2014){Zahid}, {Dima}, {Kudritzki}, {Kewley},
  {Geller}, {Hwang}, {Silverman}, \& {Kashino}}]{Zahid2014}
{Zahid}, H.~J., {Dima}, G.~I., {Kudritzki}, R.-P., {et~al.} 2014, \apj, 791,
  130

\bibitem[{{Zibetti} {et~al.}(2017){Zibetti}, {Gallazzi}, {Ascasibar},
  {Charlot}, {Galbany}, {Garc{\'{\i}}a Benito}, {Kehrig}, {de
  Lorenzo-C{\'a}ceres}, {Lyubenova}, {Marino}, {M{\'a}rquez}, {S{\'a}nchez},
  {van de Ven}, {Walcher}, \& {Wisotzki}}]{Zibetti2017}
{Zibetti}, S., {Gallazzi}, A.~R., {Ascasibar}, Y., {et~al.} 2017, \mnras, 468,
  1902

\end{thebibliography}

\end{document}